\documentclass[twocolumn]{aastex61}
\usepackage{comment}

\received{August 12, 2019}
\revised{September 16, 2019}
\accepted{September 17, 2019}

\newcommand{\kms}{\mbox{km s$^{-1}$}}
\newcommand{\Msun}{\mbox{$M_{\odot}$}}

\newcommand{\hii}{\mbox{H\,{\sc ii}}}

\newcommand{\x}{\mbox{$\times$}}

\shorttitle{The ALMA Survey of 70 $\mu$m dark High-mass clumps in Early Stages (ASHES)}
\shortauthors{Sanhueza et al.}

\begin{document}


\title{The ALMA Survey of 70 $\mu$m dark High-mass clumps in Early Stages (ASHES). \\I. Pilot Survey: Clump Fragmentation}

\correspondingauthor{Patricio Sanhueza}
\email{patricio.sanhueza@nao.ac.jp}

\author{Patricio Sanhueza}
\affil{National Astronomical Observatory of Japan, National Institutes of Natural Sciences, 2-21-1 Osawa, Mitaka, Tokyo 181-8588, Japan}

\author{Yanett Contreras}
\affil{Leiden Observatory, Leiden University, PO Box 9513, NL-2300 RA Leiden, the Netherlands}

\author{Benjamin Wu}
\affil{National Astronomical Observatory of Japan, National Institutes of Natural Sciences, 2-21-1 Osawa, Mitaka, Tokyo 181-8588, Japan}

\author{James M. Jackson}
\affil{SOFIA Science Center, USRA, NASA Ames Research Center, Moffett Field CA 94045, USA}

\author{Andr\'es E. Guzm\'an}
\affil{National Astronomical Observatory of Japan, National Institutes of Natural Sciences, 2-21-1 Osawa, Mitaka, Tokyo 181-8588, Japan}

\author{Qizhou Zhang}
\affil{Harvard-Smithsonian Center for Astrophysics, 60 Garden Street, Cambridge, MA 02138, USA}

\author{Shanghuo Li}
\affil{Shanghai Astronomical Observatory, Chinese Academy of Sciences, 80 Nandan Road, Shanghai 200030, China}
\affil{Harvard-Smithsonian Center for Astrophysics, 60 Garden Street, Cambridge, MA 02138, USA}
\affil{University of Chinese Academy of Sciences, 19A Yuquanlu, Beijing 100049, China}

\author{Xing Lu}
\affil{National Astronomical Observatory of Japan, National Institutes of Natural Sciences, 2-21-1 Osawa, Mitaka, Tokyo 181-8588, Japan}

\author{Andrea Silva}
\affil{National Astronomical Observatory of Japan, National Institutes of Natural Sciences, 2-21-1 Osawa, Mitaka, Tokyo 181-8588, Japan}

\author{Natsuko Izumi}
\affil{National Astronomical Observatory of Japan, National Institutes of Natural Sciences, 2-21-1 Osawa, Mitaka, Tokyo 181-8588, Japan}
\affil{College of Science, Ibaraki University, 2-1-1 Bunkyo, Mito, Ibaraki 310-8512, Japan}

\author{Tie Liu}
\affil{Shanghai Astronomical Observatory, Chinese Academy of Sciences, 80 Nandan Road, Shanghai 200030, People's Republic of China}

\author{Rie E. Miura}
\affil{National Astronomical Observatory of Japan, National Institutes of Natural Sciences, 2-21-1 Osawa, Mitaka, Tokyo 181-8588, Japan}

\author{Ken'ichi Tatematsu}
\affil{National Astronomical Observatory of Japan, National Institutes of Natural Sciences, 2-21-1 Osawa, Mitaka, Tokyo 181-8588, Japan}

\author{Takeshi Sakai}
\affil{Graduate School of Informatics and Engineering, The University of Electro-Communications, Chofu, Tokyo 182-8585, Japan.}

\author{Henrik Beuther}
\affil{Max Planck Institute for Astronomy, K{\"o}nigstuhl 17, 69117, Heidelberg, Germany}

\author{Guido Garay}
\affil{Departamento de Astronom\'ia, Universidad de Chile, Camino el Observatorio 1515, Las Condes, Santiago, Chile}

\author{Satoshi Ohashi}
\affil{National Astronomical Observatory of Japan, National Institutes of Natural Sciences, 2-21-1 Osawa, Mitaka, Tokyo 181-8588, Japan}
\affil{RIKEN Cluster for Pioneering Research, 2-1, Hirosawa, Wako-shi, Saitama 351-0198, Japan}

\author{Masao Saito}
\affil{National Astronomical Observatory of Japan, National Institutes of Natural Sciences, 2-21-1 Osawa, Mitaka, Tokyo 181-8588, Japan}

\author{Fumitaka Nakamura}
\affil{National Astronomical Observatory of Japan, National Institutes of Natural Sciences, 2-21-1 Osawa, Mitaka, Tokyo 181-8588, Japan}

\author{Kazuya Saigo}
\affil{National Astronomical Observatory of Japan, National Institutes of Natural Sciences, 2-21-1 Osawa, Mitaka, Tokyo 181-8588, Japan}

\author{V. S. Veena}
\affil{Physikalisches Institut, Universit\"at zu K\"oln, Z\"ulpicher Str. 77, 50937 K\"oln, Germany}

\author{Quang Nguyen-Luong}
\affil{IBM Canada, 120 Bloor Street East, Toronto, ON, M4Y 1B7, Canada}
\affil{National Astronomical Observatory of Japan, National Institutes of Natural Sciences, 2-21-1 Osawa, Mitaka, Tokyo 181-8588, Japan}

\author{Daniel Tafoya}
\affil{Department of Space, Earth and Environment, Chalmers University of Technology, Onsala Space Observatory, 439~92 Onsala, Sweden}


\begin{abstract}

The ALMA Survey of 70 $\mu$m dark High-mass clumps in Early Stages (ASHES) has been designed to systematically 
characterize the earliest stages and to constrain theories of high-mass star formation. A deep understanding of high-mass star 
formation requires the study of the clustered mode, which is the most commonly found in nature. A total of 12 massive ($>$500 \Msun), 
cold ($\leq$15 K), 3.6-70 $\mu$m dark prestellar clump candidates, embedded in infrared dark clouds (IRDCs), were carefully selected
 in the pilot survey to be observed with the 
  Atacama Large Millimeter/sub-millimeter Array (ALMA). Exploiting the unique capabilities of ALMA, we have mosaiced 
   each clump ($\sim$1 arcmin$^2$) in dust continuum and line emission with the 12 m, 7 m, and Total Power arrays at 224 GHz (1.34 mm), resulting
   in $\sim$1\farcs 2 angular resolution ($\sim$4800 AU at the average source distance of 4 kpc). As the first paper of the series, we concentrate 
   on the dust continuum emission to reveal the clump fragmentation. We have detected a total of 294 cores, from which 
   84 (29\%) are categorized as protostellar based on outflow activity or ``warm core" line emission. The remaining 210 (71\%) 
   are considered prestellar core candidates. The number of detected cores is independent of the mass sensitivity range of the observations and, on average, more
    massive clumps tend to form more cores. We find no correlation between the mass of the host clump and the most massive embedded 
    core. We find a large population of low-mass ($<$1 \Msun) cores and no high-mass ($>$30 \Msun) prestellar cores. The most massive 
    prestellar core has a mass of 11 \Msun. From the prestellar core mass function, we derive a power law index of 1.17 $\pm$ 0.10, slightly shallower than 
    the Salpeter index of 1.35. We have used the minimum spanning tree technique to characterize the separation between cores and their 
    spatial distribution, and to derive mass segregation ratios. 
    While there is a range of core masses and core separations detected in the sample, the mean separation and mean mass of cores per clump are well 
    explained by thermal fragmentation and are inconsistent with turbulent Jeans fragmentation. The core spatial distribution is well described by hierarchical subclustering 
    rather than centrally peaked clustering. There is no conclusive evidence of mass segregation. We have tested several theoretical conditions, 
    and conclude that overall, competitive accretion and global hierarchical collapse scenarios are favored over the turbulent
      core accretion scenario.

\end{abstract}


\section{Introduction}
\label{intro}

Several key questions in high-mass star formation focus on the early fragmentation of prestellar massive clumps\footnote{Consistent with 
\cite{Sanhueza17}, throughout this work we use the term ``clump'' to refer to a dense object within an IRDC with a size of 
the order $\sim$0.2--1 pc, a mass of $\sim$10$^2$--10$^3$ \Msun, and a volume density of $\sim$10$^4$--10$^5$ cm$^{-3}$ that will form a 
stellar cluster. We use the term ``core'' to describe a compact, dense object within a clump with a size of $\sim$0.01--0.1 pc, a mass of 
 $\sim$10$^{-1}$-10$^2$ \Msun, and a volume density $\gtrsim$10$^5$ cm$^{-3}$ that will form a single star or close binary system.}. 
Prestellar cores embedded in massive clumps at any evolutionary stage are rare and their observational characterization is ultimately
 needed to constrain model predictions.  Are prestellar core masses segregated with the more massive cores
  preferentially located toward the clump center? Do high-mass prestellar cores ($\gtrsim$30 \Msun) exist early on? Is the prestellar core mass function (CMF) Salpeter-like? 
  All these basic questions have not been possible to address in the past yet 
   are a necessary step before digging in to the detailed internal physics and chemistry of prestellar cores at $\lesssim$1000 AU scales, as has been recently
    done in nearby, low-mass prestellar cores \citep{Ohashi18,Caselli19}. High-mass stars form in clustered environments and the initial imprints of the core spatial
     distribution and mass segregation, as well as the prestellar CMF, found at the early clump fragmentation are important components for cluster formation simulations.
  
Theories that attempt to explain the formation mechanisms of clusters along with high-mass stars fall into two broad categories: ``clump-fed" and ``core-fed". 
In the ``clump-fed" category, competitive accretion scenarios \citep{Bonnell04,Bonnell06,Smith09,Wang10} and global hierarchical collapse
 \citep{Heitsch2008,Vazquez09,Vazquez17,Vazquez19,Ballesteros11a,Ballesteros11b,Ballesteros18} are included, which are mostly consistent with each other \citep[][see this work for a detailed discussion on the similarities and differences.]{Vazquez19}.
 These scenarios are characterized by global clump infall and simulations predict the formation of clusters along with high-mass stars. 
Fragmentation produces low-mass cores (mass $\sim$ Jeans mass) that acquire mass through gas infall from their parent structures (clumps). Those cores placed in 
preferential locations, near the center of the forming cluster gravitational potential, increase their masses to become massive enough to form high-mass stars. Given that the cores at early times have
 masses near the Jeans mass, the CMF evolves due to accretion to become the universal initial mass function (IMF)  
later on. In these ``clump-fed'' scenarios, the core distribution is expected to be hierarchical and because the cores that are the seeds of high-mass stars are near the center of the gravitational potential
 of the cluster-forming clump, 
primordial mass segregation is predicted \citep[e.g.,][]{Bonnell06}. 

Conversely, the ``core-fed" turbulent core accretion scenario \citep{McKee03} treats the
 formation of high-mass stars in isolated environments rather than as part of cluster formation, but it is supported by numerical simulations of cluster formation \cite[e.g.,][]{Krumholz12,Myers14}. In the 
 turbulent core accretion model, global infall 
 is gradual \citep{Tan06}, allowing quasi-equilibrium structures during their assembly, and does not contribute to the core mass. The core mass is fixed at the early fragmentation 
and, because the core is near virial equilibrium, the core mass is approximately constant over time. In order to form high-mass stars, high-mass prestellar cores must exist \citep{Tan13,Tan14}. Therefore, the
 turbulent core accretion theory predicts a direct relationship between the CMF and the IMF. The CMF would resemble the IMF but shifted to higher masses by an efficiency factor that would be independent
  of the core mass \citep[similar to what has been 
 postulated in nearby, low-mass star-forming regions, e.g.,][]{Alves07,Andre10,Konyves15}. No specific prediction is made on the spatial core distribution and \cite{Tan18} points out that the massive cores may or may not be at the 
 center of cluster-forming clumps (therefore, no specific prediction on primordial mass segregation). However, numerical simulations that reproduce the predicted accretion rates from this scenario find 
 primordial mass segregation \citep{Myers14}. These outlines of the high-mass star formation scenarios oversimplify their physical and chemical 
 complexity. For finer details, the following reviews by \cite{Krumholz09}, \cite{Tan14}, and \cite{Vazquez19} are suggested. 

 Comprehensive studies to address the previously posed questions and test theories backed by large samples have been historically challenging 
 mostly due to two factors. First, it is difficult to identify prestellar, massive clump candidates that can form high-mass stars. Second, after selecting 
suitable targets, the detection of the weak dust and molecular line emission of the cold, distant candidates require time-expensive 
observations at high-angular resolution, precluding systematical studies of large samples until recently. Before the ALMA era, high-angular resolution studies of massive clumps at early 
evolutionary stages mostly targeted individual regions with the Submillinter Array (SMA), Plateau de Bure Interferometer (PdBI, now NOEMA), 
Combined Array for Research in Millimeter-wave Astronomy (CARMA), and Very Large Array (VLA) in different array configurations and gas tracers that 
 make analysis taken as a whole complicated  \citep[e.g.][]{Zhang09,Pillai11,Pillai19,Wang12,Wang14,Sanhueza13,Sanhueza17,Beuther15,Lu15,Feng16a,Feng16b}.
  ALMA has finally made possible the study of large samples to achieve statistically significant conclusions in uniform fashion (e.g., similar array configurations,
   analysis strategies, and gas tracers). 

The preferred targets to study the earliest stages of high-mass star formation are infrared dark clouds (IRDCs), molecular clouds seen as dark silhouettes against the
 Galactic 8 $\mu$m mid-infrared background in Galactic plane surveys, e.g., using {\it MSX} in \cite{Simon06} and {\it Spitzer} in \cite{Peretto09}. Among IRDCs, those that 
 are also 24 and 70 $\mu$m dark are colder and denser than other IRDCs \citep{Guzman15} and are believed to trace the earliest stages of high-mass star formation
  \citep{Sanhueza13,Sanhueza17,Tan13,Contreras18}. However, lack of 24 and 70 $\mu$m  
 emission does not guarantee a complete absence of star formation activity \citep[e.g.,][]{Tan16,Feng16b}. 
  Several studies have investigated the kinematics and filamentary structure of IRDCs \citep{Busquet13,Henshaw14,Henshaw16,Foster14,Liu14,Ragan15,Contreras16,Lu18,Liu18c,Chen19}, their chemistry \citep{Sanhueza12,Sanhueza13,Sakai08,Sakai12,Sakai15,Hoq13,Miettinen14,Vasyunina14,Feng16a,Kong16,Tatematsu17}, molecular outflow content 
  \citep{Sanhueza10,Wang11,Wang14,Lu15,Kong19}, infall \citep{Sanhueza10,Contreras18,Liu18a}, magnetic fields  \citep{Pillai15,Beuther18a,Liu18a,Juvela18,Tang19}, and in the more evolved ones, ultracompact
   (UC) \hii\ regions \citep{Battersby10,Avison15}, thermal ionized jets \citep{Rosero14,Rosero16,Rosero19}, hot cores \citep{Rathborne08,Sakai13,Csengeri18},  
    and maser emission  \citep{Pillai06,Wang06,Chambers09,Yanagida14}.
   
  IRDC clumps that lack star formation indicators (UC \hii\ regions, molecular outflows, hot cores, maser emission) are prime candidates to be in the prestellar phase. Although the source 
  selection in this work is explained in detail in Section~\ref{source_selection}, the selection of prestellar massive clump candidates generally consists of the following combined 
  effort at different wavelengths: (i) categorization of prestellar/protostellar phase based on large IR surveys, GLIMPSE \citep{Benjamin03} based on {\it Spitzer}/IRAC 3--8 $\mu$m 
  emission, MIPSGAL \citep{Carey09} based on  {\it Spitzer}/MIPS 24--70 $\mu$m emission, and Hi-GAL \citep{Molinari10} using {\it Herschel}/PACS 70 $\mu$m emission, (ii) clump
   mass and temperature calculation using SED fitting of dust emission usually from Hi-GAL {\it Herschel}/SPIRE 250--500 $\mu$m and ATLASGAL using APEX 870 $\mu$m 
   \citep[e.g.,][]{Guzman15,Traficante15,Contreras17}, (iii) kinematical
    information to obtain distances and hints of active star formation (based on outlflows, chemistry, maser detection, high-temperatures) from large molecular line surveys, e.g., MALT90
     \citep{Foster11,Jackson13}, \cite{Shirley13}, \cite{Wienen15}, and RAMPS \citep{Hogge18}.
  
In this work, we present the pilot Alma Survey of 70 $\mu$m dark High-mass clumps in Early Stages (ASHES). A deep understanding of high-mass star formation
 requires the study of the clustered mode, which is the most commonly found in nature. We have therefore  
 mosaiced 12 prestellar, massive clump candidates in dust continuum and molecular line emission at $\sim$224 GHz ($\sim$1\farcs2 resolution) using the 12 m, 7m, and TP arrays of ALMA.  Here we focus
  on the clump fragmentation using the dust continuum emission to characterize the earliest stages of high-mass star formation and constrain theory. The core dynamics, based on an analysis of 
   C$^{18}$O, DCO$^+$, and N$_2$D$^+$ emission, is presented in a companion paper \citep{Contreras19}. The molecular outflow content will be presented by Li et al., in prep. ``Warm core'' 
   line emission will be presented by Izumi et al., in prep. 
   
\section{Source Selection: Prestellar (70 $\mu$m dark), High-Mass Clump Candidates}
\label{source_selection}

The identification of prestellar (70 $\mu$m dark), high-mass ($>$500 \Msun) clump candidates has substantially improved with the advent of
 {\it Spitzer} and {\it Herschel} satellites and ground-based dust continuum and molecular line surveys. For the ASHES pilot survey, 
 11 IRDC clumps were selected from the Millimetre Astronomy Legacy Team 90 GHz Survey
  \citep[MALT90;][]{Foster11,Jackson13,Foster13}. MALT90 was built on the ATLASGAL 870 $\mu$m catalogues \citep{Schuller09,Contreras13}, from 
  which a sample of 3246 high-mass clumps was selected for follow-up in 16 spectral lines. The first MALT90 line catalogue was presented in 
   \cite{Rathborne16} and several studies have taken advantage of the molecular line data \citep[e.g.,][]{Hoq13,Miettinen14,He15,He16,Yu15,Stephens15,Stephens16,Contreras16,Jackson18,Jackson19,Li19a,Li19b}.
    By combining {\it Herschel} and ATLASGAL dust continuum emission observations, \cite{Guzman15} derived temperatures and column densities for the MALT90 survey targets. 
    After determining clump kinematical distances \citep{Whitaker17}, masses and 
    number densities were calculated by \cite{Contreras17}. With all these vast ancillary multi-wavelength data sets, 
    we made a careful selection of prestellar clumps candidates that will potentially form high-mass stars. 

In \cite{Guzman15}, we first identify IR-dark clumps from 3.6 to 70 $\mu$m in { {\it Spitzer}/{\it Herschel} (see Figures~\ref{IR-plots1},~\ref{IR-plots2},~\ref{IR-plots3}, and~\ref{IR-plots4}). The presence of IR compact
 emission indicates embedded sources in the protostellar phase, while their absence makes the clump a prestellar candidate.  From 3246 sources,
  only 83 sources fulfill the latter 
  requirement. This small fraction of potentially prestellar clumps demonstrates the rarity, and presumably, short lifetime of the high-mass prestellar phase. To ensure
   the selection of the best prestellar candidates with sufficient mass to form high-mass stars,  we impose additional selection criteria, 
   clumps must have: (1) dust temperatures equal to or lower than the average temperature of the 70 $\mu$m-dark sub-sample, i.e., $\leq$15 K, (2) masses larger than 
   500 \Msun, (3) have number densities $\gtrsim$10$^{4}$ cm$^{-3}$, and (4) molecular line
    emission from MALT90 consistent with cold gas, i.e., no shock (SiO) or hot core (HC$_3$N, CH$_3$CN, and HNCO) emission.
     To ensure good spatial resolution, an additional constraint is that the targets are closer than 5.5 kpc. Only 18  
     sources satisfy these conditions and 11 were observed in this pilot survey. The 12$^{\rm th}$ target in the ASHES pilot survey is G028.273--00.167, which is in the 
     first quadrant and was not covered in MALT90. This IRDC satisfies all previous requirements and has been well studied in the past by 
     \cite{Sanhueza12,Sanhueza13,Sanhueza17}. Key physical properties for all 12 IRDC clumps are listed in Table~\ref{tbl-clumps}: Columns (1-3) contain clump names with 
     their coordinates, Columns (4-5) contain the $V_{\rm lsr}$ and velcity dispersion of the gas ($\sigma$) determined by using high-density tracers with critical densities $>$10$^5$ cm$^{-3}$, and 
     Columns (6-8) are the clump properties considered for target selection (see further details in the notes of Table~\ref{tbl-clumps}). Columns (9-12) are described in the following paragraph.

\begin{deluxetable*}{lccccccccccc}
\tabletypesize{\footnotesize}
\tablecaption{Physical Properties of the Prestellar, High-Mass Clump Candidates \label{tbl-clumps}}
\tablewidth{0pt}
\tablehead{
\colhead{IRDC\tablenotemark{a}} & \multicolumn{2}{c}{\underline {~~~~~~~~~~Position\tablenotemark{b}~~~~~~~~~~}}     & \colhead{$V_{\rm lsr}$}  & \colhead{$\sigma$\tablenotemark{c}}  &\colhead{Dist.}     &  
\colhead{$T_{\rm dust}$}  & \colhead{Mass} & \colhead{$R_{\rm cl}$}   & \colhead{$M_{\rm cl}$} & \colhead{$\Sigma_{\rm cl}$} & \colhead{$n_{\rm cl}({\rm H_2})$}\\
\colhead{Clump} &  \colhead{$\alpha$(J2000)} & \colhead{$\delta$(J2000)}& \colhead{(km s$^{-1}$)}& \colhead{(km s$^{-1}$)}&\colhead{(kpc)}                 &\colhead{(K)} & \colhead{(\Msun)} & \colhead{(pc [\arcsec])} & \colhead{(\Msun)} & \colhead{(gr cm$^{-2}$)} & \colhead{($\times$10$^4$ cm$^{-3}$)}\\
\colhead{(1)} &  \colhead{(2)} & \colhead{(3)}& \colhead{(4)}&\colhead{(5)}   &\colhead{(6)}              &\colhead{(7)} & \colhead{(8)} & \colhead{(9)} & \colhead{(10)} & \colhead{(11)} & \colhead{(12)}\\
}
\startdata
G010.991--00.082 & 18:10:06.65 & $-$19.27.50.7                           & 29.5  	&	1.27	& 3.7   & 12.0 &  2230 &      0.49 (27) &  1810	& 0.50 	& 	 5.3\\
G014.492--00.139 & 18:17:22.03 & $-$16.25.01.9                            & 41.1   	&	1.68	& 3.9   & 13.0 &  5200 &    0.44  (23)   & 3120	& 1.1 	& 	13\\
G028.273--00.167 & 18:43:31.00 & $-$04.13.18.1                            & 80.0   	&	0.81	& 5.1  & 12.0 & 1520 &      0.59 (24)  &1520 	& 0.28 	& 	2.4\\
G327.116--00.294 & 15:50:57.18 & $-$54.30.33.6                            &  $-$58.9 &	0.56	& 3.9  & 14.3 &  580 &      0.39 (20)   & 580	& 0.26 	&	3.5\\
G331.372--00.116 & 16:11:34.10 & $-$51.35.00.1                             &  $-$87.8 &	1.29	& 5.4  & 14.0 & 1640 &      0.63 (24)  & 1230	& 0.20 	&	1.7\\
G332.969--00.029 & 16:18:31.61 & $-$50.25.03.1                            & $-$66.6  &	1.41	& 4.4 &  12.6 & 730 &       0.59 (28)   & 530	& 0.10 	& 	0.9\\
G337.541--00.082 & 16:37:58.48 & $-$47.09.05.1                            & $-$54.6 &	2.01	& 4.0 &  12.0 & 1180 &        0.42 (22) & 1040	& 0.40 	& 	5.0\\
G340.179--00.242  & 16:48:40.88 & $-$45.16.01.1                           & $-$53.7 &	1.48	& 4.1 &  14.0 &  1470 &        0.74 (37) & 1020	& 0.12 	&	0.9\\
G340.222--00.167  & 16:48:30.83 & $-$45.11.05.8                           & $-$51.3 &		3.04	& 4.0  &  15.0 &  760 &        0.36 (19) & 720	& 0.38 	&	5.5\\
G340.232--00.146  & 16:48:27.56 & $-$45.09.51.9                           & $-$50.8 &	1.23	& 3.9  &  14.0 &  710 &        0.48 (25) & 520	& 0.15 	& 	1.7\\
G341.039--00.114 & 16:51:14.11 & $-$44.31.27.2                            & $-$43.0 &		0.97	& 3.6  &  14.3 & 1070 &        0.47 (27) & 850	& 0.26 	&	2.9\\
G343.489--00.416 & 17:01:01.19 & $-$42.48.11.0                            & $-$29.0 &		1.00	& 2.9 &  10.3 & 810 &        0.42 (29)   & 790	& 0.30 	&	3.8\\
\enddata
\tablecomments{Properties in Column (6), (7), and (8) were used for source selection. Clump properties for G028.273--00.167, also known as G028.23--00.19, were derived by
 \cite{Sanhueza12,Sanhueza13}. Clump properties for G010.991--00.082 and G014.492--00.139 were calculated using the column densities from \cite{Guzman15} and the distances  
 derived according to \cite{Whitaker17}.
 Clump properties for the remaining 9 clumps 
were derived and presented in a series of works by the MALT90 team: \cite[][temperatures]{Guzman15}, \cite[][V$_{\rm lsr}$]{Rathborne16}, \cite[][masses]{Contreras17}, and
 \cite[][distances]{Whitaker17}. Due to multiple velocities along the line of sight, and based in the C$^{18}$O emission, the masses of G331.372--00.116 and G332.969--00.029 could be lower in   
$\sim$25\% and $\sim$50\%, respectively.}
\tablecomments{Properties in Column (9), (10), (11), and (12) are used for clump analysis through this work. $R_{\rm cl}$ was derived from Gaussian fitting to the dust continuum 
emission from ATLASGAL and $M_{\rm cl}$ scaled from Column (8) using the integrated flux derived in the Gaussian fitting. The clump surface density, column (11), is calculated as  $\Sigma_{\rm cl}=M_{\rm cl}/\pi R_{\rm cl}^2$. The volume density, Column (12), was calculated assuming a spherical clump of radius $R_{\rm cl}$ and using the
  molecular weight per hydrogen molecule ($\mu_{\rm H_2}$) of 2.8. }
 \tablenotetext{a}{By replacing G in the IRDC name for AGAL, the name of the source has the same nomenclature as in the ATLASGAL catalog \citep{Schuller09}.} 
 \tablenotetext{b}{Phase center for ALMA mosaics. Due to the positioning 
 of the mosaic, the phase center and the ATLASGAL catalog coordinates are slightly different in few arcsecs.}
\tablenotemark{c}{Velocity dispersion was obtained using NH$_2$D J$_{K_a,K_b}$ = $1_{1,1}-1_{0,1}$ emission for G028.273--00.167, HNC J = 1-0 emission for G337.541--00.082 and G340.222--00.167, and  N$_2$H$^+$ J = 1-0 emission for the remaining 9 clumps. All these three molecular tracers have critical densities $>$10$^{5}$ cm$^{-3}$ \citep{Sanhueza12}.}
\end{deluxetable*}

\begin{figure*}
\begin{center}
\includegraphics[angle=0,scale=0.45]{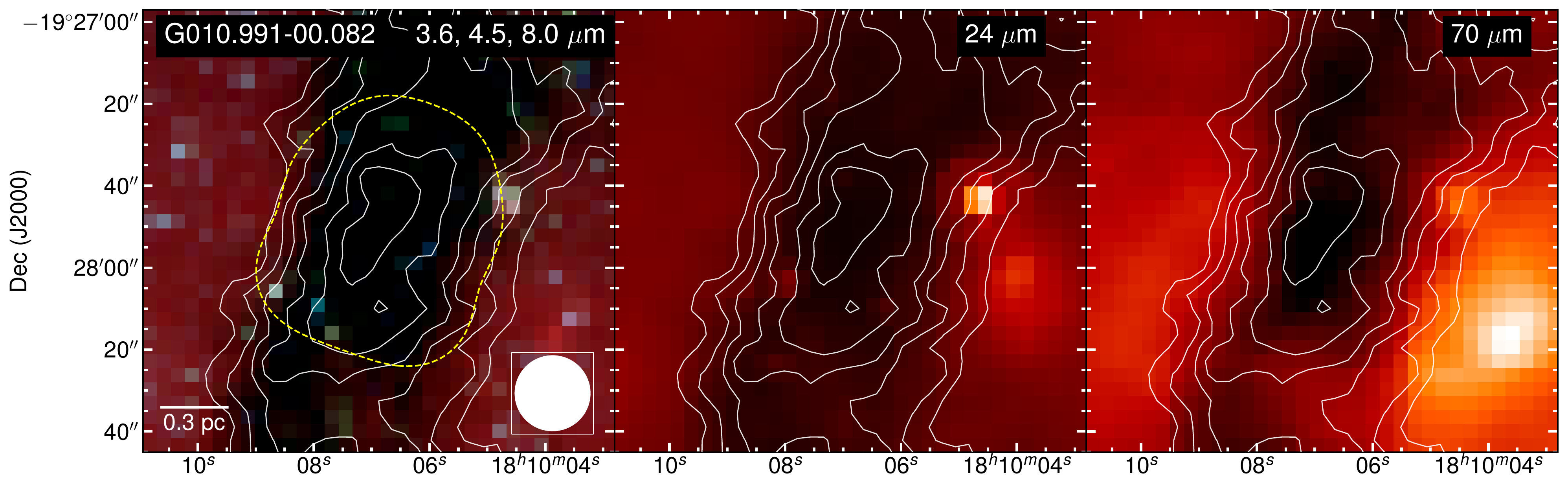}\vspace{-0.1cm}
\includegraphics[angle=0,scale=0.45]{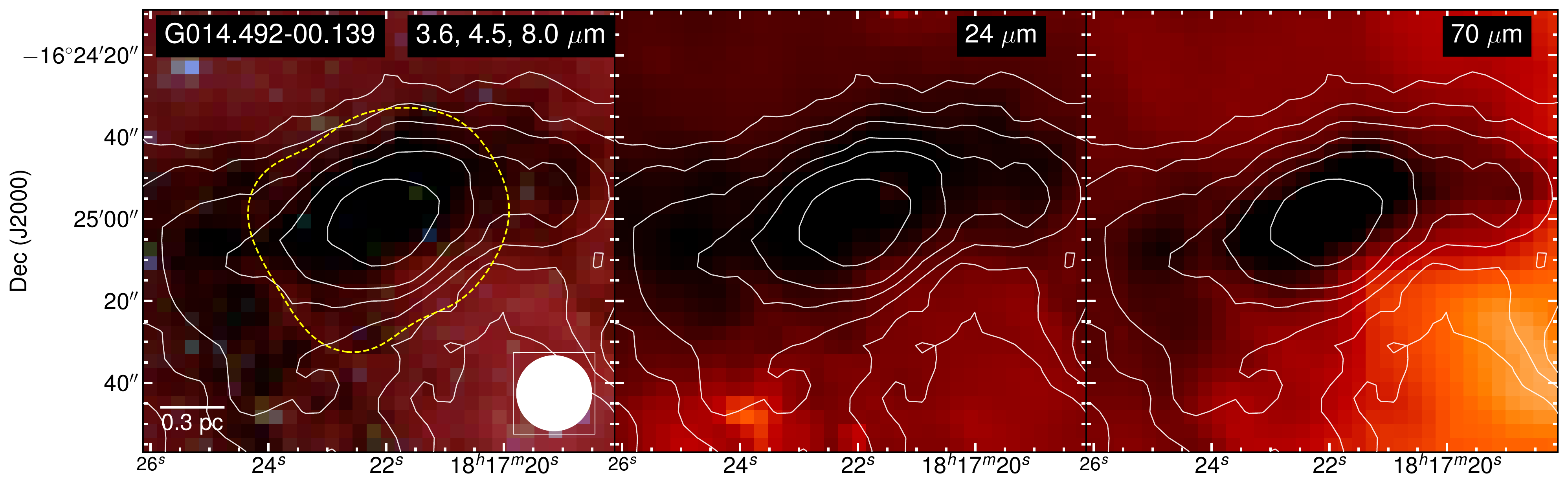}\vspace{-0.1cm}
\includegraphics[angle=0,scale=0.45]{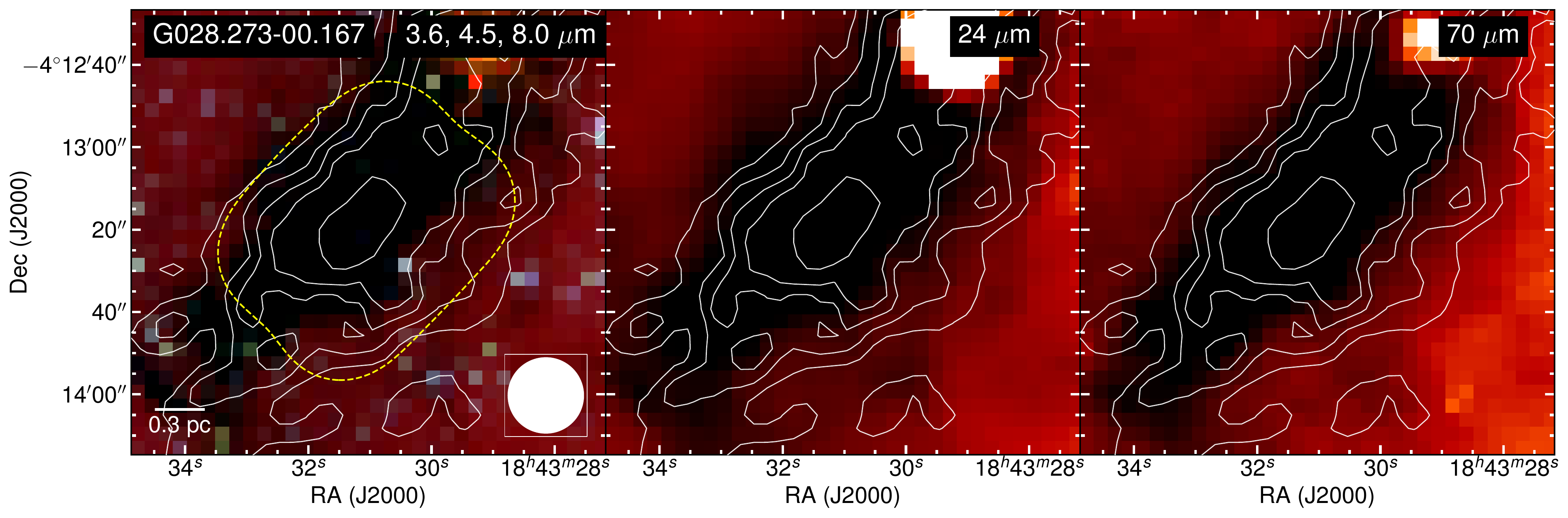}
\end{center}
\caption{ {\it Spitzer} and {\it Herschel} IR images for observed IRDC clumps overlaid with 870 $\mu$m dust continuum emission from the ATLASGAL survey  
(19\farcs2; shown on the bottom right of the first panel) in white contours. Left: {\it Spitzer}/IRAC 3-color (3.6 $\mu$m in blue, 4.5 $\mu$m in green, and 8.0 $\mu$m in red) image. Dashed 
yellow contour delineates the area mosaiced with ALMA. Center:
 {\it Spitzer}/MIPS 24 $\mu$m image. Right: {\it Herschel}/PACS 70 $\mu$m image. Contour levels for the 870 $\mu$m dust continuum emission are:
  3, 5, 7, 9, 12, and 15 \x\ $\sigma$, with $\sigma = 71.8$ mJy beam$^{-1}$, for G010.991--00.082; 
  3, 6, 9, 13, 17, and 25 \x\ $\sigma$, with $\sigma = 82.7$ mJy beam$^{-1}$, for G014.492--00.139; and 
  3, 5, 7, 9, and 12 \x\ $\sigma $, with $\sigma = 64.9$ mJy beam$^{-1}$, for G028.273--00.167.}
\label{IR-plots1}
\end{figure*}

\begin{figure*}
\begin{center}
\includegraphics[angle=0,scale=0.45]{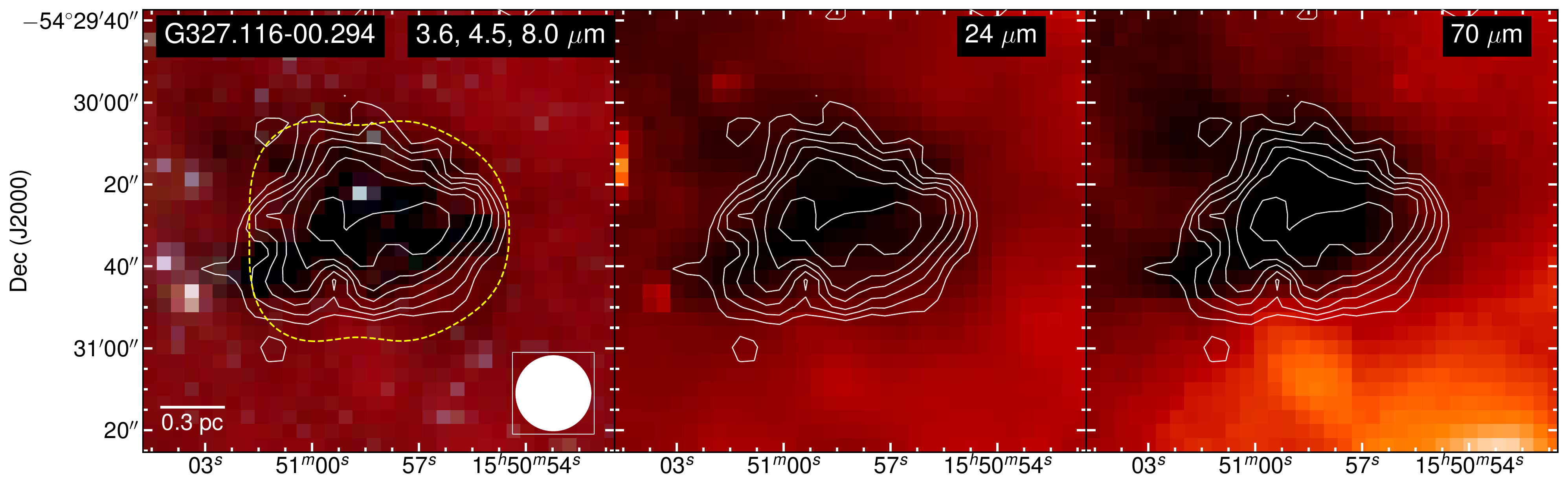}\vspace{-0.1cm}
\includegraphics[angle=0,scale=0.45]{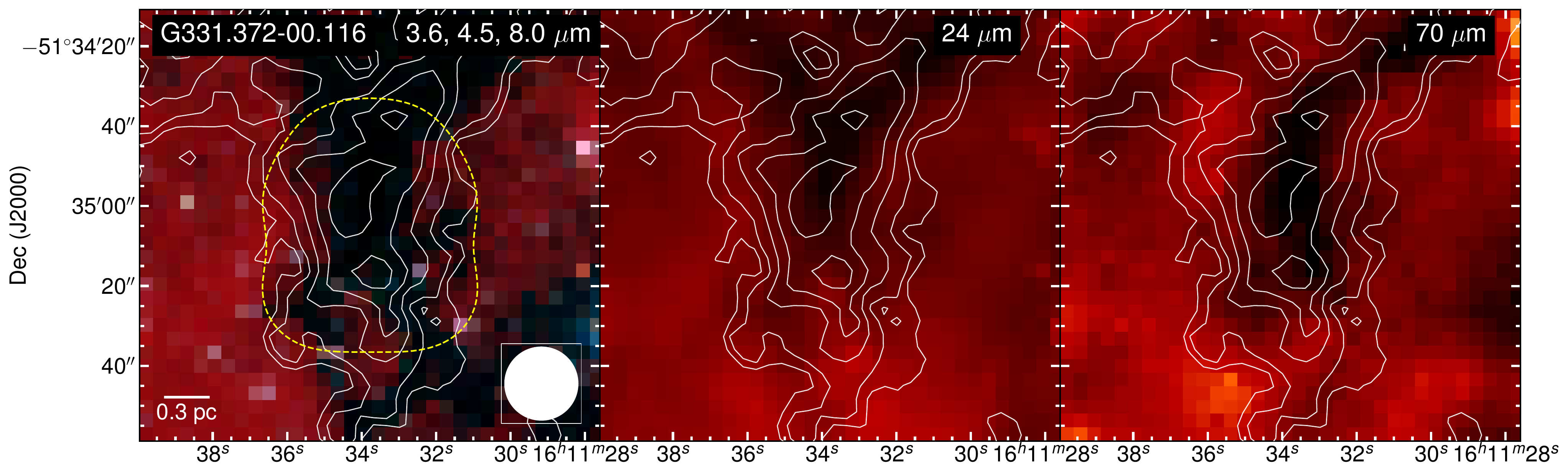}\vspace{-0.1cm}
\includegraphics[angle=0,scale=0.45]{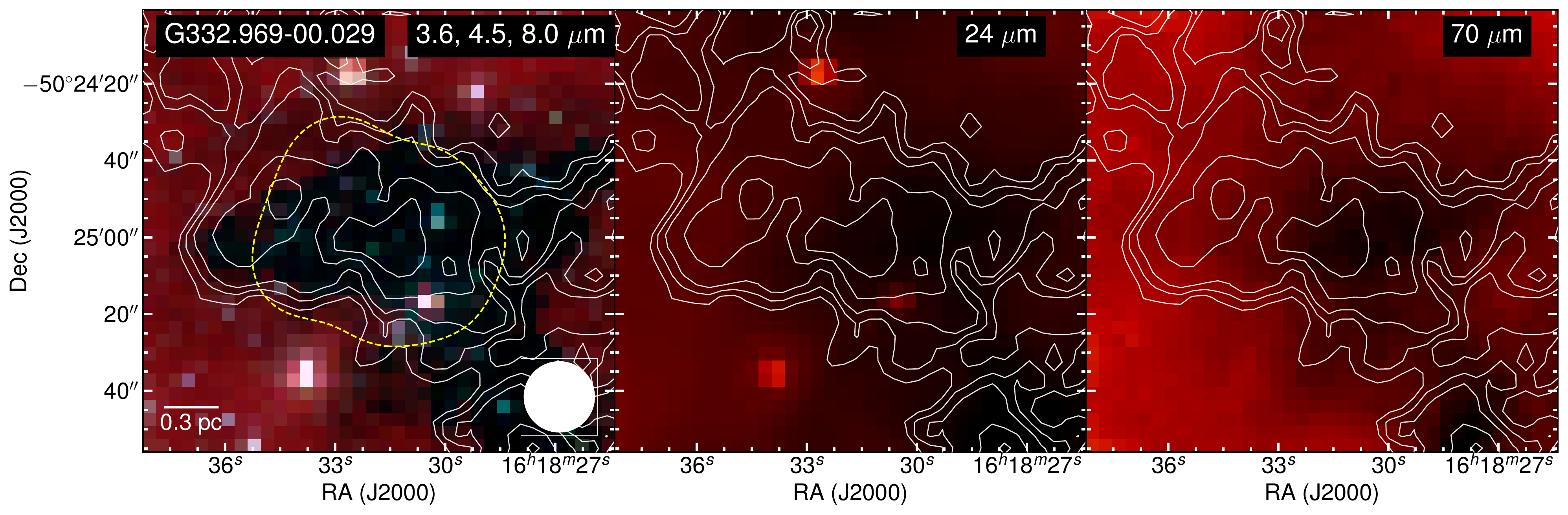}
\end{center}
\caption{Same as in Figure~\ref{IR-plots1}, except for contour levels for the 870 $\mu$m dust continuum emission, which are:
  3, 4, 5, 6, 7, and 9 \x\ $\sigma$, with $\sigma = 70.9$ mJy beam$^{-1}$, for G327.116--00.294; 
  3, 4, 6, 8, and 10 \x\ $\sigma$, with $\sigma = 56.5$ mJy beam$^{-1}$, for G331.372--00.116; and 
  3, 4, 5, 7, and 9 \x\ $\sigma $, with $\sigma = 46.9$ mJy beam$^{-1}$, for G332.969--00.029.}
\label{IR-plots2}
\end{figure*}

\begin{figure*}
\begin{center}
\includegraphics[angle=0,scale=0.45]{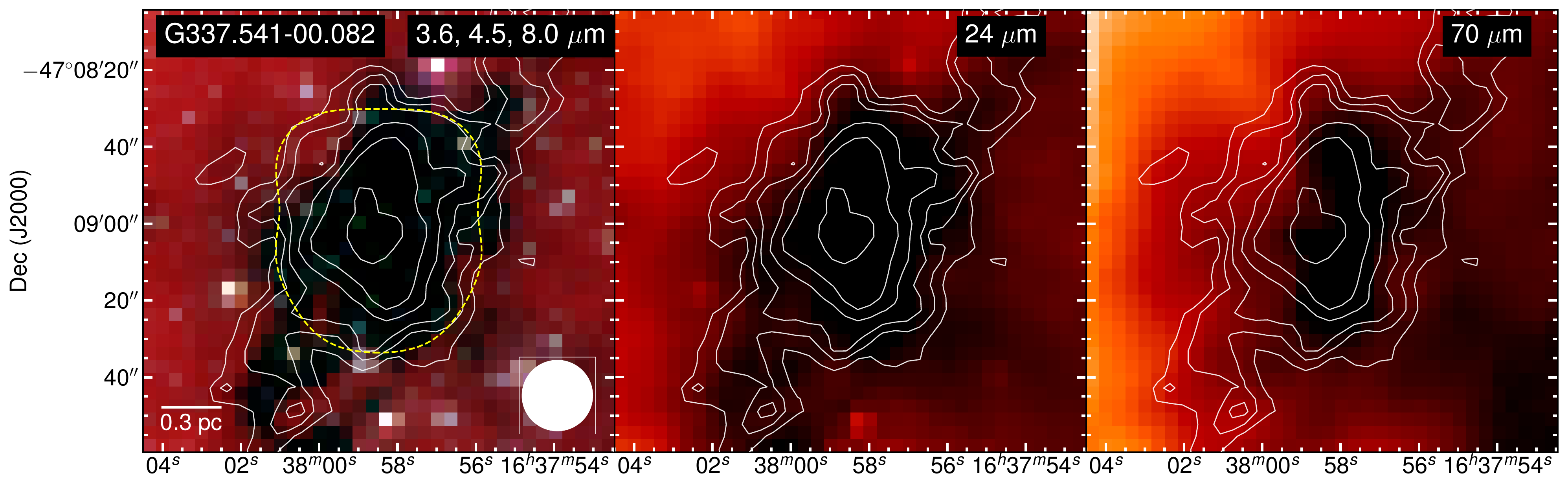}\vspace{-0.1cm}
\includegraphics[angle=0,scale=0.45]{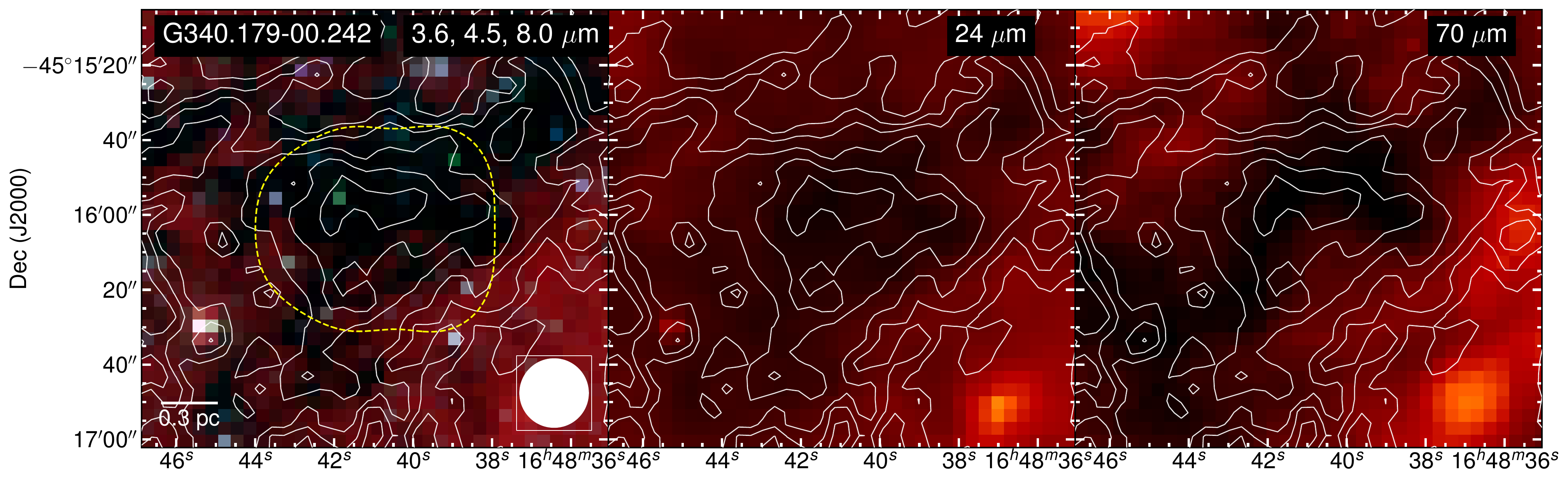}\vspace{-0.1cm}
\includegraphics[angle=0,scale=0.45]{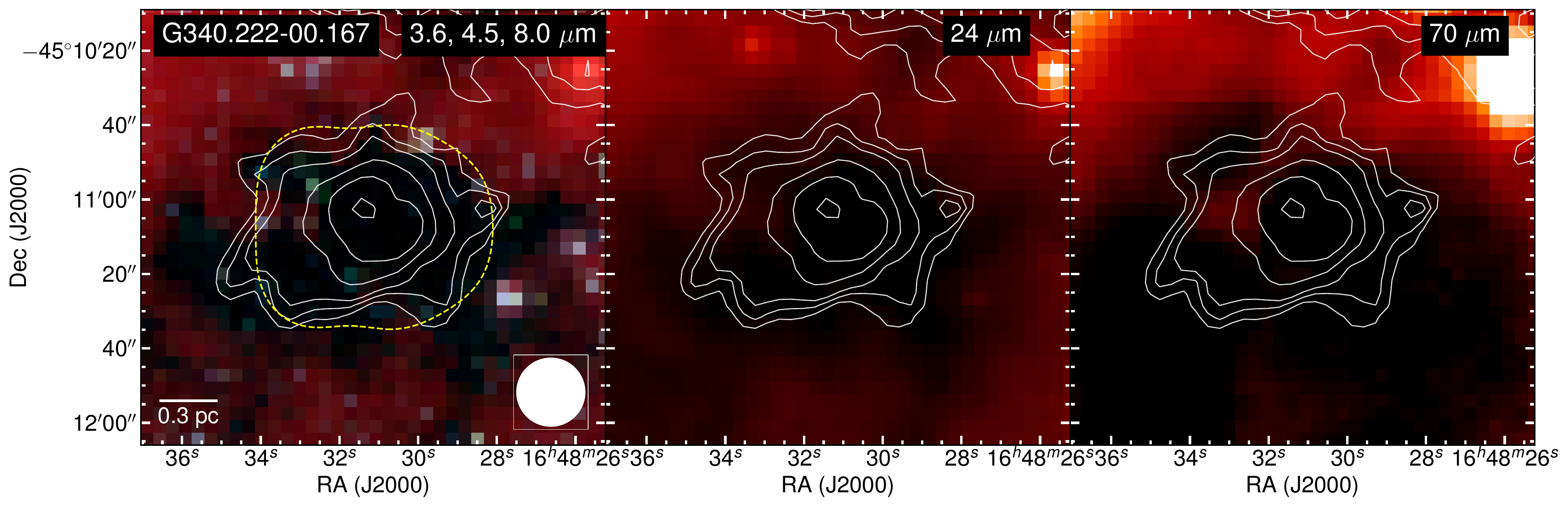}
\end{center}
\caption{Same as in Figure~\ref{IR-plots1}, except for contour levels for the 870 $\mu$m dust continuum emission, which are:
  3, 4, 5, 7, 9, and 12 \x\ $\sigma$, with $\sigma = 66.3$ mJy beam$^{-1}$, for G337.541--00.082; 
  3, 4, 5, 7, 9, and 11 \x\ $\sigma$, with $\sigma = 57.3$ mJy beam$^{-1}$, for G340.179--00.242; and 
  3, 4, 5, 7, 9, and 12 \x\ $\sigma $, with $\sigma = 65.7$ mJy beam$^{-1}$, for G340.222--00.167.}
\label{IR-plots3}
\end{figure*}

\begin{figure*}
\begin{center}
\includegraphics[angle=0,scale=0.45]{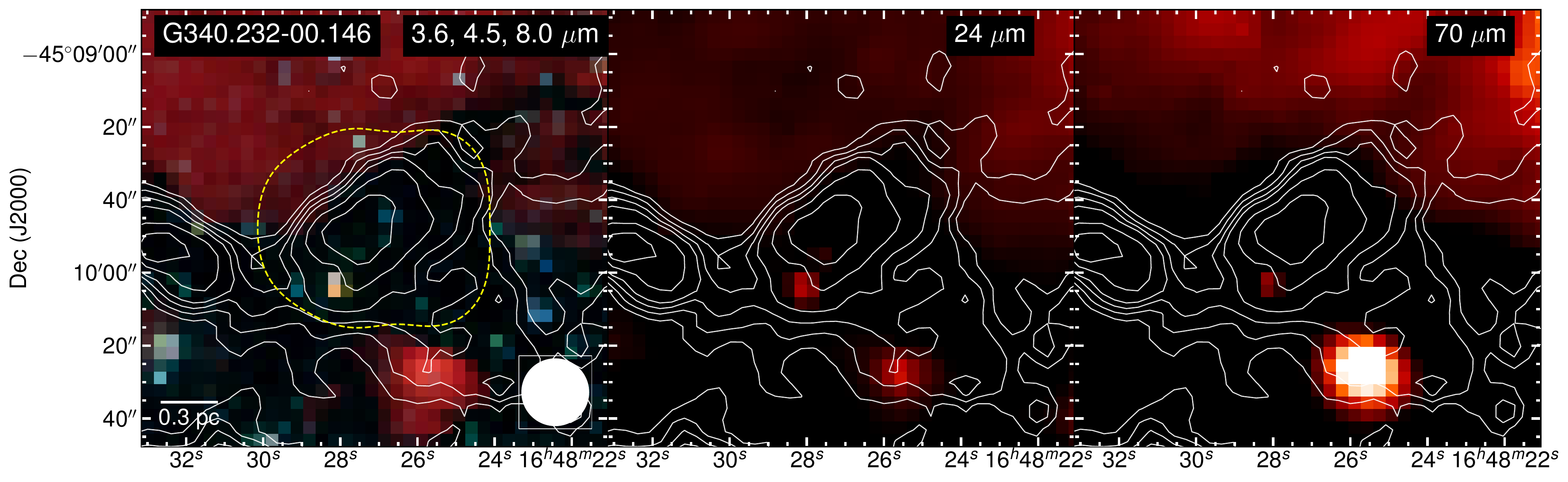}\vspace{-0.1cm}
\includegraphics[angle=0,scale=0.45]{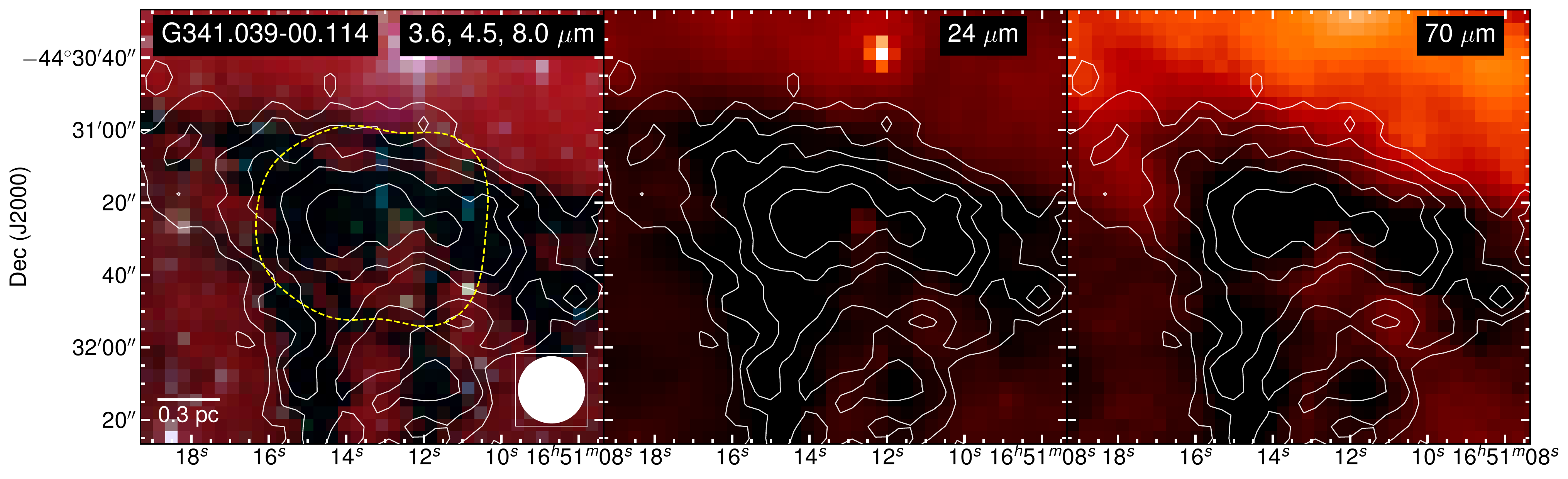}\vspace{-0.1cm}
\includegraphics[angle=0,scale=0.45]{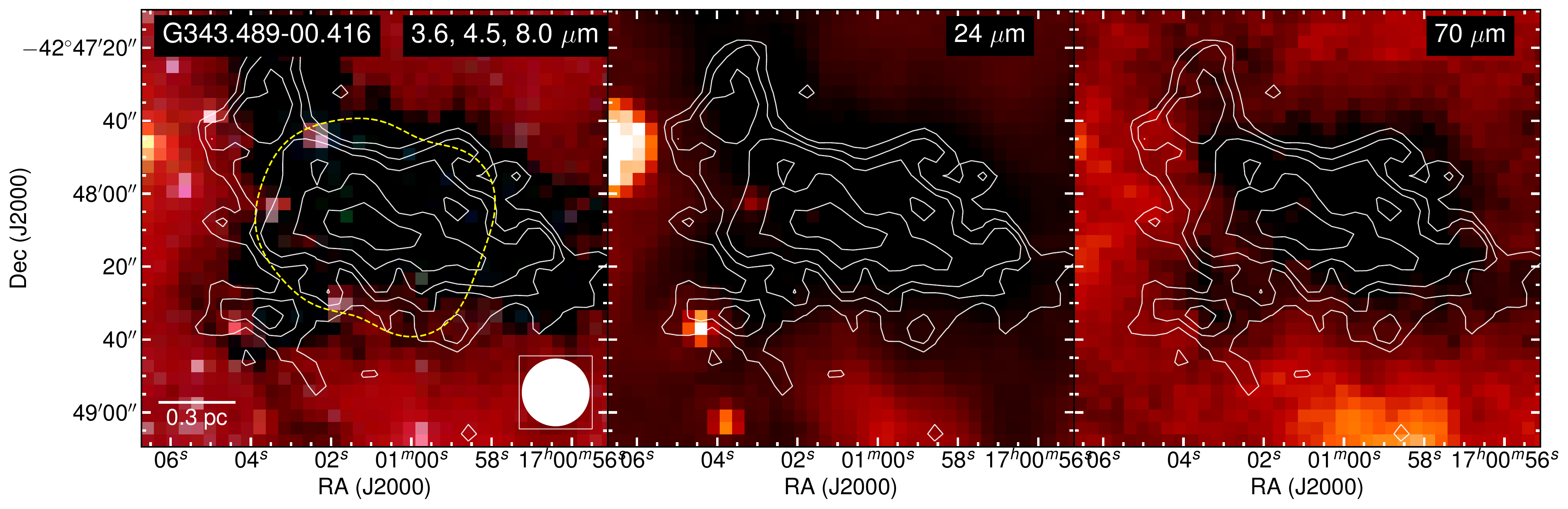}
\end{center}
\caption{Same as in Figure~\ref{IR-plots1}, except for contour levels for the 870 $\mu$m dust continuum emission, which are:
  3, 4, 5, 6, 8, and 10 \x\ $\sigma$, with $\sigma = 65.1$ mJy beam$^{-1}$, for G340.232--00.146; 
  3, 5, 7, 9, and 12 \x\ $\sigma$, with $\sigma = 52.2$ mJy beam$^{-1}$, for G341.039--00.114; and 
  3, 4, 5, 7, and 9 \x\ $\sigma $, with $\sigma =53.9$ mJy beam$^{-1}$, for G343.489--00.416.}
\label{IR-plots4}
\end{figure*}

While the source selection was based on the MALT90 properties, we refine the size and mass of the clumps to be more representative of the observed region with ALMA.
We determine clump sizes by performing Gaussian fitting to the ATLASGAL 870 $\mu$m dust emission maps and define $R_{\rm cl}$, Column (9) in Table~\ref{tbl-clumps}, 
as the geometric mean of the semi-major and semi-minor FWHM. Consequently, we scale the MALT90 mass to a new clump mass $M_{\rm cl}$, Column (10) in Table~\ref{tbl-clumps}, 
based on the measured integrated flux from the Gaussian fitting and the flux inside the mask defining the MALT90 source. $R_{\rm cl}$ and $M_{\rm cl}$, which also define 
the surface density ($\Sigma_{\rm cl}$) and volume density ($n_{\rm cl}({\rm H_2})$) in Column (11-12) in Table~\ref{tbl-clumps}, will be used throughout this work.
     
     We note that all of our target clumps had a single velocity component in MALT90 data, while the sensitive C$^{18}$O J=2-1 ALMA observations reveal in most 
      clumps more than one velocity component along the line of sight. Based on the integrated
  intensity of C$^{18}$O, \cite{Contreras18} estimate that the mass of G331.372--00.116 could be 75\% of the value previously reported by
   \cite{Contreras17} using {\it Herschel} observations. We have checked the C$^{18}$O J=2-1 emission in the remaining 11 clumps and confirmed that, except for 
   G332.969--00.029 which could have its mass reduced by $\sim$50\%, clumps have contamination of $<$10\% of the mass derived using  {\it Herschel}
    observations (which is within the $\sim$50\% uncertainty  of mass determination). 

Assuming a star-cluster formation efficiency of 18\% \citep{Lada03}, the least massive clump (G340.232--00.146 with $M_{\rm cl}$ = 520 \Msun) should form a stellar cluster of 94 \Msun. 
Following Equations (1) and (2)\footnote{In \cite{Sanhueza17}, Equation (2) has a typo that overestimated the maximum stellar mass in $\sim$10\%.
 The correct version is added in Appendix~\ref{app} in the present work. The lowest mass regime of the \cite{Kroupa01} formulation for the IMF has also been added.} in \cite{Sanhueza17}, 
 based on the empirical relation from \cite{Larson03} and the IMF from \cite{Kroupa01}, we estimate that G340.232--00.146 should form a high-mass star of 8-9 \Msun. The most massive clump, 
 G014.492--00.139 ($M_{\rm cl}$ = 3120 \Msun), 
 with a stellar cluster of 562 \Msun\ should form a high-mass star of 21-29 \Msun. All IRDC clumps are above the empirical high-mass star formation thresholds from \cite{Kauffmann10}, \cite{Urquhart14}, and 
 \cite{He15}. \cite{Kauffmann10} suggest that clumps with masses larger than $m_{\rm lim}$ = 580 \Msun\ (r/pc)$^{1.33}$, where $r$ is the source radius, are currently forming or will likely form high-mass stars. The 
  values for $m_{\rm lim}$ range from 150 to 390 \Msun, all lower than the clump masses, which indicate that it is highly likely that the clumps will form high-mass stars. Both \cite{Urquhart14} and \cite{He15} propose that
  high-mass stars form in clumps with $\Sigma_{\rm clump} > 0.05$ gr cm$^{-2}$. All $\Sigma_{\rm cl}$ listed in Table~\ref{tbl-clumps} satisfy this threshold as well. 
  Therefore, each source selected for this pilot survey exhibits the necessary physical properties 
 likely to form a stellar cluster hosting at least one high-mass star. Thus, this overall sample is suitable for the characterization 
 of the earliest stages of high-mass star formation.

\section{Observations}
\label{alma-obs}

Observations of the 12 IRDCs were carried out with the Atacama Large Millimeter/sub-millimeter 
Array (ALMA) on different days during Cycle 3 (Project ID: 2015.1.01539.S; PI: Sanhueza) and a resubmission for Cycle 4
 (Project ID: 2016.1.01246.S; PI: Sanhueza). The data sets consist of observations in band 6 ($\sim$224 GHz; 1.34 mm) with
  the main 12 m array, the Atacama Compact 7 m Array (ACA; Morita Array), and total power (TP). Table~\ref{obs-param} summarizes
   all observational parameters. 
   
   Total 12 m array time on source per mosaic for 
  sources that were observed in one execution is $\sim$16 min (first six sources in Table~\ref{obs-param}), while sources that
   were observed in two executions have a total 12 m array time per mosaic of $\sim$25 min. Total 7 m array observing time per mosaic is
    $\sim$50 min, except for the three first sources in Table~\ref{obs-param} that were observed longer ($\sim$100 min). 
     Some sources were observed in different configurations, 
   resulting in different angular resolution (baselines are listed in Table~\ref{obs-param}). To have a more uniform data set, uv-taper was used in those observations with more extended 
   baselines in order to achieve a similar synthesized beam of $\sim$1\farcs 2 for every source (see Table~\ref{obs-param} for 
   individual values). This angular resolution corresponds to a physical size of 4800 AU (0.023 pc) at the average source distance of 4 kpc. 
   At 224 GHz, the primary
      beam of the 12 m array and ACA are 25\farcs 2 and 44\farcs 6, respectively. These observations are sensitive to angular scales
       smaller than $\sim$11\arcsec, and  $\sim$19\arcsec, respectively.  

\begin{deluxetable*}{lccccc}
\tabletypesize{\footnotesize}
\tablecaption{Observational Parameters \label{obs-param}}
\tablewidth{0pt}
\tablehead{
\colhead{IRDC} &    \colhead{rms Noise\tablenotemark{a}} &   Beam Size\tablenotemark{a}                      & \colhead{Baselines\tablenotemark{b}}    & \colhead{Configuration}     & \colhead{Number of }  \\
\colhead{Clump} &    \colhead{(mJy beam$^{-1}$)}                  & \colhead{(\arcsec\ $\times$ \arcsec)}           &\colhead{(m)}                                         &\colhead{}                            &\colhead{Antennas\tablenotemark{c}} \\
}
\startdata
G010.991--00.082 &  0.115						& 1.29 $\times$ 0.86                 			& 15 -- 330  						& C36-1   				& 41 (9 -- 10)  \\
G014.492--00.139 &  0.168						& 1.29 $\times$ 0.85	             		         & 15 -- 330  						&  C36-1  				& 41 (9 -- 10)   \\
G028.273--00.167 &  0.164						& 1.28 $\times$ 1.20	                  		& 15 -- 462  						& C36-2/3				& 41 (8 -- 10)   \\
G327.116--00.294 &  0.089						& 1.32 $\times$ 1.11                  		& 15 -- 330 						& C36-1				& 48  (8)  \\
G331.372--00.116 &  0.083						& 1.34 $\times$ 1.09                  		& 15 -- 330 						& C36-1				& 48  (8)   \\
G332.969--00.029 &  0.080						& 1.35 $\times$ 1.08                  		& 15 -- 330 						& C36-1				& 48  (8)   \\
G337.541--00.082 &  0.068						& 1.29 $\times$ 1.18                  		&  15 -- 639 						&  C36-2/3 -- C40-1		& 41 -- 43 (8 -- 9) \\
G340.179--00.242 &  0.094						& 1.41 $\times$ 1.29                  		& 15 -- 704						&   C36-2/3 -- C40-4		& 36 -- 41 (8 -- 9)  \\
G340.222--00.167 &  0.112						& 1.40 $\times$ 1.28                  		& 15 -- 704						&   C36-2/3 -- C40-4		& 36 -- 41 (8 -- 9)  \\
G340.232--00.146 &  0.139						& 1.39 $\times$ 1.26 	         		& 15 -- 704 						&   C36-2/3 -- C40-4		& 36 -- 41 (8 -- 9) \\
G341.039--00.114 &  0.070						& 1.30 $\times$ 1.18 	         		& 15 -- 639 						&  C36-2/3 -- C40-1		& 41 -- 43 (8 -- 9)  \\
G343.489--00.416 &  0.068						& 1.30 $\times$ 1.18                  		& 15 -- 639 						&  C36-2/3 -- C40-1		& 41 -- 43 (8 -- 9)  \\
\enddata
 \tablenotetext{a}{Continuum sensitivity and synthesized beam in the combined 12 and 7 m data sets.}
  \tablenotetext{b}{For the 7 m array, the baselines range from 8 -- 48 m.}
   \tablenotetext{c}{Values in parenthesis refer to the number (or range) of antennas for the 7 m array. Ranges are given when there are more than one execution block with different number of antennas.}
\end{deluxetable*}

With one exception, IRDCs were observed in continuum and line emission in Nyquist sampled 10-pointing and 3-pointing mosaics with the
 12 m array and the ACA, respectively. IRDC G028.273--00.167 was observed with 11 and 5 pointings, respectively. Within the 20\% 
power point, a 10-pointing mosaic corresponds to 0.97 arcmin$^2$ (1.06 arcmin$^2$ for IRDC G028.273--00.167), which is 
equivalent to an effective FOV of $\sim$1\arcmin\ per target. By using mosaics, we assure coverage of a large area of clumps, 
as defined by single-dish continuum observations. The same correlator setup was used for all sources. The continuum emission
 was produced by averaging the line-free channels in visibility space. All images have 512 $\times$ 512 pixels, with a pixel size of 
 0\farcs 2. To mitigate artefacts produced by the extended emission from IRDCs, we used {\sc tclean} and its {\sc multi-scale} imaging option
  with scales values of 0, 5, 15, and 25 times the pixel size. Using Briggs weighting with a robust parameter of 0.5, the
  1$\sigma$ rms noise for the continuum emission is on average 0.10 mJy beam$^{-1}$ (see Table~\ref{obs-param} for each individual source). 

At least 10 different molecular lines were included in the spectral setup (N$_2$D$^+$ J=3-2, 
DCN J=3-2, DCO$^+$ J=3-2, CCD J=3-2, $^{13}$CS J=5-4, SiO J=5-4, C$^{18}$O J=2-1, 
CO J=2-1, H$_2$CO J=3-2, and CH$_3$OH J=4-3). The line sensitivity for the first six lines is  $\sim$9.5 mJy beam$^{-1}$ per channel of 0.17 \kms, while for the 
last four lines is $\sim$3.5 mJy beam$^{-1}$ per channel of 1.3 \kms\ (we note these channels correspond to the spectral resolution and not to the raw channel size which is 
half of the spectral resolution, e.i., $\sim$0.085 and $\sim$0.65 \kms, respectively). We defer 
the analysis of all molecular lines for future papers. In this work, we analyze the dust continuum emission. We only use
 qualitative information of line emission for the classification of the evolutionary sequence of the cores (CO, SiO, H$_2$CO,
  and CH$_3$OH) and the determination of multiple velocity 
components on the line of sight (C$^{18}$O).  

Calibration was carried out using the {\sc CASA} software package version 4.5.3, 4.6, and 4.7, while imaging was done using {\sc CASA} 5.4 \citep{McMullin07}. All
 images presented in this paper are not primary beam corrected; but all fluxes are measured on the primary beam corrected images. We note that 
 all targets were also observed with the Total Power (TP) antennas. However, TP antennas do not provide continuum emission and are therefore
  not used in this work. 

\begin{figure*}
\begin{center}
\centering
\includegraphics[angle=0,scale=0.43]{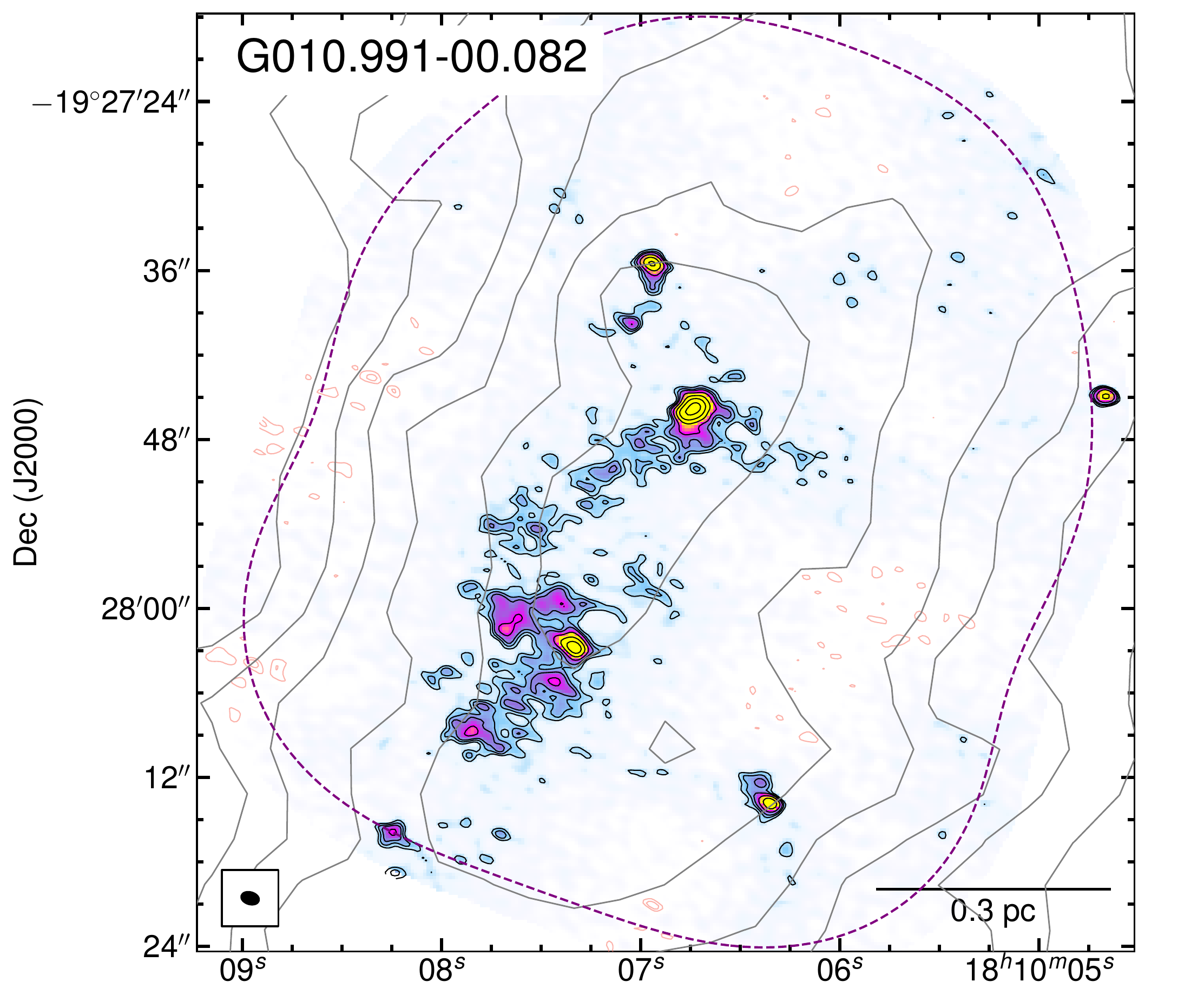}\hspace{-0.6cm}
\includegraphics[angle=0,scale=0.43]{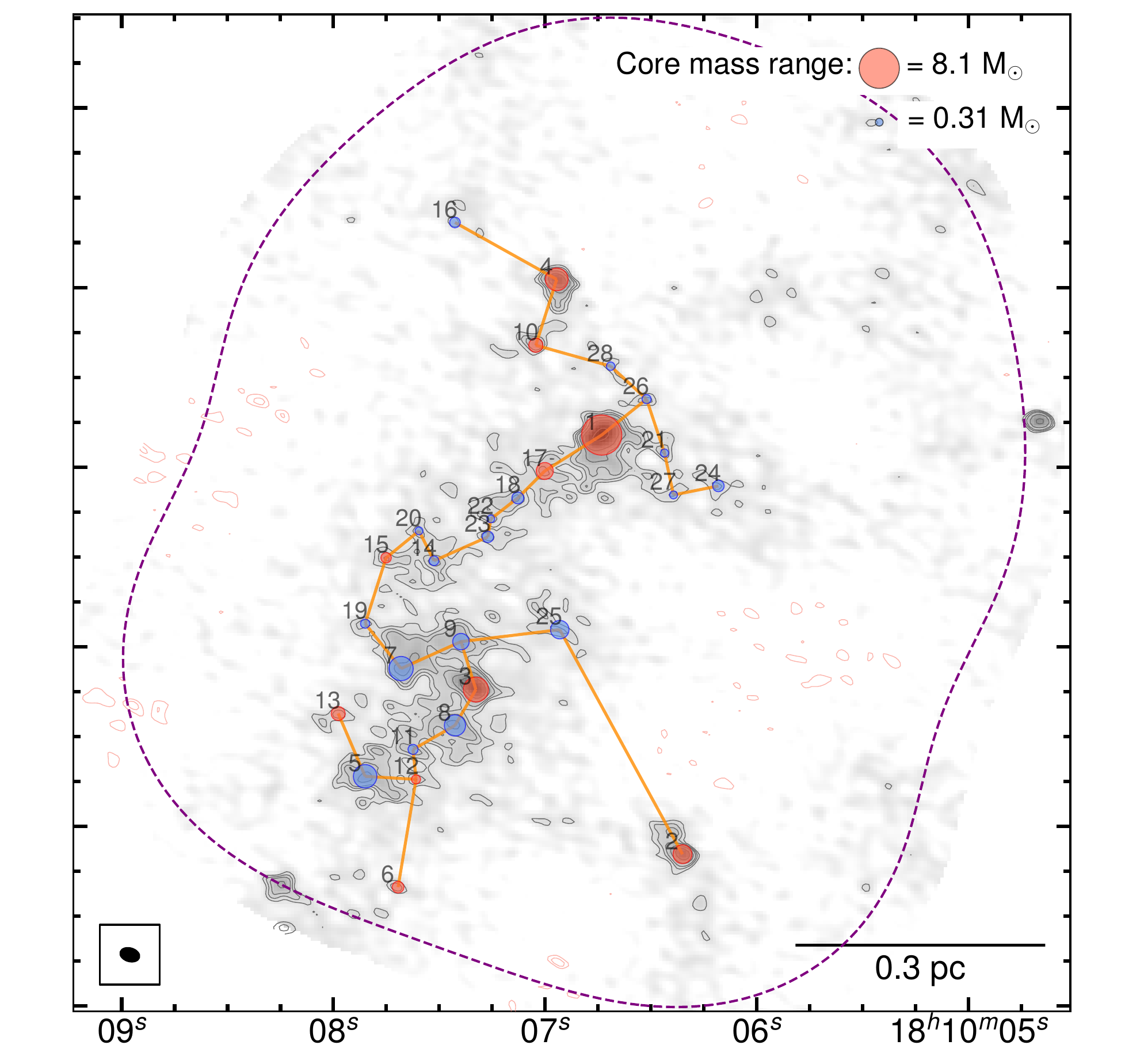}
\includegraphics[angle=0,scale=0.43]{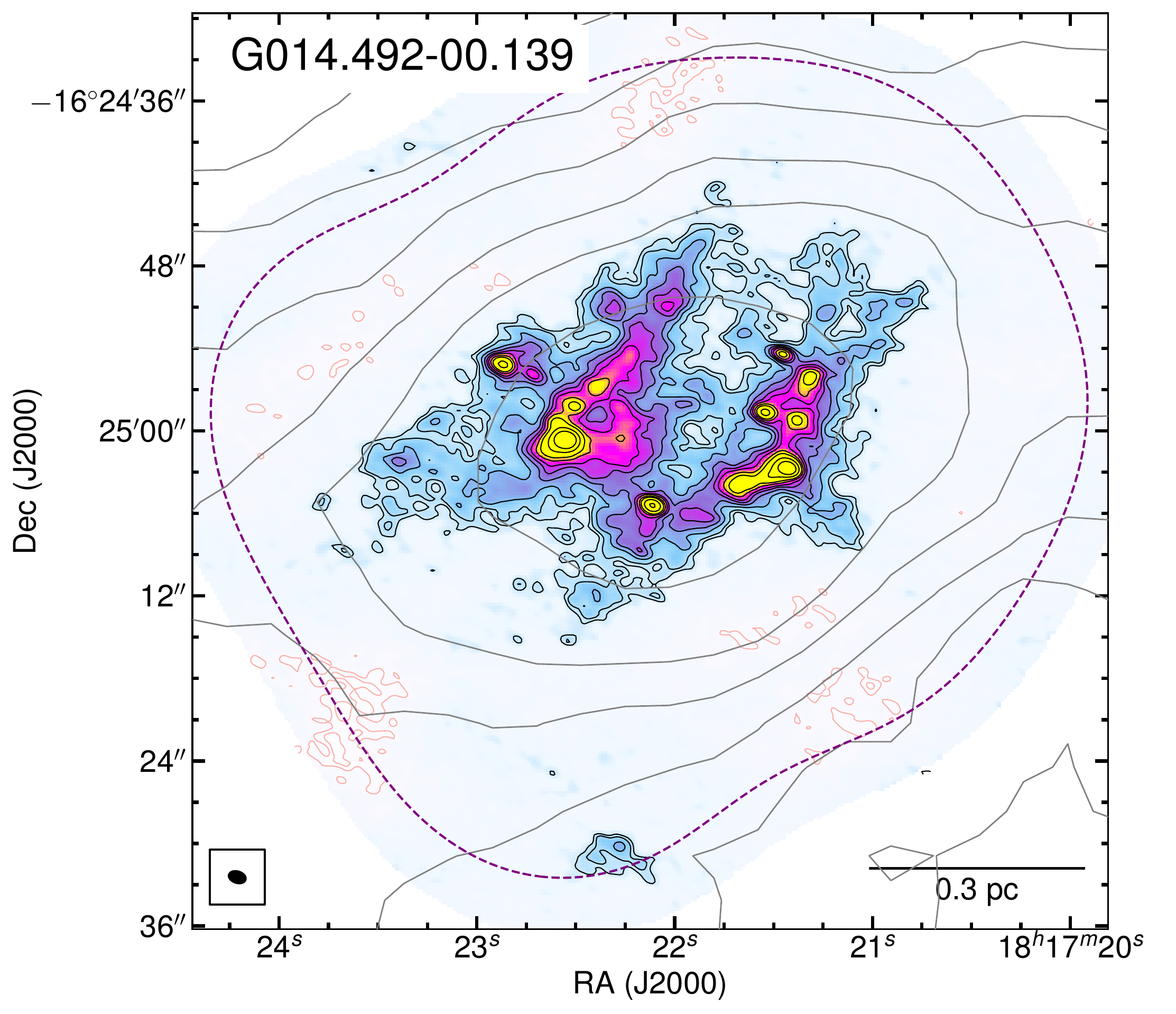}\hspace{-0.6cm}
\includegraphics[angle=0,scale=0.43]{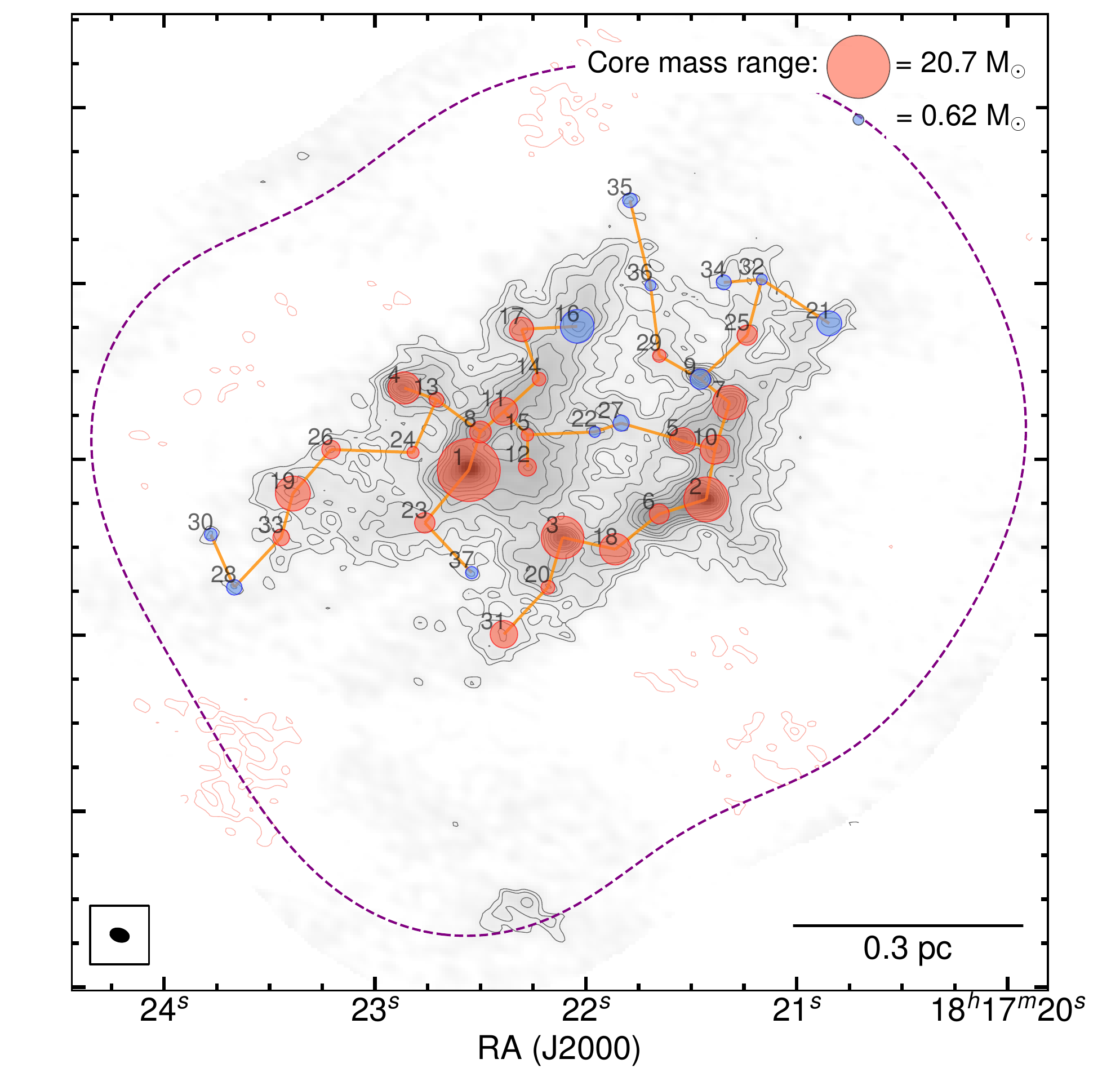}
\end{center}
\caption{ALMA 1.34 mm dust continuum emission for two IRDC clumps. Dashed 
purple contour delineates the area mosaiced with ALMA. 
Left: color image and small-scale contours correspond to the ALMA dust continuum emission (12 and 7 m array combined).
ALMA contour levels are -4, -3, 3, 4, 5, 7, 10, 14, and 20 $\times$ $\sigma$, with $\sigma = 0.115$ mJy beam$^{-1}$, for G010.991--00.082 (1\farcs1 
angular resolution); and 
-4, -3, 3, 4, 6, 8, 10, 12, 15, 18, 25, and 35 $\times$ $\sigma$, with $\sigma = 0.168$ mJy beam$^{-1}$, for G014.492--00.139  (1\farcs1 
angular resolution).  Grey and red contours mark the positive and negative levels, respectively. Synthesized beams are displayed at the bottom left in each panel. 
Large-scale contours delineate the single-dish dust continuum emission from ATLASGAL (contour levels are the same as in the correponding Figure~\ref{IR-plots1},~\ref{IR-plots2},~\ref{IR-plots3}, or~\ref{IR-plots4}).
Right: gray-scale image and contours correspond to the ALMA dust continuum emission (12 and 7 m array combined). Blue circles show the positions 
of the prestellar core candidates, while the red circles show the positions of the protostellar cores (see Section~\ref{evo_stage}). The circle size is
 proportional to the core mass and the range of mass values is displayed on the top, right side of the panel. Numbers near each core correspond to the core name (ALMA1, ALMA2, ...), where the prefix ALMA has been dropped for 
 clarity. Orange segments show the outcome from the minimum spanning tree (MST; Section~\ref{mst_sec}), which corresponds to 
 the set of straight lines that connects cores in a way that minimize the sum of the lengths.
}
\label{ALMA_cont1}
\end{figure*}

\begin{figure*}
\begin{center}
\centering
\includegraphics[angle=0,scale=0.41]{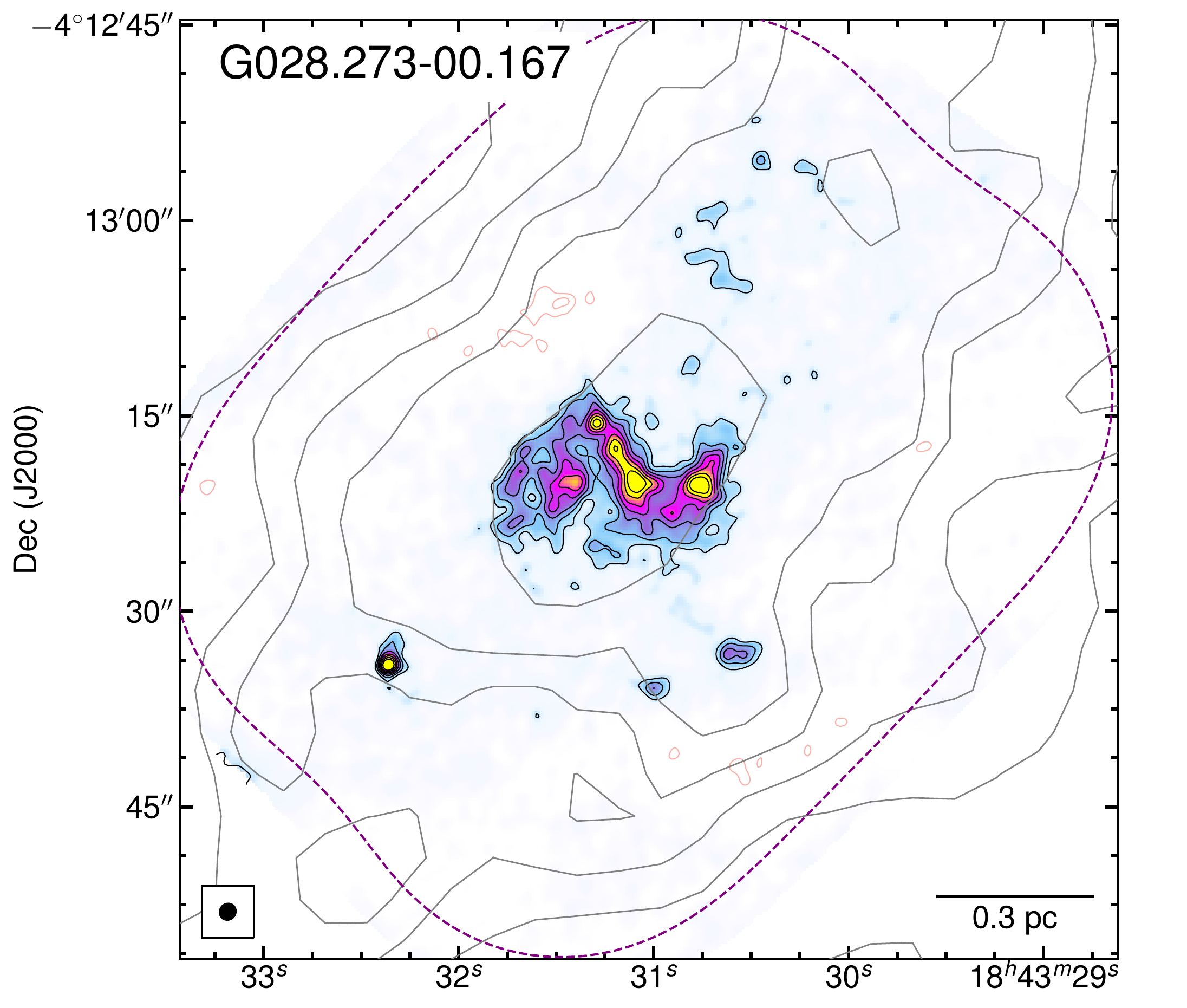}\hspace{-0.6cm}
\includegraphics[angle=0,scale=0.41]{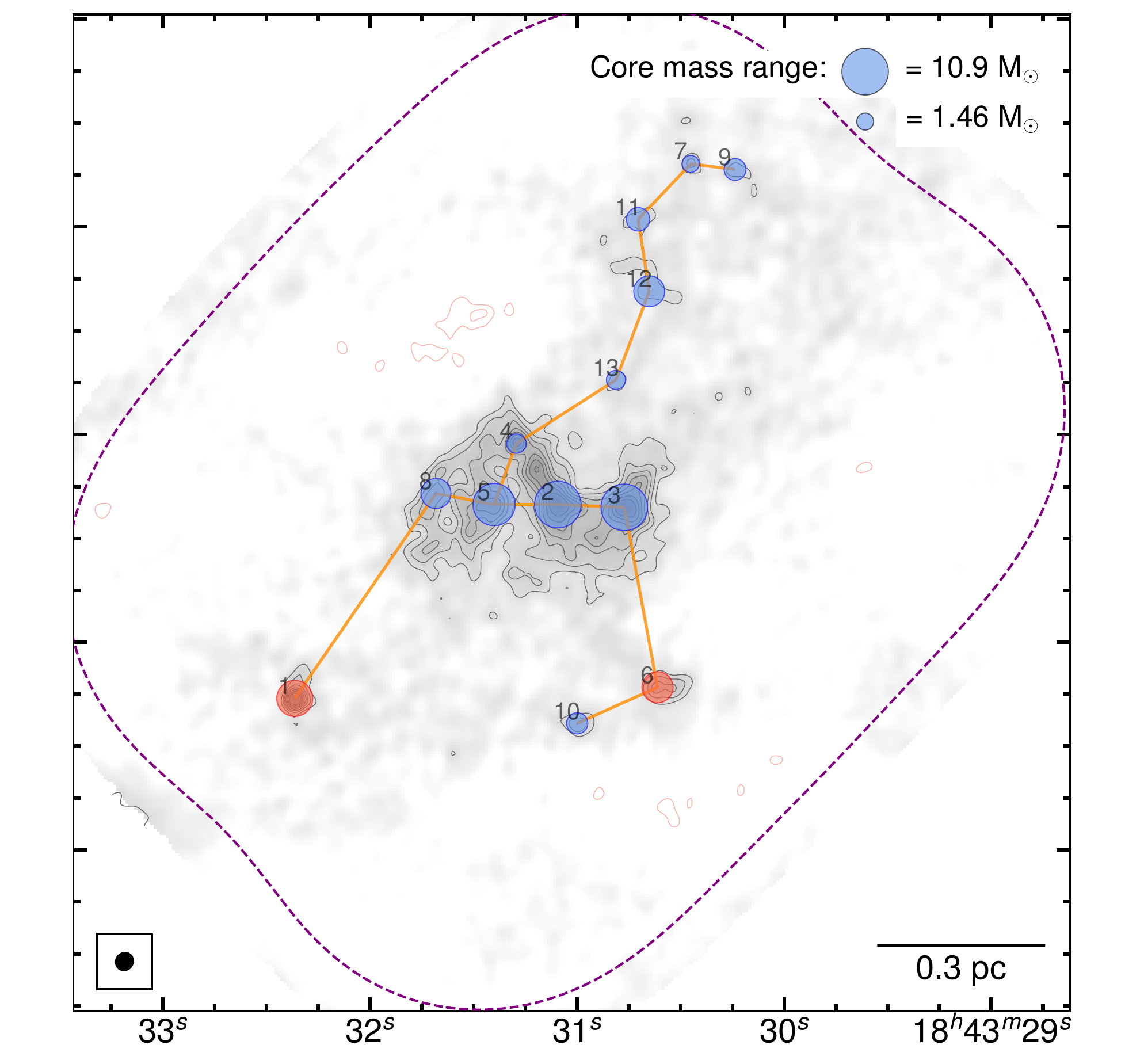}
\includegraphics[angle=0,scale=0.41]{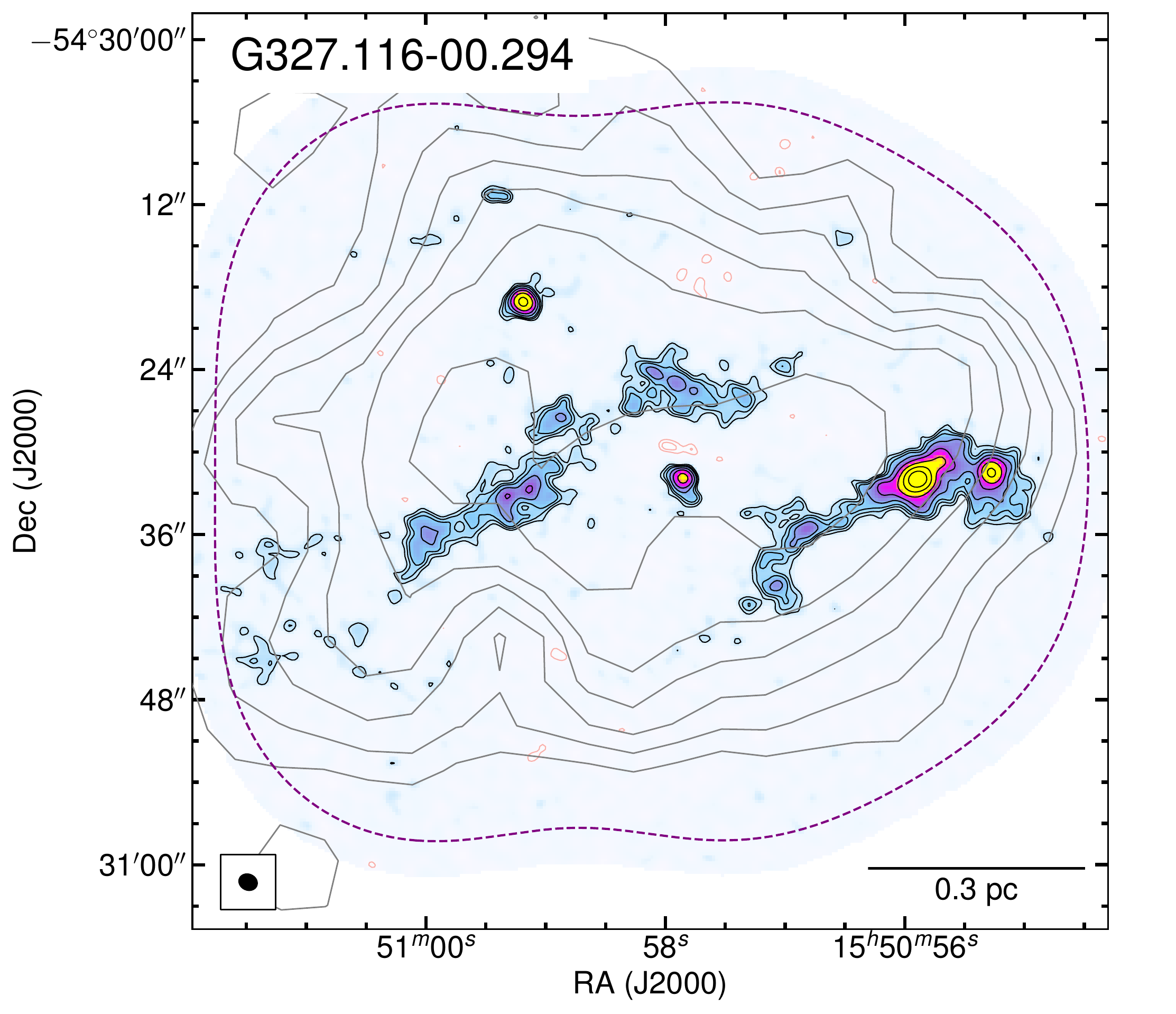}\hspace{-0.6cm}
\includegraphics[angle=0,scale=0.41]{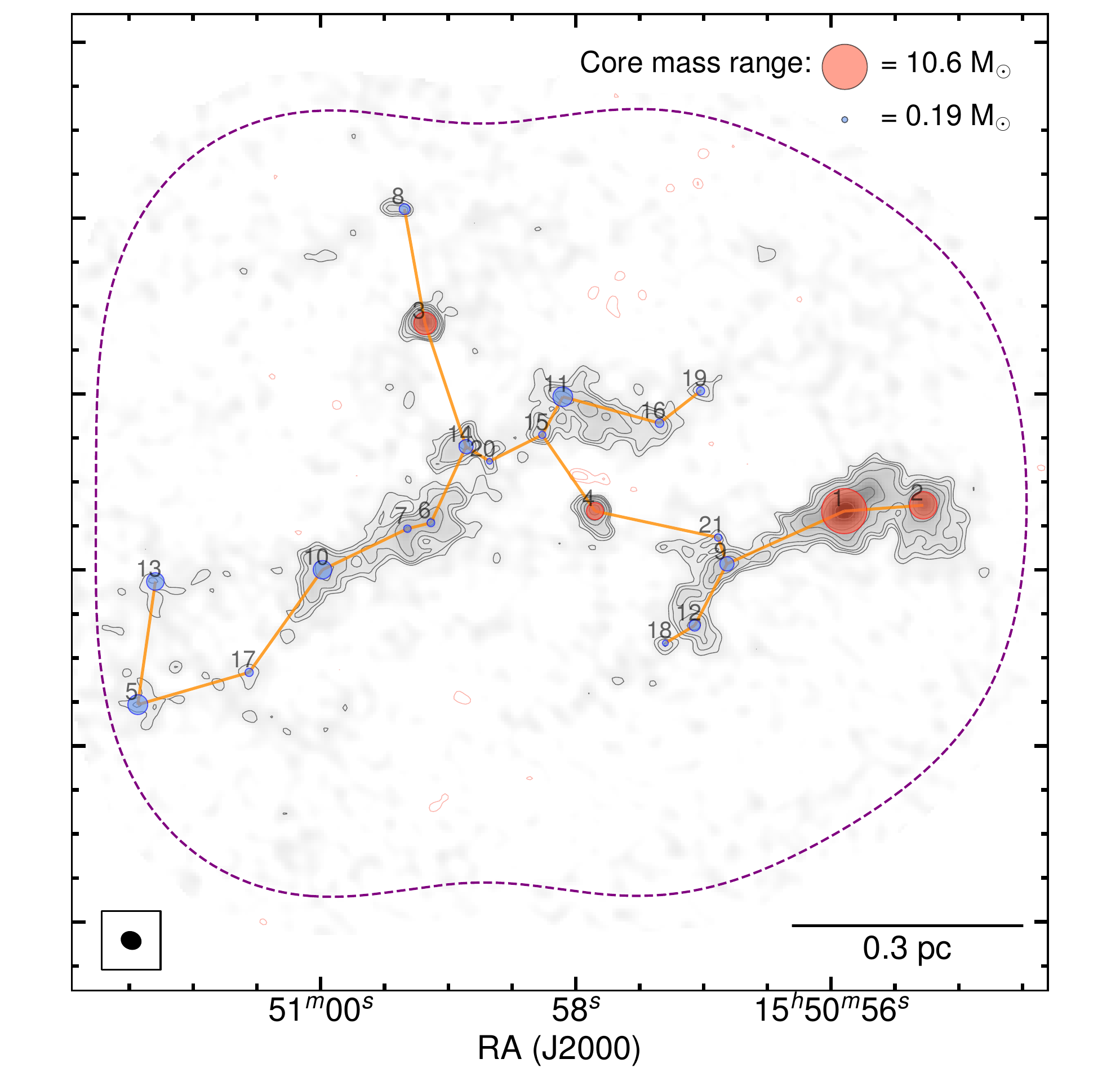}
\end{center}
\caption{Same as in Figure~\ref{ALMA_cont1}, except for ALMA contour levels of -4, -3, 3, 4, 5, 6, 7, 8, and 9 $\times$ $\sigma$, with $\sigma = 0.164$ mJy beam$^{-1}$,
 for G028.273--00.167 (1\farcs2 angular resolution); and -4, -3, 3, 4, 5, 7, 10, 15, 23, and 35 $\times$ $\sigma$, with $\sigma = 0.089$ mJy beam$^{-1}$, for
  G327.116--00.294  (1\farcs2 angular resolution).
}
\label{ALMA_cont2}
\end{figure*}

\begin{figure*}
\begin{center}
\centering
\includegraphics[angle=0,scale=0.41]{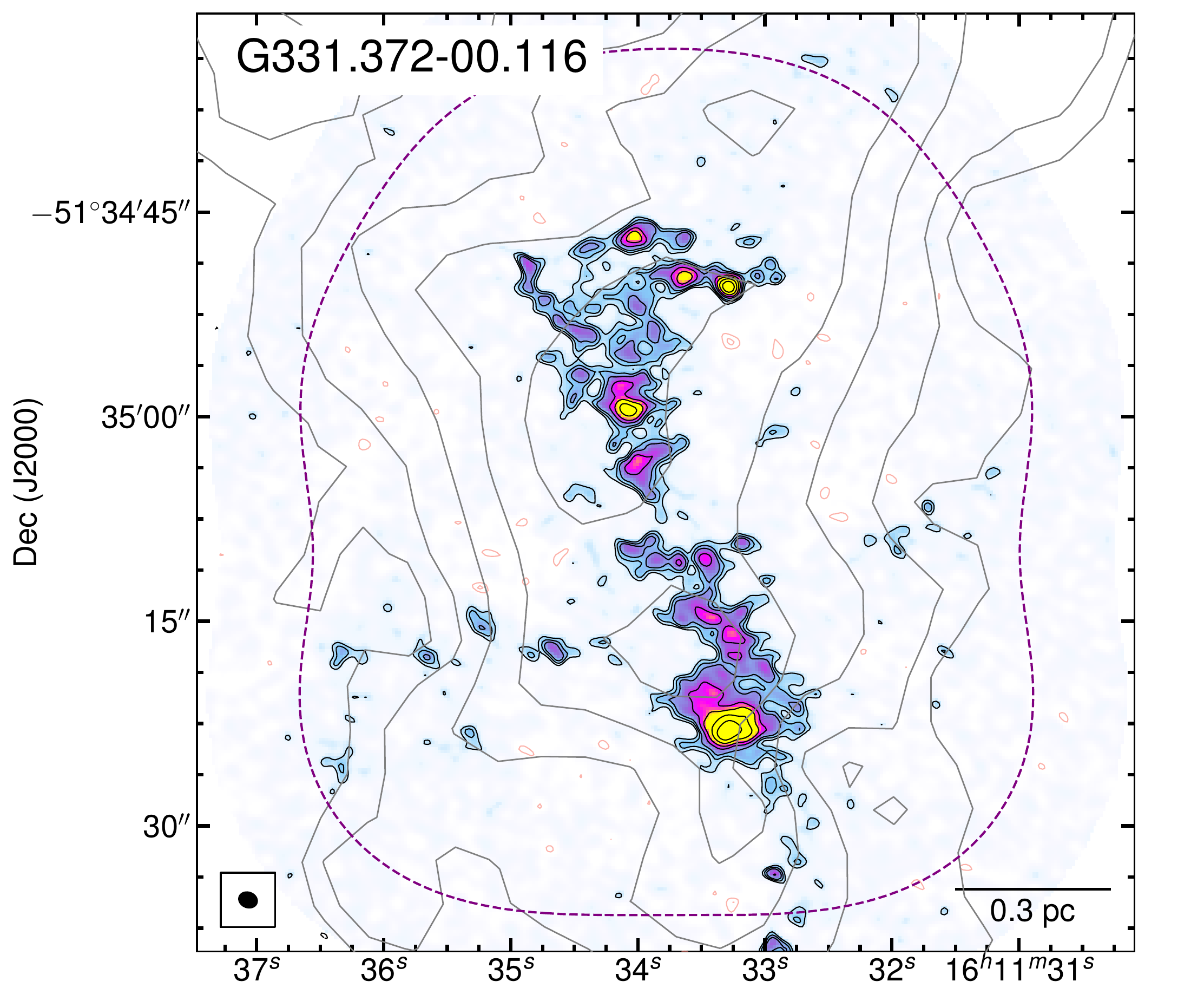}\hspace{-0.6cm}
\includegraphics[angle=0,scale=0.41]{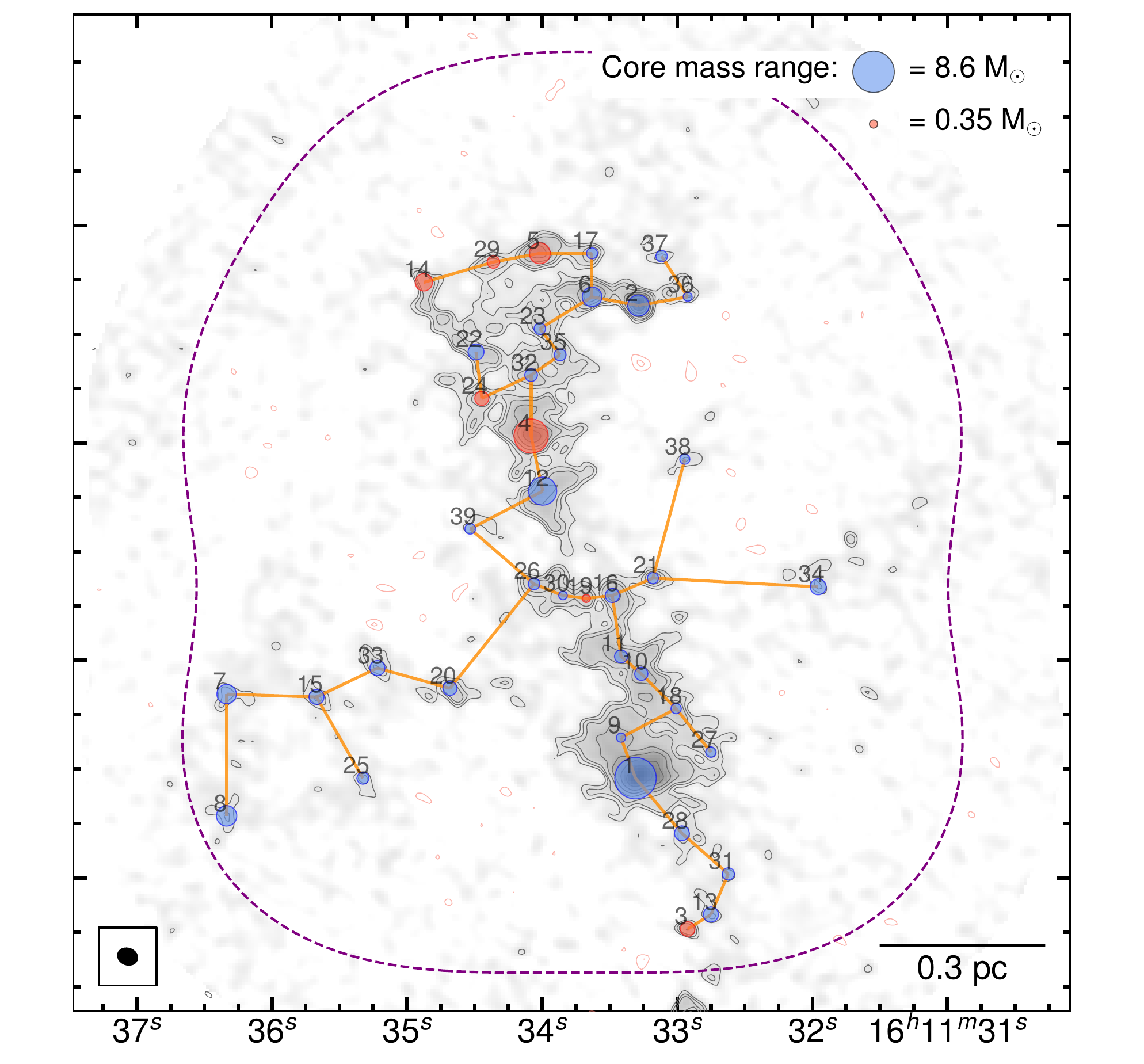}
\includegraphics[angle=0,scale=0.41]{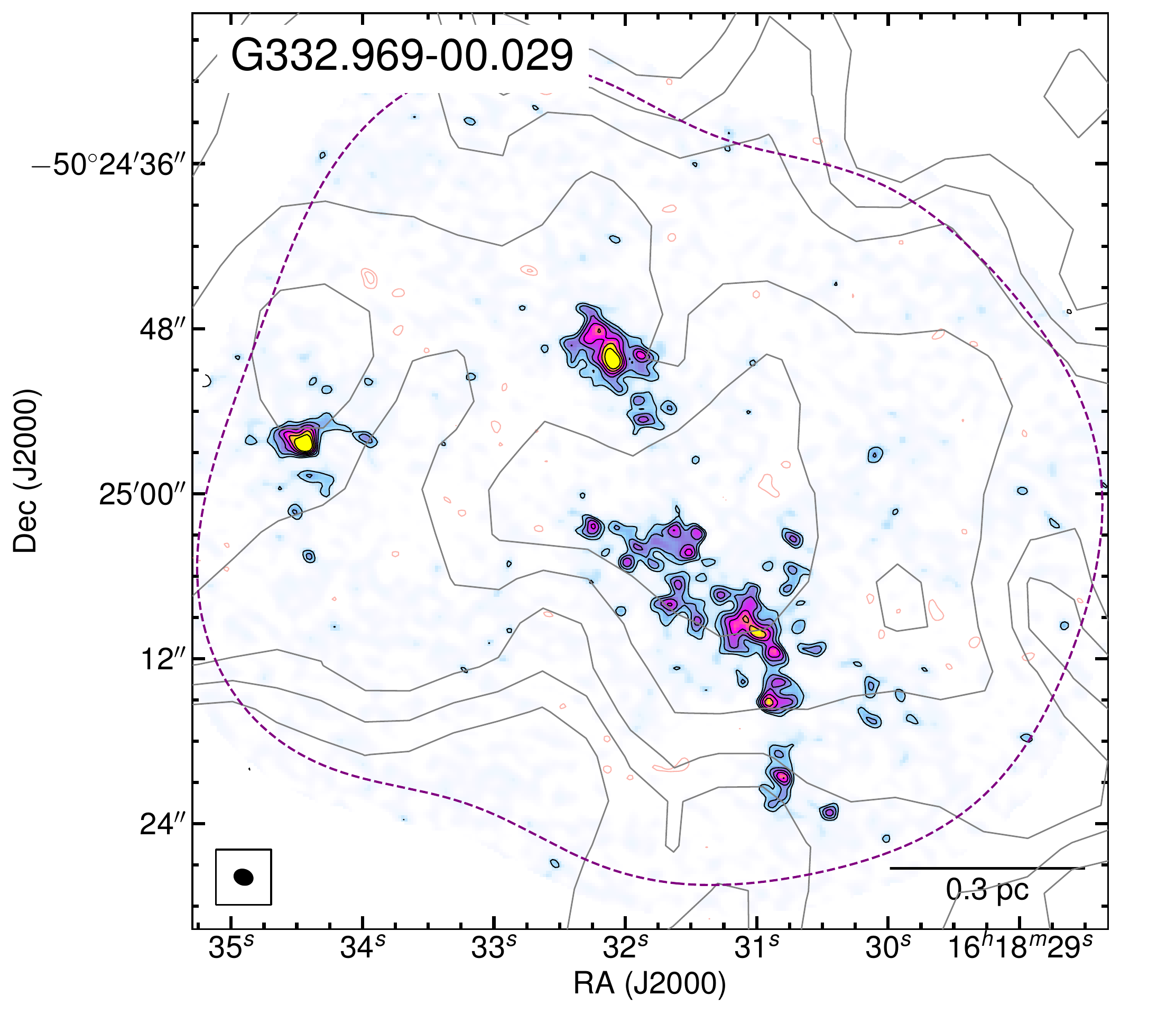}\hspace{-0.6cm}
\includegraphics[angle=0,scale=0.41]{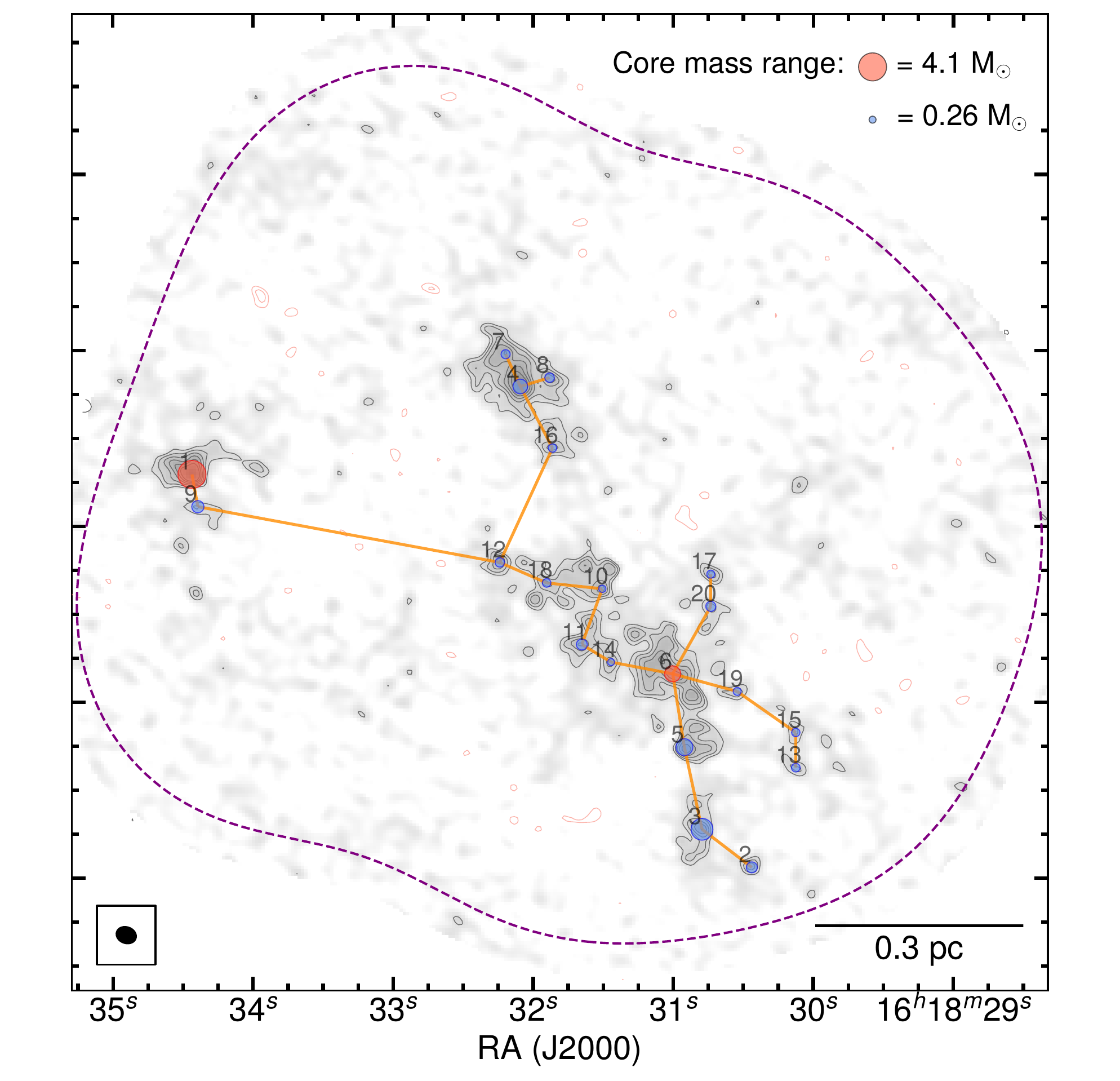}
\end{center}
\caption{Same as in Figure~\ref{ALMA_cont1}, except for ALMA contour levels of -4, -3, 3, 4, 5, 7, 9, 12, and 16 $\times$ $\sigma$, with $\sigma = 0.083$ mJy beam$^{-1}$,
 for G331.372--00.116 (1\farcs2 angular resolution); and -4, -3, 3, 4, 5, 6, 7, and 8 $\times$ $\sigma$, with $\sigma = 0.080$ mJy beam$^{-1}$, for
  G332.969--00.029 (1\farcs2 angular resolution).}
\label{ALMA_cont3}
\end{figure*}

\begin{figure*}
\begin{center}
\centering
\includegraphics[angle=0,scale=0.41]{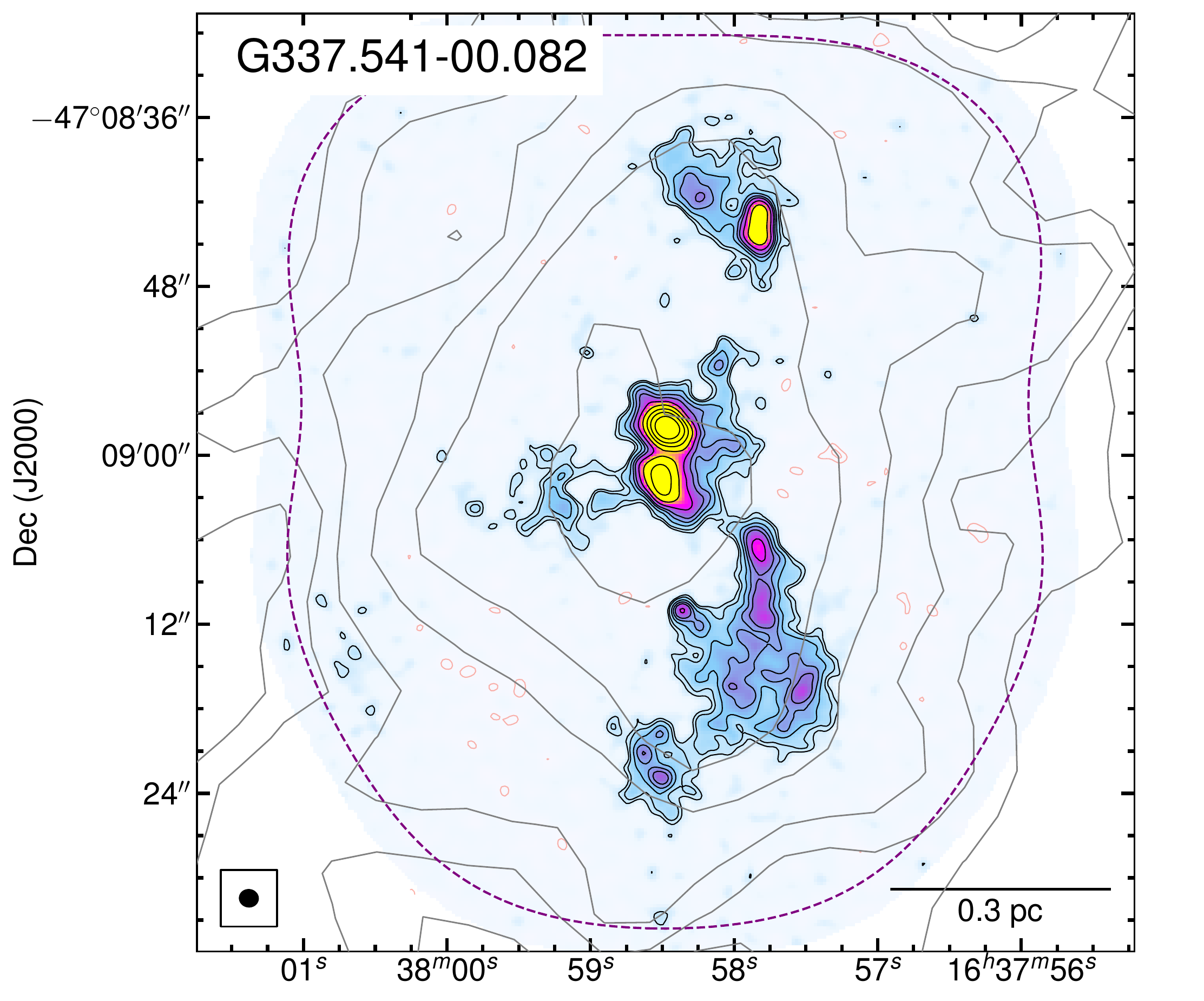}\hspace{-0.6cm}
\includegraphics[angle=0,scale=0.41]{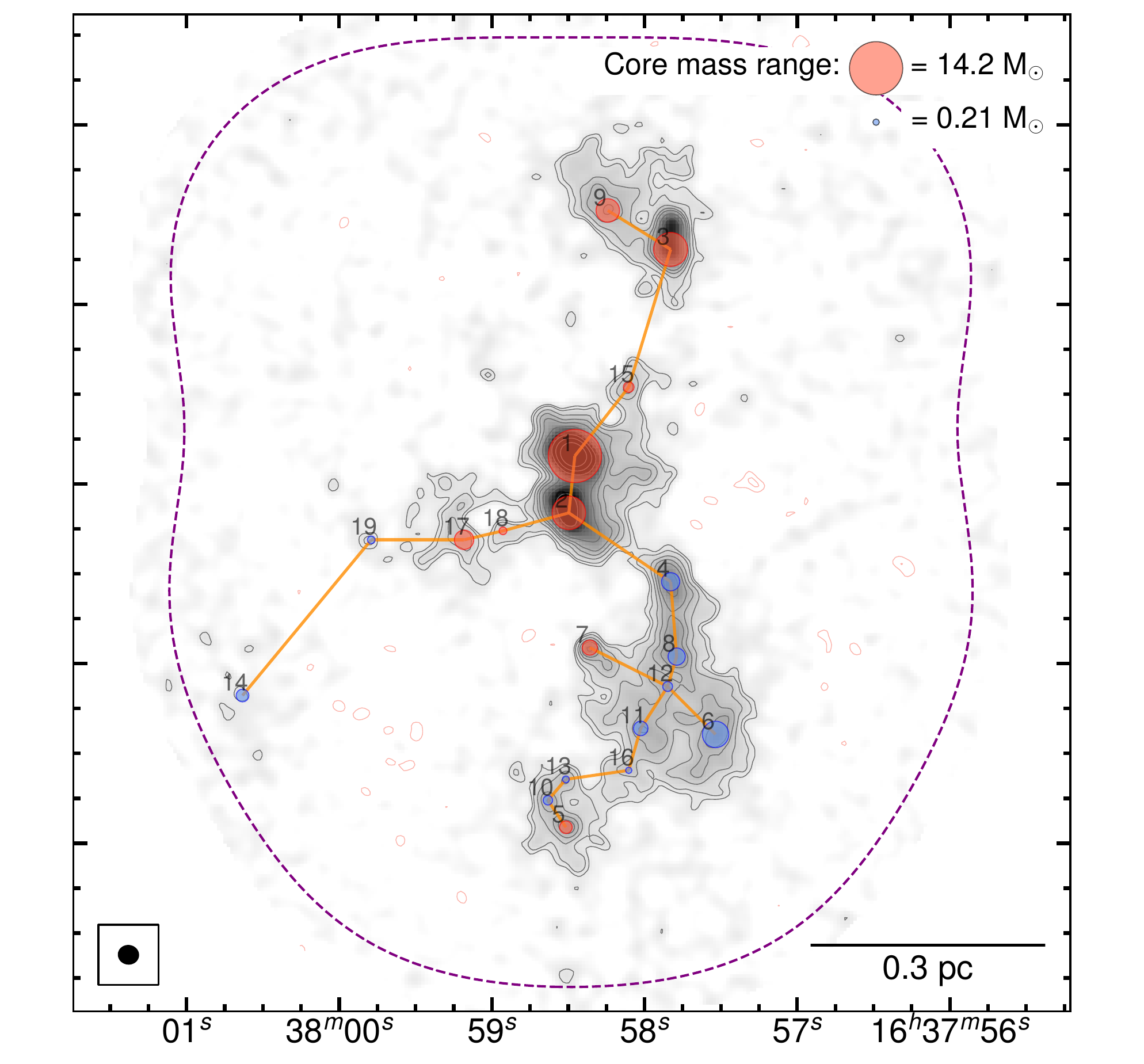}
\includegraphics[angle=0,scale=0.41]{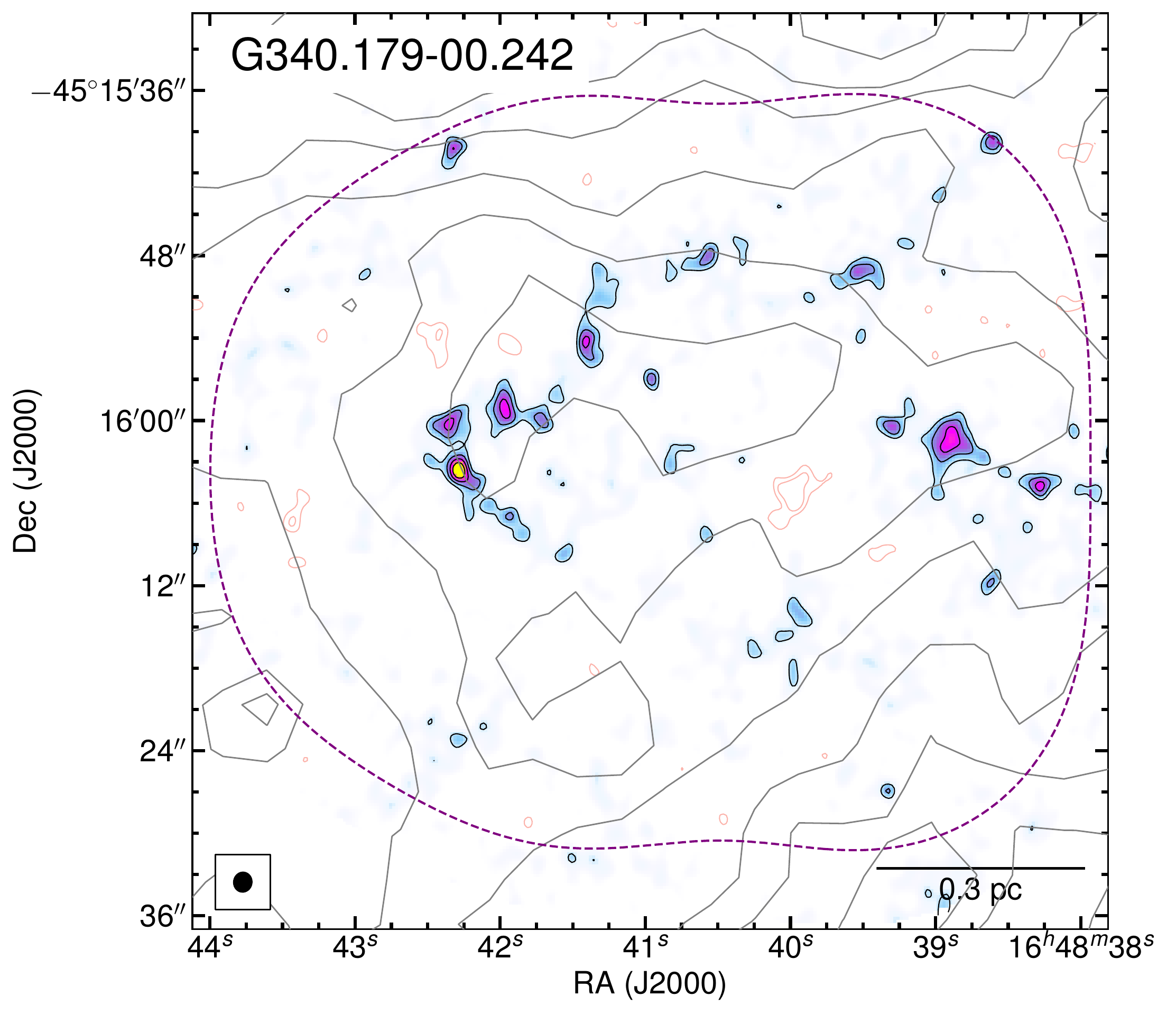}\hspace{-0.6cm}
\includegraphics[angle=0,scale=0.41]{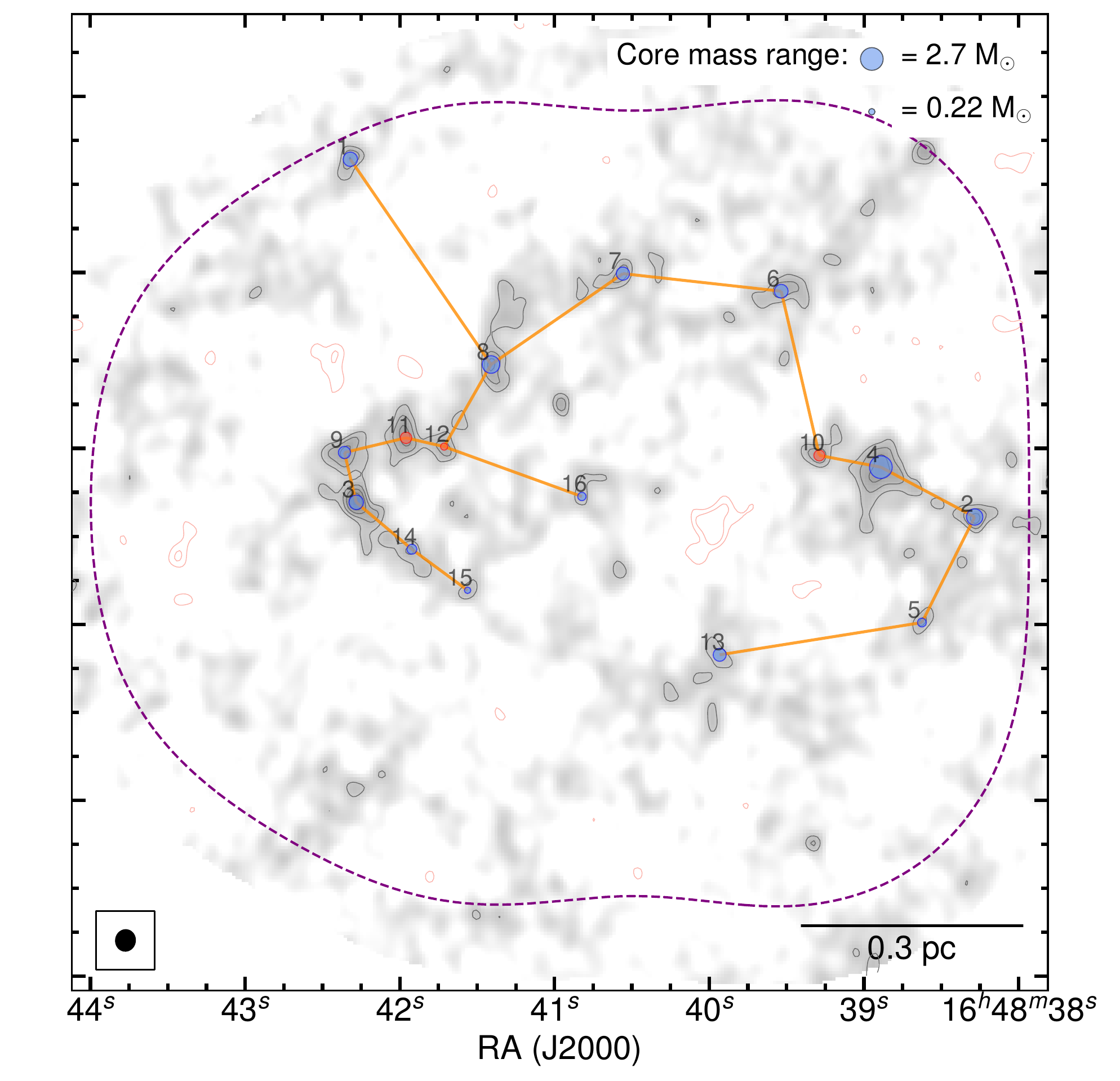}
\end{center}
\caption{Same as in Figure~\ref{ALMA_cont1}, except for ALMA contour levels of -4, -3, 3, 4, 6, 8, 10, 14, 20, 30, 45, and 75 $\times$ $\sigma$, with $\sigma = 0.068$ mJy beam$^{-1}$,
 for G337.541--00.082 (1\farcs2 angular resolution); and -4, -3, 3, 4, 5, and 6 $\times$ $\sigma$, with $\sigma = 0.094$ mJy beam$^{-1}$, for
  G340.179--00.242 (1\farcs3 angular resolution).
  }
\label{ALMA_cont4}
\end{figure*}

\begin{figure*}
\begin{center}
\centering
\includegraphics[angle=0,scale=0.41]{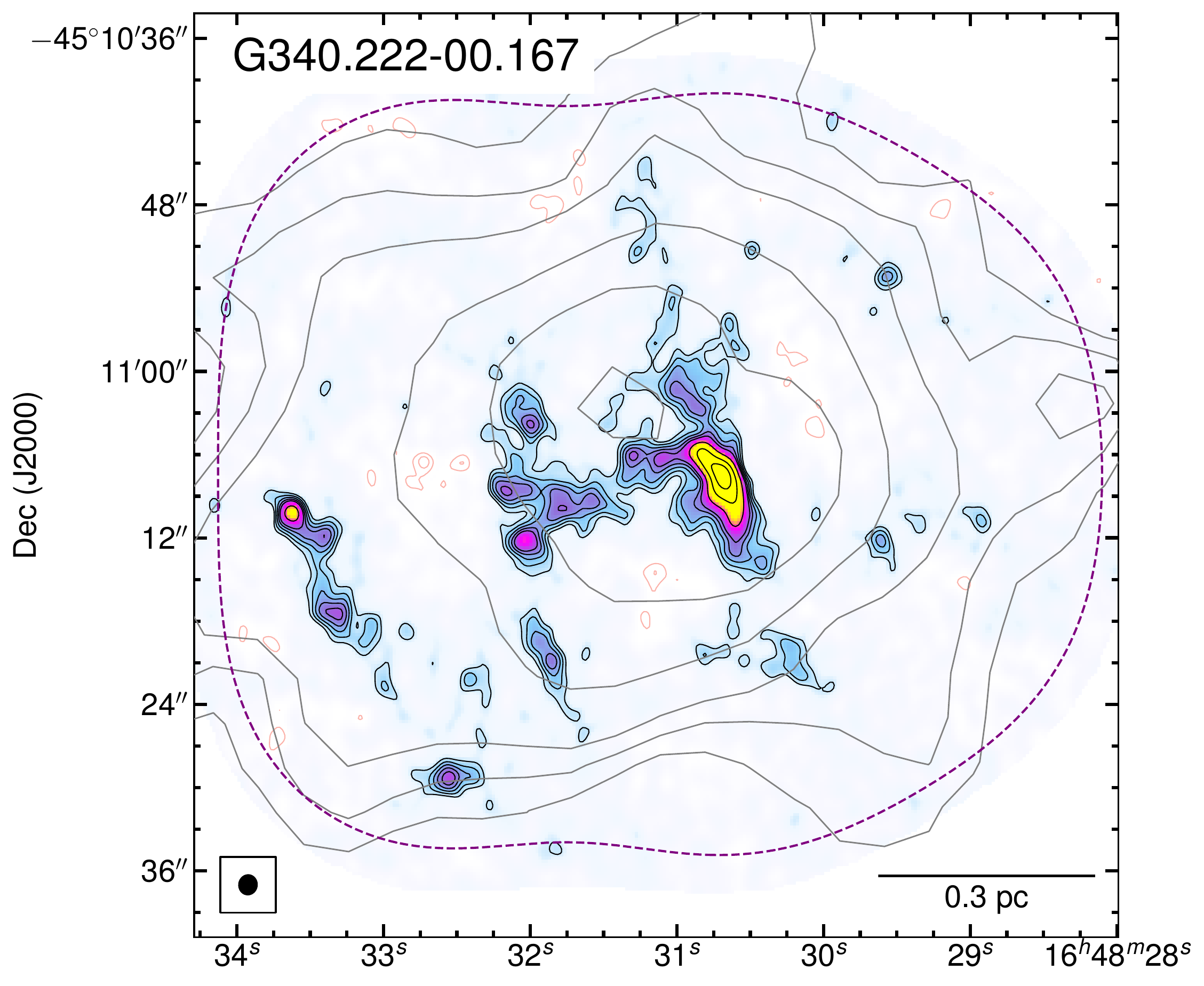}\hspace{-0.65cm}
\includegraphics[angle=0,scale=0.41]{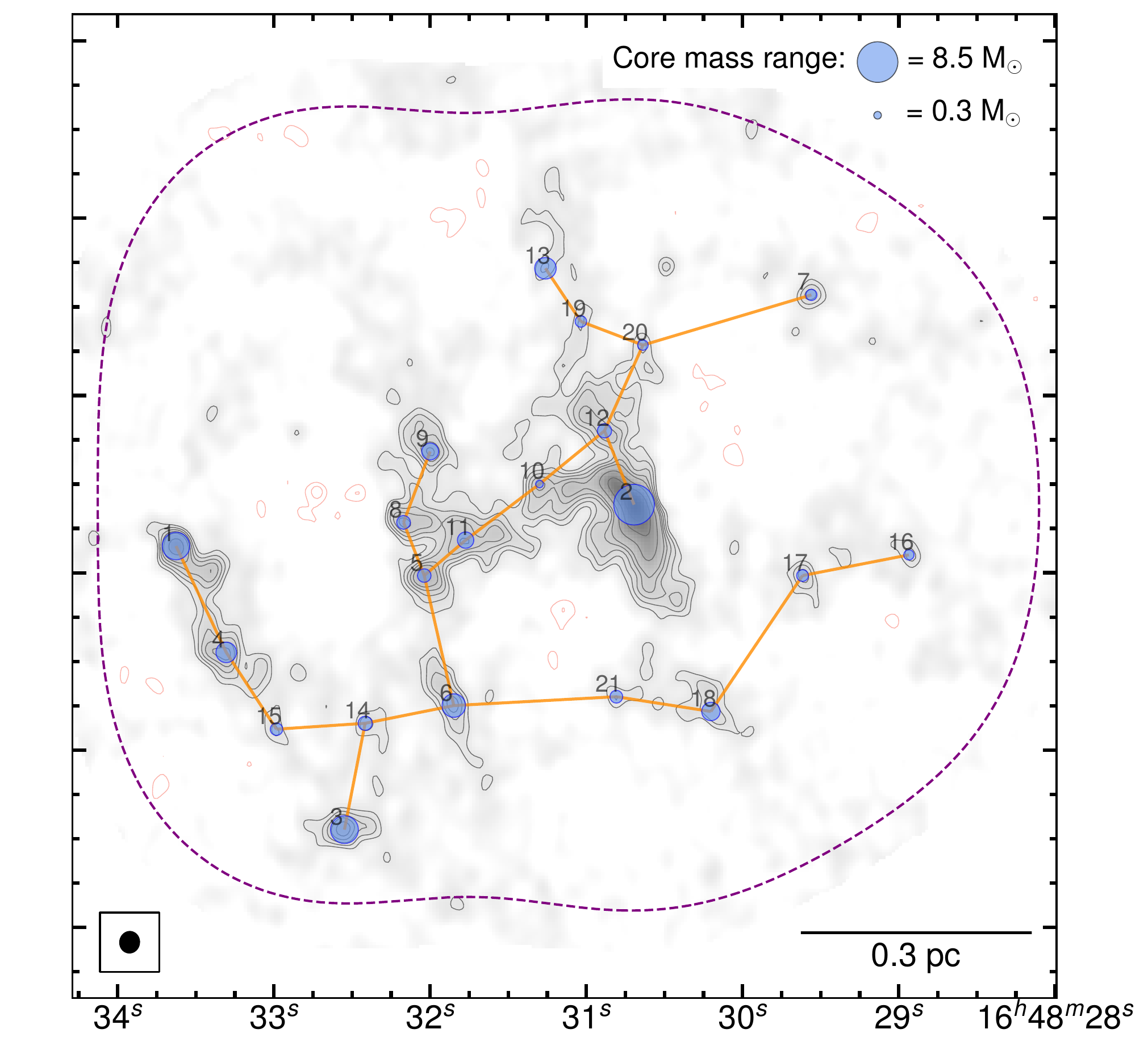}
\includegraphics[angle=0,scale=0.41]{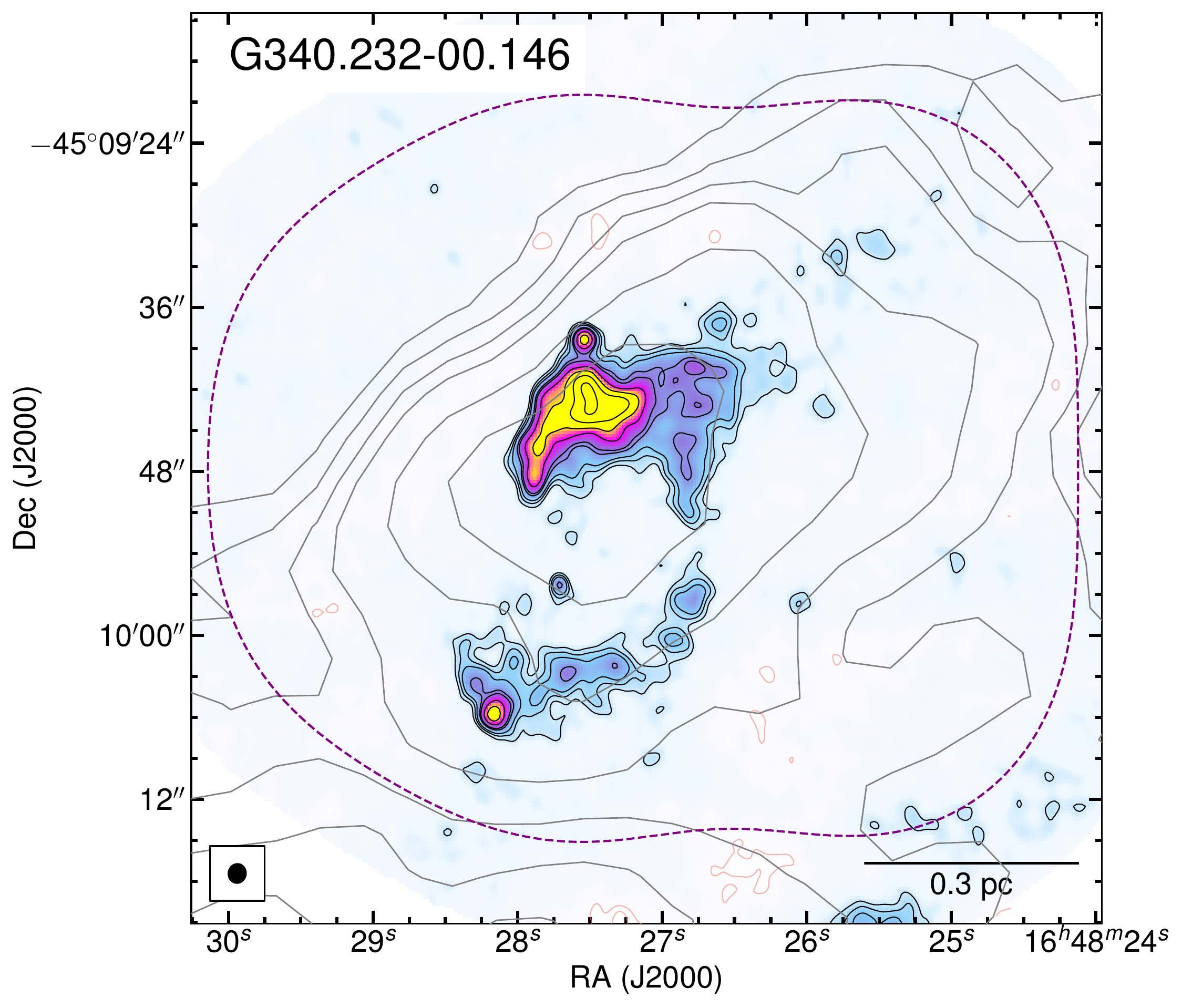}\hspace{-0.65cm}
\includegraphics[angle=0,scale=0.41]{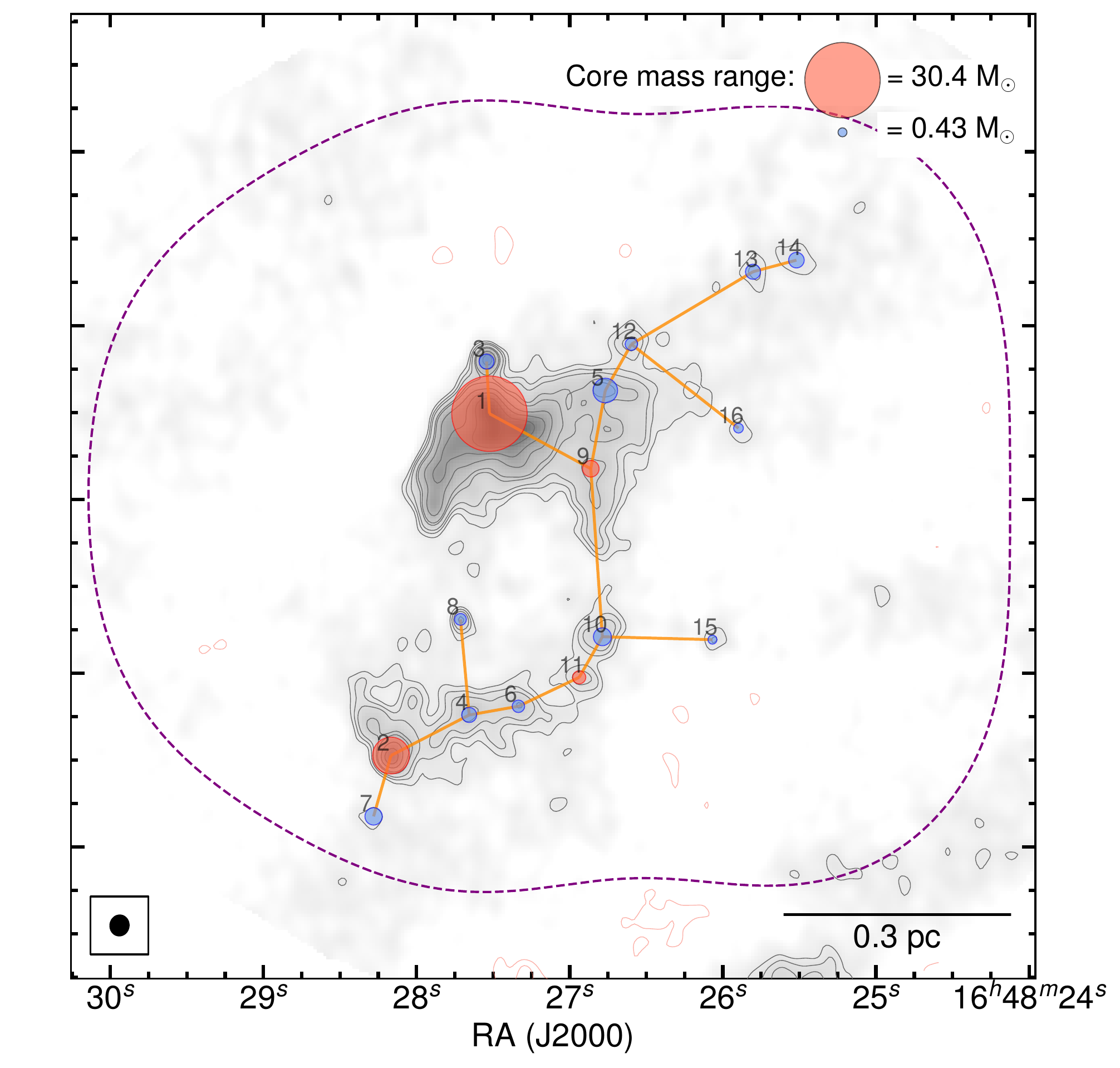}
\end{center}
\caption{ Same as in Figure~\ref{ALMA_cont1}, except for ALMA contour levels of -4, -3, 3, 4, 5, 6, 7, 10, 14, and 18 $\times$ $\sigma$, with $\sigma = 0.112$ mJy beam$^{-1}$,
 for G340.222--00.167 (1\farcs3 angular resolution); and -4, -3, 3, 4, 5, 7, 8, 11, 14, 18, and 23 $\times$ $\sigma$, with $\sigma = 0.139$ mJy beam$^{-1}$, for
  G340.232--00.146  (1\farcs3 angular resolution).
}
\label{ALMA_cont5}
\end{figure*}

\begin{figure*}
\begin{center}
\centering
\includegraphics[angle=0,scale=0.42]{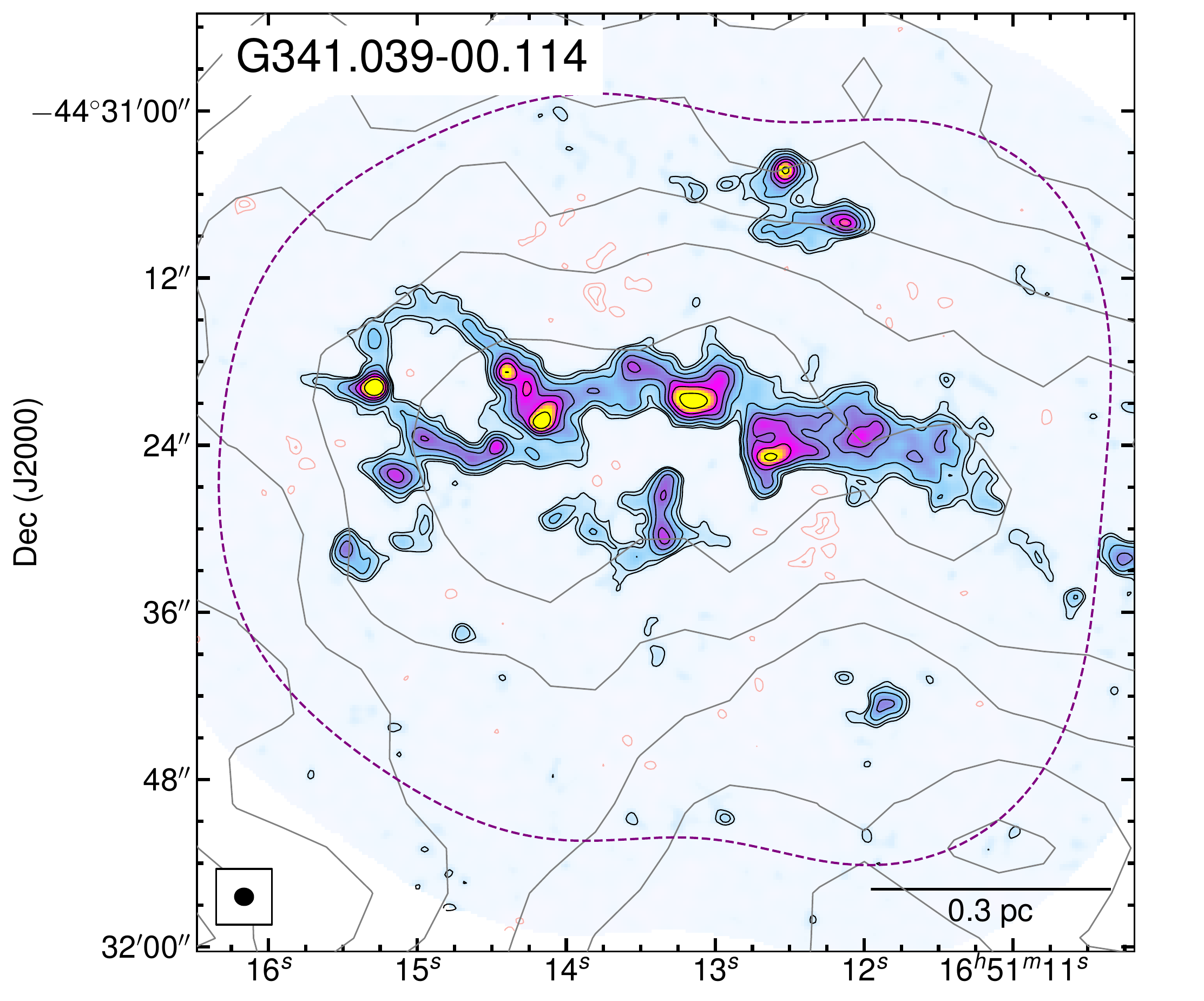}\hspace{-0.6cm}
\includegraphics[angle=0,scale=0.42]{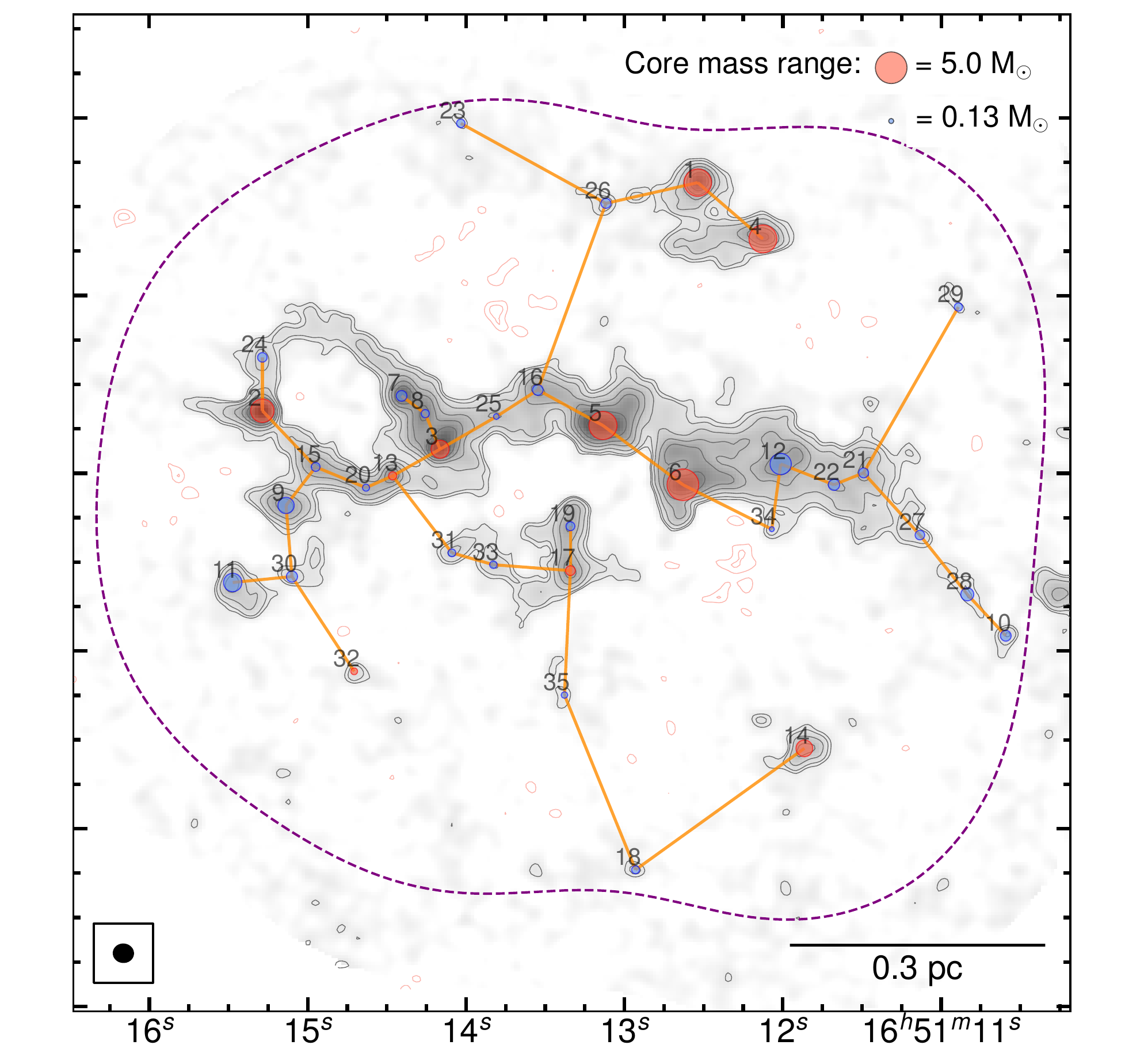}
\includegraphics[angle=0,scale=0.42]{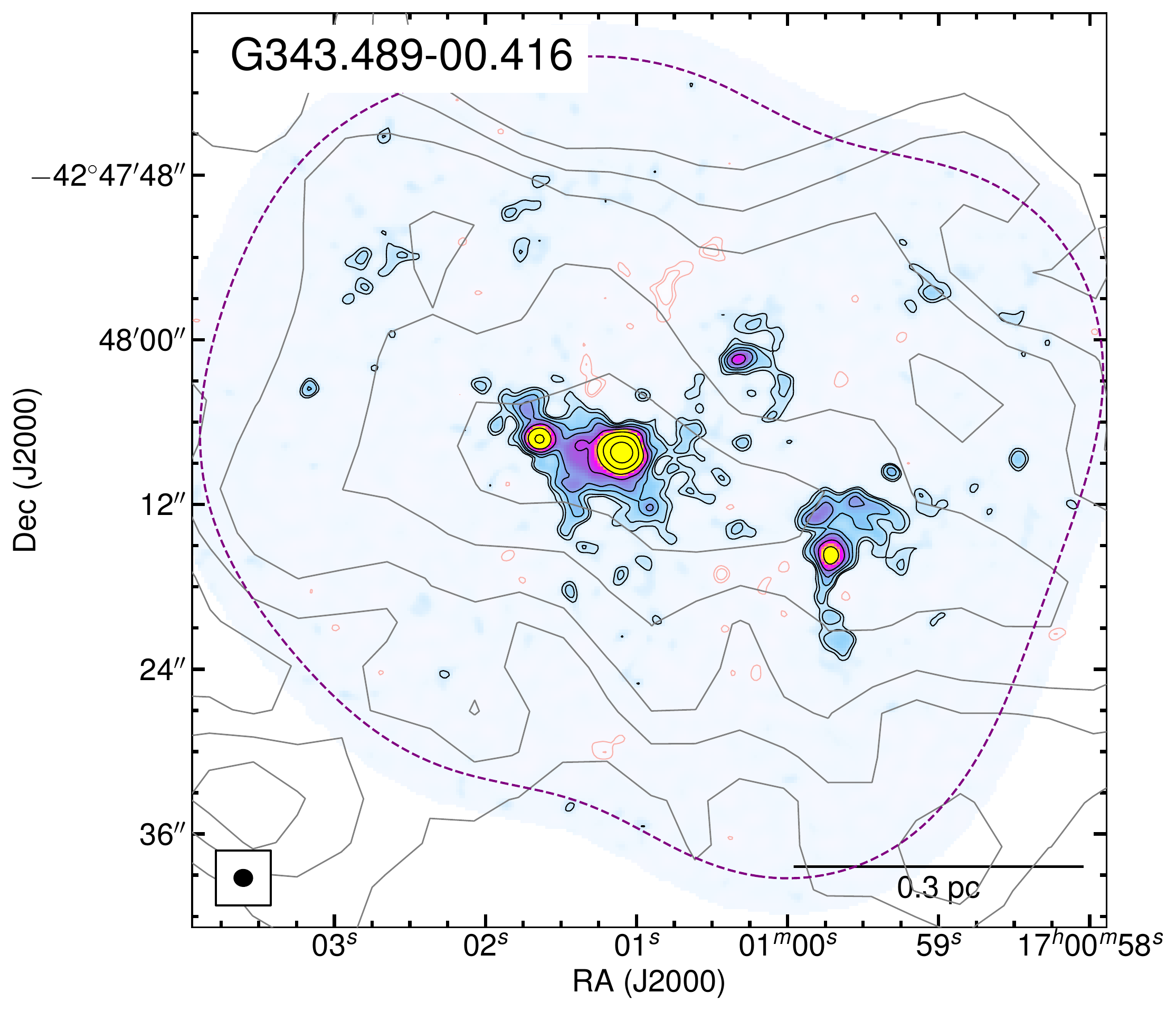}\hspace{-0.6cm}
\includegraphics[angle=0,scale=0.42]{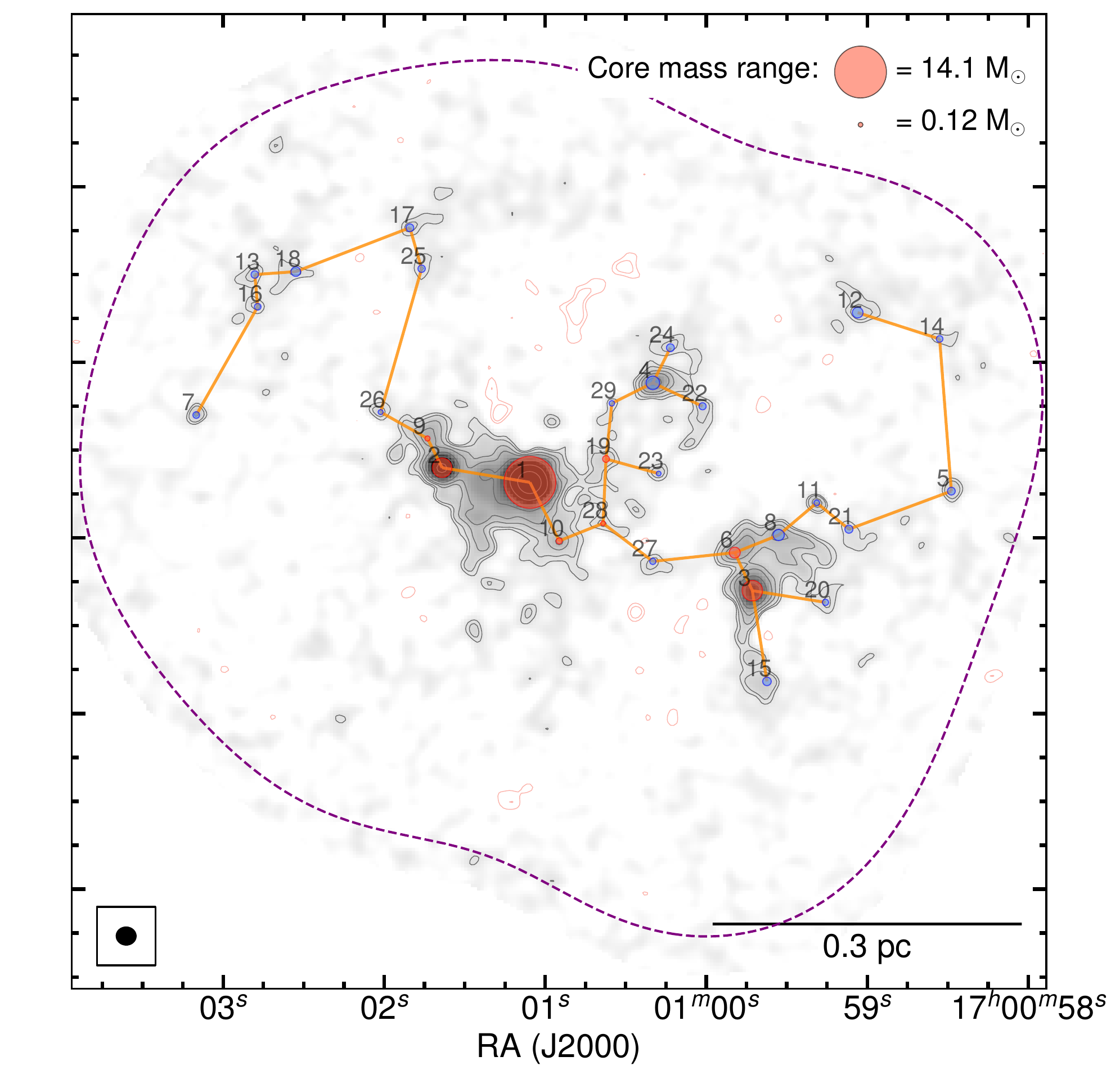}
\end{center}
\caption{Same as in Figure~\ref{ALMA_cont1}, except for ALMA contour levels of -4, -3, 3, 4, 6, 9, 12, 16, and 22 $\times$ $\sigma$, with $\sigma = 0.070$ mJy beam$^{-1}$,
 for G341.039--00.114 (1\farcs2 angular resolution); and -4, -3, 3, 4, 6, 8, 12, 20, 40, and 100 $\times$ $\sigma$, with $\sigma = 0.068$ mJy beam$^{-1}$, for
  G343.489--00.416 (1\farcs2 angular resolution).
}
\label{ALMA_cont6}
\end{figure*}

\section{Results}

\subsection{Dust Continuum Emission}

Figures~\ref{ALMA_cont1},~\ref{ALMA_cont2},~\ref{ALMA_cont3},~\ref{ALMA_cont4},~\ref{ALMA_cont5},~and~\ref{ALMA_cont6}
 show the 1.34 mm dust continuum images of the combined 12 and 7 m 
arrays. For comparison, the 870 $\mu$m dust continuum emission from the single-dish survey ATLASGAL 
is overlaid.  
ALMA dust continuum emission was successfully detected in all 12 targets. The small scale structure resolved 
with ALMA presents different morphologies and is roughly in agreement with the single-dish emission delineated by 
ATLASGAL. Some sources are filamentary (e.g., G331.372--00.116 and G341.039--00.114), while others are rather clumpy  
(e.g., G028.273--00.167 and G340.232--00.146). 

Integrating the flux over the compact and extended emission, the combined data sets (12 + 7 m) have between 1.1 and 7.1 (on average 2.6)
 times more flux than the 12 m alone images. All dust continuum images show more structures in the combined data sets (12 + 7 m) and,
  unless is explicitly stated, all analyses will be carried out on the combined data sets. In the absence of continuum emission observations with
   single-dish at 1.34 mm, the 870  $\mu$m emission was scaled by assuming a dust emissivity spectral index ($\beta$) of 1.5 to estimate
    how much flux is recovered by ALMA. Consistent with SMA/ALMA observations in other IRDC studies \citep[e.g.,][]{Wang14,Sanhueza17,Liu18b},
     between 10 and 33\% (average of 21\%) of the single-dish emission is recovered. This relatively low flux recovery likely indicates that dust and 
     gas in the clumps is distributed on large scales ($\gtrsim$20\arcsec). Therefore, most of the mass at the earliest stages of high-mass star 
     formation is diffuse and not (yet) confined in cores. 

\subsection{Extraction of Core's Properties}

To measure the integrated flux,  peak flux, size, and position of cores from the dust continuum images, we have
 adopted the dendrogram technique \citep{Rosolowsky08}. An intensity threshold of 2.5$\sigma$, step of 1.0$\sigma$,
  and a minimum number of pixels equal to those contained in half of each synthesized beam were used to define 
   the smaller structures called ``leaves'', which are defined as cores ($\sigma$ equal to the rms noise in Table~\ref{obs-param}). 
   Finally, cores with integrated flux densities smaller than 3.5$\sigma$ were filtered to eliminate spurious  detections. After core
      identification, all fluxes were corrected by the primary beam response. A total of 301 cores were detected in the 12 
    IRDC clumps (an average of 25 cores per clump).  On average, $\sim$26 cores per arcmin$^2$ are detected, which
     corrected by the clump distance  implies $\sim$18 cores per pc$^2$. The number of cores identified in each IRDC 
     ranges from 13 to 41. The broad range may indicate differences in the nature of each clump or just be related to 
     the mass sensitivity, which depends on the flux sensitivity, temperature, and distance to the source (see section~\ref{core_prop} for 
     the derivation of core mass). There is only a weak correlation between the flux sensitivity (rms in Table~\ref{obs-param}) and the
      number of cores identified, with a Spearman's rank correlation coefficient\footnote{The Spearman's rank correlation is a
       non-parametric measure of the monotonicity of the relationship between two variables. The advantage of the Spearman's correlation
         over others, e.g. Pearson correlation, is that is not constrained to only linear correlations and does not require Gaussian distributions
          to the data. The Spearman's coefficient, $\rho_s$, ranges from -1 to 1, with 0 indicating no correlation. The value of 1 
  implies exact increasing monotonic relation between two quantities, while -1 implies an exact decreasing monotonic relation. To interpret the Spearman's rank correlation,
   the following is usually applied to assess the significance of different $\rho_s$ values: $|\rho_s| \geq 0.5$ means strong correlation, $0.5 > |\rho_s| \geq 0.3$ means moderate 
   correlation, $0.3 > |\rho_s| \geq 0.1$ means weak correlation, and  $ 0.1 > |\rho_s|$ no correlation \citep{Cohen88}.} of $-0.16$. The number of detected cores is uncorrelated with 
   distance, with  $\rho_s$ equal to $-0.06$. As seen 
   in Figure~\ref{sigma_core-number}, the  3.5$\sigma$ point-source mass sensitivity has no correlation with the number of cores identified over
    this threshold, with $\rho_s$ equal to -0.09. Therefore, the core detection is independent of the mass sensitivity range proved by the observations. 
      Figure~\ref{sigma_core-number} also shows that more
      massive clumps tend to fragment into more cores than less massive clumps. The group of 6 clumps with a below-average ($<$25) core 
      count has an average clump mass of $\sim$770 \Msun, while the group above the average has 
      an average mass of 1560 \Msun. Table~\ref{tbl-dendro} displays position, peak flux, integrated flux, and radius for each individual core derived from 
     dendrograms. The radius corresponds to half of the geometric mean between the deconvolved major and minor axes of the ellipse determined 
     via dendrograms. All fluxes are primary beam corrected. Cores are named ALMA1, ALMA2, ALMA3... in order of descending peak intensity. Among all clumps, seven 
      cores are located at the edge of the images ($\sim$20-30\% power point) where flux measurements are more uncertain. They have been excluded
       from the forthcoming analyses in Section~\ref{disscu}. However, their properties are still listed in Table~\ref{tbl-dendro}.

\begin{deluxetable*}{lcccccccc}
\tabletypesize{\footnotesize}
\tablecaption{Core Parameters Obtained from Dendrograms \label{tbl-dendro}}
\tablewidth{0pt}
\tablehead{
\colhead{IRDC} & \colhead{Core} & \multicolumn{2}{c}{\underline {~~~~~~~~~~~~Position~~~~~~~~~~~~}} & \colhead{Peak} & \colhead{Integrated} & \colhead{Radius}  & \colhead{Core} & \colhead{Notes\tablenotemark{b}}  \\
\colhead{Clump} & \colhead{Name} & \colhead{$\alpha$(J2000)} & \colhead{$\delta$(J2000)}           & \colhead{Flux}   &\colhead{Flux}            &\colhead{} &
 \colhead{Classification\tablenotemark{a}} & \colhead{} \\
\colhead{} &  \colhead{} & \colhead{}& \colhead{}&\colhead{(mJy beam$^{-1}$)}                & \colhead{(mJy)} & \colhead{(\arcsec)} & \colhead{} & \colhead{} \\
}
\startdata
G010.991-00.082 & ALMA1 & 18:10:06.66 & -19:27:44.5 & 2.70 & 12.63 & 1.35 & 3 & 0 \\
G010.991-00.082 & ALMA2 & 18:10:06.37 & -19:28:13.1 & 2.33 & 2.80 & 0.50 & 3 & 0 \\
G010.991-00.082 & ALMA3 & 18:10:07.33 & -19:28:01.5 & 2.27 & 4.91 & 0.71 & 1 & 0 \\
G010.991-00.082 & ALMA4 & 18:10:06.93 & -19:27:34.5 & 1.90 & 4.04 & 0.77 & 3 & 0 \\
G010.991-00.082 & ALMA5 & 18:10:07.77 & -19:28:07.7 & 1.40 & 4.33 & 0.83 & 0 & 0 \\
\enddata
(This table is available in its entirety in a machine-readable form in the online
journal. A portion is shown here for guidance regarding its form and content.)
 \tablenotetext{a}{Core classification ranges from 0 to 3 meaning: 0 = prestellar candidate, 1 = only molecular outflow emission is detected, 2 = only warm core line emission is detected, and  
 3 = both protostellar indicators are detected.} 
 \tablenotetext{b}{Cores indicated with 0 are used in the analysis in Section~\ref{disscu}, while cores indicated with 1 are not used because they locate near the edge of the images (7 cores; properties are still given here for completeness).}
\end{deluxetable*}

\begin{figure}[!h]
\begin{center}
\includegraphics[angle=0,scale=0.4]{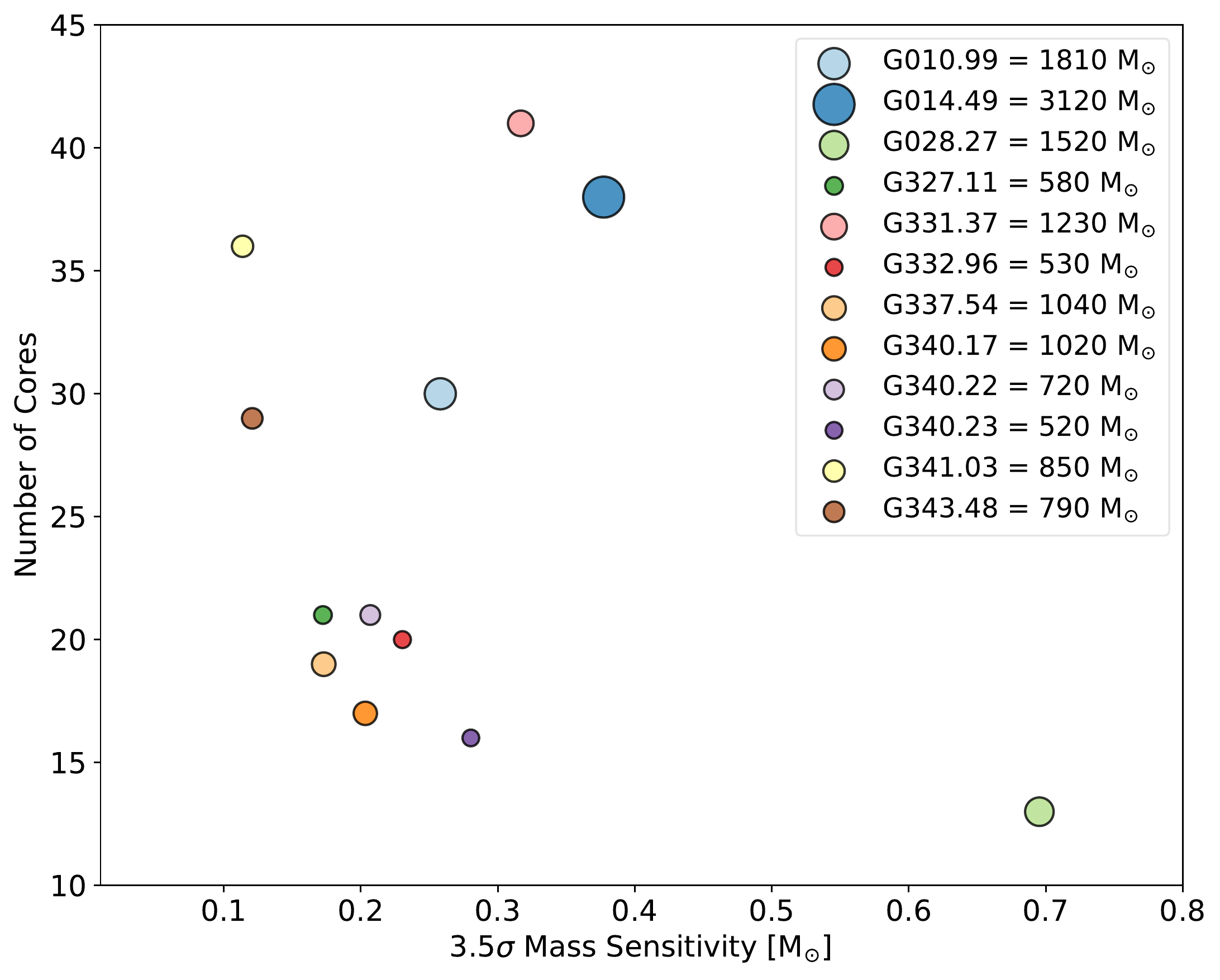}
\end{center}
\caption{Number of cores detected by clump against the 3.5$\sigma$ mass sensitivity. The value of sigma corresponds to 
the point source sensitivity at the clump's temperature. The size of the circles depends on the mass of the clump as shown 
on the label. Core detection is independent of mass sensitivity, ignoring the outlier at the bottom, right. Massive
 clumps tend to fragment more than less massive clumps.}
\label{sigma_core-number}
\end{figure}

Using the same set of input parameters for dendrograms in the 12 m alone images, a total of 242 cores were 
detected (20\% less than in the combined images). By adding the more extended flux recovered by the 7 m array, dust emission in the combined 
images increases S/N ratios above the 3.5$\sigma$ threshold allowing  the detection of more cores. On average, cores detected in the combined
 images have higher integrated fluxes by a factor of 1.6 (with $\sim$75\% of integrated fluxes increasing by a factor lower than 2). The inclusion of 
 the 7 m array, with its maximum recoverable scale of $\sim$19\arcsec\ is thus key to recover the flux from 1-2\arcsec\ cores. The cores sizes are 
 much smaller ($\lesssim$10\%) than the maximum recoverable scale achieved in our observations and only the diffuse, lower density intra-clump 
 emission is filtered out.   

\subsection{Evolutionary Stage of the Cores}
\label{evo_stage}

Because the clumps in this study have no emission detected at {\it Spitzer} wavelengths nor 
at 70 $\mu$m emission from $\it Herschel$, all detected ALMA cores would be prestellar 
candidates if we had no molecular line information at high-angular resolution. Therefore, 
the evolutionary stage of the cores was assessed by systematically searching for 
molecular outflows and/or ``warm core" line emission. 

In 52 (17\%) cores, molecular outflows were evident in the CO, SiO, and/or H$_2$CO lines (Li et al., in prep.).  If outflows were
 detected in any of these tracers, the core was classified as protostellar. In 
this work, we refer as ``warm core" tracers to those molecular transitions with upper 
energy levels ($E_u$) larger than 22 K (defined by the $E_u$ of the observed deuterated molecules), which are temperatures lower
than  those from typical ``hot core" tracers ($\gtrsim$100 K).
 Cores with ``warm core" line emission would be in an evolutionary stage prior to the ``hot core" phase typically found in 
 high-mass star formation. Therefore, if an ALMA core is associated with any of the two H$_2$CO warm transitions J = 3$_{2,2}$-2$_{2,1}$ ($E_u/k$ = 68.09 K) 
 and J = 3$_{2,1}$-2$_{2,0}$ ($E_u/k$ = 68.11 K), or the CH$_3$OH J$_k$ = 4$_{2,2}$-3$_{1,2}$  ($E_u/k$ = 45.46 K) line, it was classified as protostellar
  (total of 62 cores, 21\% of the whole sample). We note, however, that these transitions with high $E_u/k$ could be sub-thermally excited and not really tracing 
  star formation activity. Therefore, adopting their detection as a star formation indicator works as a strict limit that contributes to obtain a pristine prestellar core sample.  
 The 26 (9\%) cores with both outflow and warm core lines are presumably the more evolved. Cores with an absence of both molecular outflow and warm core tracers 
 were categorized as prestellar. From the total of 301 cores, 213 (71\%) are classified as 
 prestellar, while 88 (29\%) as protostellar. Table~\ref{tbl-dendro} includes a description for each 
individual core: if molecular outflows and warm core tracers are detected, or if the 
core is prestellar. In the core classification column (Table~\ref{tbl-dendro}) a 0 is given for prestellar cores, while for the protostellar cores a 1 is given when molecular
 outflow emission alone is detected, a 2 
when warm core line emission is detected, and a 3 when both protostellar indicators are detected (which would correspond to the most evolved cores in the sample). 
 Excluding the cores located at the edges, for the discussion in Section~\ref{disscu}, we have 294 cores in total, with 210 (71\%) prestellar candidates. 

Based on this classification scheme, half of the clump sample shows evidence for some star formation activity, with 
$<$20\% of cores having signs of star formation. Among them, only one clump (G340.222--00.167) 
 seems completely prestellar. Considering that G340.222--00.167 is the most compact IRDC in the sample, this may indicate that the G340.222--00.167 is young and maybe is still accreting mass to become
  a larger more massive IRDC. The most evolved clumps are G014.492--00.139 and 
G337.541--00.082, with $\gtrsim$50\% of cores classified as protostellar. We therefore suggest that most of 70 $\mu$m dark clumps 
indeed have nascent, but deeply embedded, star formation activity. However,  this star formation activity is, at the current evolutionary stage,
 apparently only from low-mass protostars that may become high-mass in the future, as discussed in the following section. 

\section{Discussion}
\label{disscu}

\subsection{Core Physical Properties}
\label{core_prop}

The mass of the cores was computed assuming optically thin dust continuum emission as follows:
\begin{equation}
M_{\rm core} = \mathbb{R}~\frac{F_\nu D^2}{\kappa_\nu B_\nu (T)}~,
\label{eqn-dust-mass}
\end{equation}
where $F_\nu$ is the measured integrated source flux, $\mathbb{R}$ is the gas-to-dust mass ratio, 
$D$ is the distance to the source, $\kappa_\nu$ is the dust opacity per gram of dust, and $B_\nu$ is
 the Planck function at the dust temperature $T$. A value of 0.9 cm$^2$ g$^{-1}$ is adopted for
 $\kappa_{1.3 mm}$, which corresponds to the opacity of dust grains with thin ice
 mantles at gas densities of 10$^6$ cm$^{-3}$ \citep{Ossenkopf94}. In the absence of dust temperature 
 measurements at high angular resolution ($\sim$1\arcsec), we have adopted the clump's dust temperature 
 derived by \cite{Guzman15} using {\it Herschel} and APEX telescopes. Nevertheless, given the early evolutionary 
 stage of the clumps and the lack of hot cores, it is expected that the dust temperature throughout each cluster member  
 does not strongly vary.  A gas-to-dust mass ratio of 100 was 
 assumed in this work. The number density, $n$(H$_2$), was calculated by assuming a spherical core and using the
  molecular weight per hydrogen molecule ($\mu_{\rm H_2}$) of 2.8. Masses, number densities, surface densities ($\Sigma$ = $M_{\rm core}/(\pi r^2)$), 
  and peak column densities ($N_{\rm peak}(\rm H_2)$) for all cores are listed in Table~\ref{tbl-core_prop}. The average
   core parameters for each clump are summarized in Table~\ref{tbl-av-cores}.  
   
    In spite of dust emission being the most reliable method for mass determination of star-forming cores, there are still several sources of uncertainty.
   \cite{Sanhueza17} searched in the literature for possible values of $\mathbb{R}$ and $\kappa_\nu$, finding the extreme possible values. Assuming the possible values are distributed in a 
   uniform way between the extreme values, the standard deviation can be estimated \citep[see details in][]{Sanhueza17}. For $\mathbb{R}$,  1$\sigma$ = 23 corresponds to 
   23\% of uncertainty of the adopted value of 100. For $\kappa_\nu$, 1$\sigma$ = 0.25 cm$^2$ g$^{-1}$ corresponds to a 28\% uncertainty in the adopted value of 0.9 cm$^2$ g$^{-1}$. 
   Both $\mathbb{R}$ and $\kappa_\nu$ combined add an ``intrinsic" uncertainty of 32\% to the mass determination. Considering an absolute flux uncertainty of 10\% for ALMA observations 
   in band 6\footnote{Absolute flux uncertainty quoted for band 6 in the ALMA proposal guide.}, a dust temperature uncertainty $<$20\%, and a distance uncertainty of $\sim$10\%, we estimate mass, number density, and surface density uncertainties of $\sim$50\%.

 Figure~\ref{clump_core_mass} shows the core masses for each clump. Core masses range from 0.12 to 30.4 \Msun\ and
  8 cores have masses larger than 10 \Msun. There is no correlation between the clump mass and the maximum core mass, with 
  a Spearman's rank correlation coefficient, $\rho_s$, of 0.08. 
  Therefore,  at the earliest stages of fragmentation traced in the present study, there is no preference for more massive clumps to form
   the most massive cores.  In Figure~\ref{clump_core_mass}, the most massive prestellar core in each clump is marked
    with a black cross. In four clumps, the most massive core is a prestellar core.

\begin{deluxetable*}{lcccccc}
\tabletypesize{\footnotesize}
\tablecaption{Calculated Properties for the Whole Core Sample \label{tbl-core_prop}}
\tablewidth{0pt}
\tablehead{
\colhead{IRDC}  & \colhead{Core} & \colhead{Mass}  &  \colhead{Radius} &  \colhead{$n$(H$_2$)}                 &   \colhead{$\Sigma$}             &                         \colhead{$N_{\rm peak}(\rm H_2)$}\\
\colhead{Clump} & \colhead{Name}     & \colhead{(\Msun)}    &\colhead{(AU)} &\colhead{($\times$10$^6$ cm$^{-3}$)}& \colhead{(g cm$^{-2}$)}  &   \colhead{($\times$10$^{23}$ cm$^{-2}$)}     \\
}
\startdata
G010.991-00.082 & ALMA1 & 8.09 & 5000 & 2.0 & 0.91 & 5.63\\
G010.991-00.082 & ALMA2 & 1.79 & 1840 & 8.7 & 1.49 & 4.86\\
G010.991-00.082 & ALMA3 & 3.15 & 2620 & 5.3 & 1.29 & 4.73\\
G010.991-00.082 & ALMA4 & 2.59 & 2870 & 3.3 & 0.89 & 3.95\\
G010.991-00.082 & ALMA5 & 2.77 & 3080 & 2.9 & 0.82 & 2.92\\
\enddata
(This table is available in its entirety in a machine-readable form in the online
journal. A portion is shown here for guidance regarding its form and content.)
\tablecomments{ $n$(H$_2$), $\Sigma$, and $N(\rm H_2)$ correspond to number density, surface density, 
and peak column density, respectively.} 
\end{deluxetable*}

\begin{figure}[h!]
\begin{center}
\includegraphics[angle=0,scale=0.39]{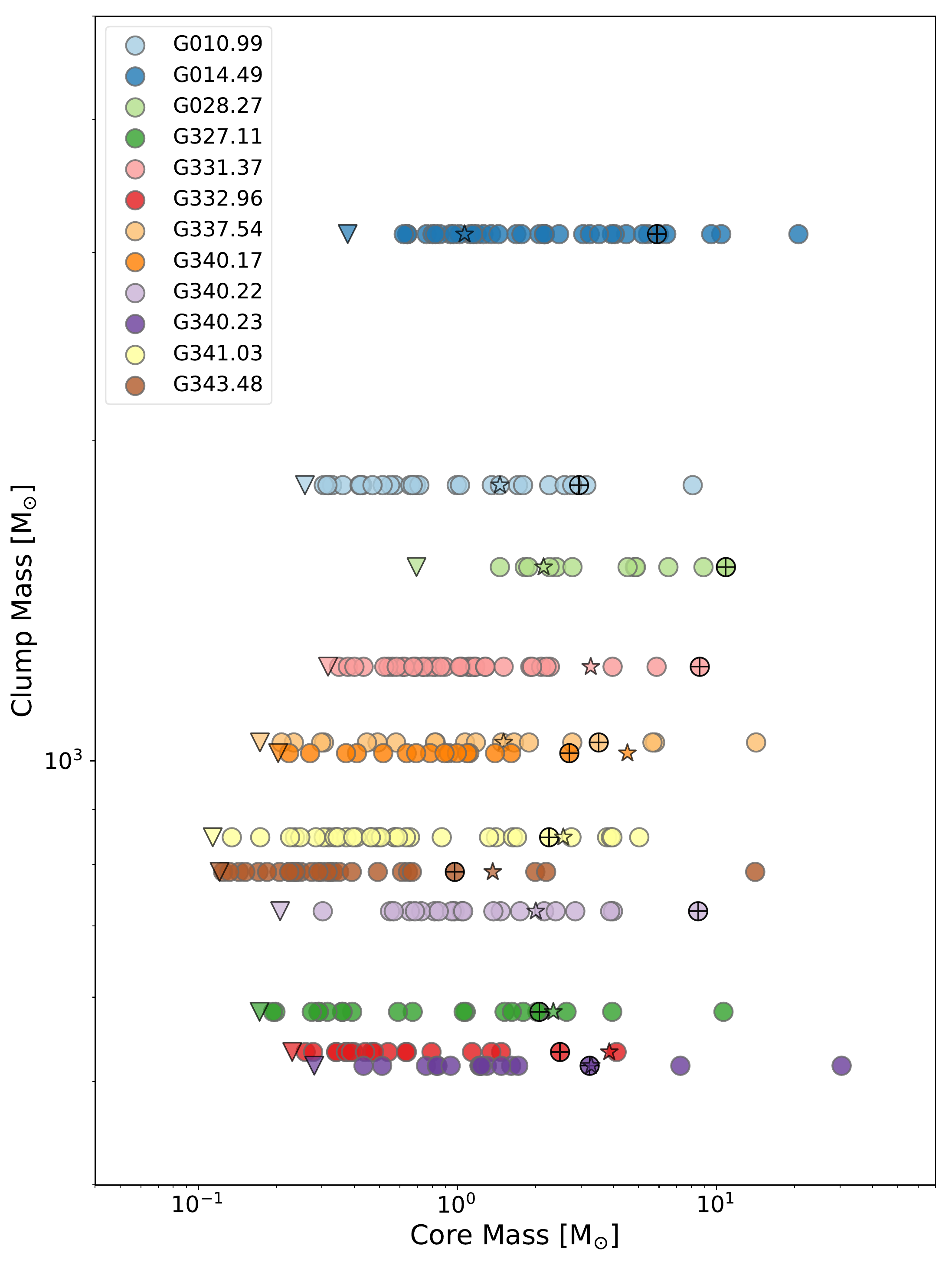}
\end{center}
\caption{Clump masses against the core masses. Triangle signs indicate the 3.5$\sigma$ level above which cores 
are defined. Stars signs show the value of the Jeans mass of each clump. Encircled plus signs indicate the most massive 
prestellar core in each clump. No correlation between the clump mass and the maximum core mass 
is found. A large population of low-mass cores ($<$1 \Msun) is detected. The range of core masses is well explained 
by thermal Jeans fragmentation, without the need of invoking turbulent Jeans fragmentation.}
\label{clump_core_mass}
\end{figure}

  Figure~\ref{four_plots} shows the  core distribution of sizes, peak column densities, surfaces densities, and volume densities as a function of the 
core mass. The purpose of these plots is to show the distribution of the core properties at the earliest stage of high-mass star formation. The radii
 strongly correlate with mass, $\rho_s$ equal to 0.71, and the correlation persists per individual clumps (see Figure~\ref{four_rad_mass} in Appendix~\ref{appimages}). We refrain 
 from calculating correlation factors to other physical properties due to their intrinsic correlation on physical quantities (e.g., flux, mass, distance).
Most peak column densities ($\sim$80\%) are between 6\x10$^{22}$ and 3\x10$^{23}$ cm$^{-2}$. The bulk of cores ($\sim$90\%) have surface densities between 0.1 and
 1 g cm$^{-2}$. A non-negligible number	 of 31 cores ($\sim$10\%) have $\Sigma$ values larger than 1 g cm$^{-2}$. This value has been suggested by \cite{Krumholz08}
  to be the minimum necessary (but not sufficient) to halt fragmentation and allow the formation of high-mass stars. Volume densities are rather high, with more than 50\% having values
   larger than 10$^{6}$ cm$^{-3}$. The effect of assuming 30 K  for protostellar cores, instead of the clump temperatures that are about a factor 2 lower, can be seen in the Appendix~\ref{appimages}, Figure~\ref{four_30K}.
   
\begin{figure*}
\begin{center}
\includegraphics[angle=0,scale=0.75]{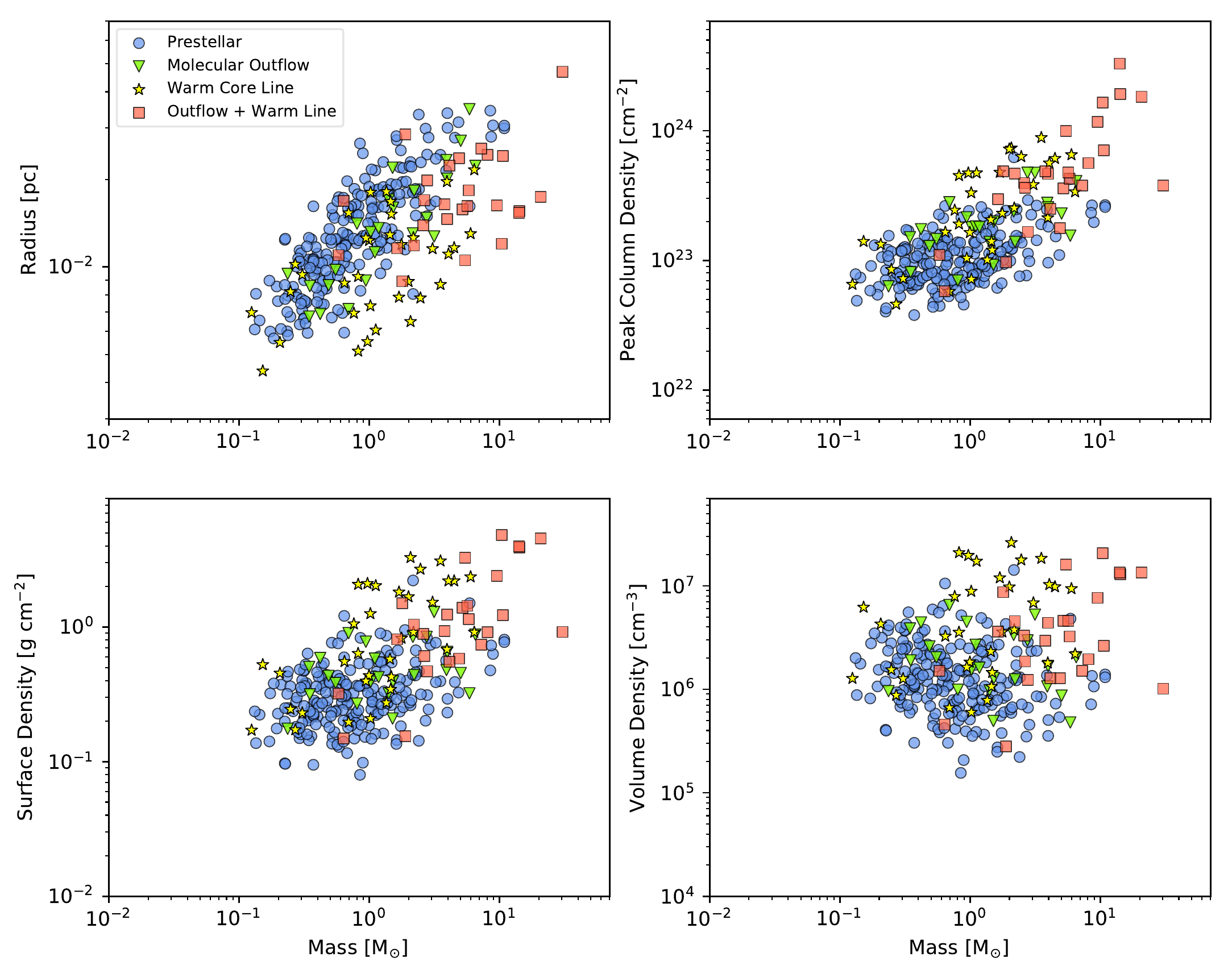}
\end{center}
\caption{Radius, peak column, surface density, and volume density of cores against the core mass color-coded by protostellar activity (prestellar, molecular outflow only, warm core line only, and 
both protostellar indicators; see Section~\ref{evo_stage}). The purpose of these scatter plots is to show the distribution 
of core properties. }
\label{four_plots}
\end{figure*}

\begin{deluxetable*}{lcccccccccc}
\tabletypesize{\footnotesize}
\tablecaption{Overall Properties per Clump of the Embedded ALMA Cores \label{tbl-av-cores}}
\tablewidth{0pt}
\tablehead{
\colhead{IRDC} & \colhead{1$\sigma$ Mass} & \colhead{Number}    & \multicolumn{2}{c}{\underline {~~~Core Mass~~~}}   & \multicolumn{5}{c}{\underline {~~~~~~~~~~~~~~~~~~~~~~~~~~~~~~~~~~Mean~~~~~~~~~~~~~~~~~~~~~~~~~~~~~~~~~~}}   & \colhead{Number of} \\
\colhead{Clump} &        \colhead{Sensitivity}            &  \colhead{of Cores}   & \colhead{Min.} & \colhead{Max.}  &\colhead{Mass}   &\colhead{Radius} &\colhead{$n$(H$_2$)}    &  \colhead{$\Sigma$}            &  \colhead{$N_{\rm peak}(\rm H_2)$} &  \colhead{Pre-/Proto-stellar}\\
\colhead{} &            \colhead{(\Msun)}               &     \colhead{}              & \colhead{(\Msun)}     & \colhead{(\Msun)}        &\colhead{(\Msun)}&\colhead{(AU)} &\colhead{($\times$10$^6$ cm$^{-3}$)}& \colhead{(g cm$^{-2}$)}  &   \colhead{($\times$10$^{23}$ cm$^{-2}$)}      & \colhead{Cores}\\
}
\startdata
G010.991--00.082 & 0.074 & 28 & 0.31 & 8.1 & 1.35 & 2370 & 2.9 & 0.57 & 2.02 & 18/10 \\
G014.492--00.139 & 0.108 & 37 & 0.62 & 20.7 & 3.30 & 2290 & 8.9 & 1.63 & 4.79 & 12/25 \\
G028.273--00.167 & 0.199 & 13 & 1.46 & 10.9 & 4.93 & 4810 & 1.5 & 0.57 & 1.91 & 11/2 \\
G327.116--00.294 & 0.049 & 21 & 0.19 & 10.6 & 1.54 & 2940 & 1.6 & 0.39 & 1.54 & 17/4 \\
G331.372--00.116 & 0.091 & 39 & 0.35 & 8.6 & 1.40 & 3460 & 1.0 & 0.29 & 0.95 & 32/7 \\
G332.969--00.029 & 0.066 & 20 & 0.26 & 4.1 & 0.87 & 2670 & 1.2 & 0.28 & 0.95 & 18/2 \\
G337.541--00.082 & 0.049 & 19 & 0.21 & 14.2 & 2.29 & 2840 & 2.7 & 0.67 & 2.66 & 10/9 \\
G340.179--00.242 & 0.058 & 16 & 0.22 & 2.7 & 0.91 & 3710 & 0.6 & 0.18 & 0.66 & 13/3 \\
G340.222--00.167 & 0.059 & 21 & 0.30 & 8.5 & 1.79 & 4100 & 0.8 & 0.27 & 1.01 & 21/0 \\
G340.232--00.146 & 0.080 & 16 & 0.43 & 30.4 & 3.44 & 3510 & 1.6 & 0.46 & 1.53 & 12/4 \\
G341.039--00.114 & 0.032 & 35 & 0.13 & 5.0 & 1.09 & 2520 & 1.9 & 0.39 & 1.50 & 25/10 \\
G343.489--00.416 & 0.035 & 29 & 0.12 & 14.1 & 0.92 & 1810 & 2.9 & 0.51 & 2.53 & 21/8 \\
\enddata
\tablecomments{Total of 294 cores with 210 prestellar candidates. $n$(H$_2$), $\Sigma$, and $N(\rm H_2)$ correspond to number density, surface density, 
and peak column density, respectively.} 
\end{deluxetable*}

  \subsection{Low-mass Core Population}
  
  Notably, a large population of low-mass cores ($<$1 \Msun) is detected, contrary to what has been observed with ALMA at similar mass
   sensitivity in more evolved star-forming regions (e.g., {IRDC G28.34+0.06}, \citealt{Zhang15}; {G11.92-0.61}, \citealt{Cyganowski17}).  
  From the total of 294 cores, 159 cores (54\%) have masses $<$1 \Msun.  We find that 56\% of the core population with masses $<$1 \Msun\
   (55\% for $<$2 \Msun) are located outside a circle of 25\farcs2 diameter (equivalent to the primary beam of the ALMA 12 m antenna) 
   centered on the ATLASGAL position. With a single pointing observation, 
  \cite{Zhang15} find a lack of a widespread low-mass protostellar population and suggest that low-mass
   protostars form after high-mass stars. However, \cite{Kong18a} observe the same IRDC on a large mosaic revealing cores previously
    undetected, which may suggest that mapping a larger area plays an important role in detecting a low-mass population of cores. 
    This may be the case in the work by \cite{Cyganowski17}, which indeed find a widespread population of low-mass cores ($\sim$1-2 \Msun).     
    Based on different approaches, \cite{Foster14} and later \cite{Pillai19} suggest that low-mass stars may form earlier or coevally with
     high-mass stars. \cite{Foster14} observe an IRDC using deep near-infrared observations and discover a distributed population of low-mass protostars. 
     Part of the area is covered with ALMA \citep{Sakai13,Sakai15,Sakai18,Yanagida14} and most of the low-mass protostars revealed in near IR 
     have no counterpart in 1.3 mm dust continuum emission. The  low-mass protostars may presumably be a relatively older population with no 
     significant envelope to be detected by ALMA. Using CO J=2-1 outflow emission, \cite{Pillai19} infer that low-mass 
    protostars have formed before or coevally with high-mass cores. 
    In our work, which samples a greater number of clumps, cover a lager mosaic area per clump, and recovers extended flux using the 7 m array, 
    we find  a large, widespread population of low-mass cores ($<$1 \Msun). This suggests that the seeds of high-mass stars form and evolve together
     with the seeds of low-mass stars.

\subsection{Lack of High-Mass Prestellar Cores}

Table~\ref{tbl-most-massive-cores}  presents a list of the cores with masses larger than 10 \Msun. When a clump has no cores with masses larger than 
10 \Msun, the most massive core is listed. Half of clumps have cores with masses above 10 \Msun\ and two of them, G014.492--00.139 and 
G028.273--00.167, have two. Except for the two cores in G028.273--00.167, all cores with
 masses larger than 10 \Msun\ are protostellar. All cores in Table~\ref{tbl-most-massive-cores} are resolved or barely resolved. All eight cores with masses 
 larger than  10 \Msun\  have surface densities $\gtrsim$0.8 g cm$^{-2}$, similar to values found in the most massive cores embedded in more evolved 
 IRDCs \cite[e.g.,][]{Tan13,Kong17}. Of these eight cores, four of them have extreme volume densities of few times 10$^7$ cm$^{-3}$ and peak 
 column densities of few times 10$^{24}$ cm$^{-2}$.
 
 Following the discussion from \cite{Sanhueza17}, the definition of a bonafide ``high-mass prestellar core'' is rather vague.  \cite{Longmore11} suggest that 
 in order to form an O-type star through the direct collapse of a core, the core should have of the order of 100 \Msun. This is consistent with the simulations 
 of \cite{Krumholz07}, in which a high-mass star of 9 \Msun\ is formed from a turbulent, virialized core of 100 \Msun. \cite{Tan14} suggest
 that a high-mass prestellar core should contain $\sim$100 core Jeans masses. Another important piece of information in the definition of a high-mass 
 prestellar core is that $\sim$80\% of high-mass stars are found in binary systems \citep{Kouwenhoven05,Chini12} and that the majority of the high-mass
  systems contain pairs of similar mass. Combining all this information, it seems clear that a high-mass prestellar core should have several tens of solar masses. 
In this work, we define a high-mass core as a core with a mass larger than $\sim$30 \Msun. This definition is consistent with the star formation efficiency\footnote{We note, 
however, that in clump-feed star formation scenarios, star formation efficiencies are larger than 100\%\ for the cores forming high-mass stars. This is because cores start with 
masses lower than 8 \Msun\ and end forming a high-mass star ($>$8 \Msun)} of 30\% derived by
 \cite{Alves07} in the Pipe dark cloud \citep[also tentatively determined in the Cygnus X complex by][]{Bontemps10}, assuming that the initial mass function is a
  direct product of the core mass function as stated for the turbulent core accretion model, e.g., \cite{Tan14}. The adopted value of 30 \Msun\ is also consistent with the core Jeans 
  mass determined for the most massive prestellar cores detected in this sample. The prestellar cores with masses of $\sim$11 \Msun\ (density $\sim$$1.4\times10^6$ cm$^{-3}$) have a
   Jeans mass of $\sim$0.3 \Msun. Therefore, the most massive prestellar cores contain only $\sim$40 core Jeans masses. In order to reach 100 Jeans masses \citep{Tan14}, these cores would 
  need instead a mass of 30 \Msun\  (maintaining the same density). 
 
 In the sample observed in the pilot survey, there are no high-mass prestellar cores. Remarkably, high-mass prestellar cores are inexistent even adopting higher star formation efficiencies 
 of 40-50\%. The most massive core (30.4 \Msun), located in G340.232--00.146, 
 shows evidence of protostellar activity, based on warm-core line emission and molecular outflows. However, this core is rather large (radius of
  $\sim$10,000 AU) and after visual inspection of the dendrogram leaf structure, it seems likely that higher angular resolution observations will reveal 
  a more fragmented structure with smaller condensations. 

\begin{deluxetable*}{lcccccccc}
\tabletypesize{\footnotesize}
\tablecaption{Properties of Most Massive Cores \label{tbl-most-massive-cores}}
\tablewidth{0pt}
\tablehead{
\colhead{IRDC}  & \colhead{Core}  & \colhead{Mass} & \colhead{Mass/$M_J$}  &  \colhead{Radius} &  \colhead{$n$(H$_2$)}                 &   \colhead{$\Sigma$}             &                       \colhead{$N_{\rm peak}(\rm H_2)$} &  \colhead{Core}\\
\colhead{Clump} & \colhead{Name} & \colhead{(\Msun)}     & \colhead{}    &\colhead{(AU)} &\colhead{($\times$10$^6$ cm$^{-3}$)}& \colhead{(g cm$^{-2}$)}  &   \colhead{($\times$10$^{23}$ cm$^{-2}$)}      & \colhead{Classification}\\
}
\startdata
G010.991--00.082 & ALMA1  & 8.1 & 5.5 & 5000 & 2.0 & 0.91 & 5.63 & protostellar \\
G014.492--00.139 & ALMA1  &  20.7 & 19.4 & 3590 & 13.5 & 4.52 & 18.30 & protostellar \\
G014.492--00.139 & ALMA2  &  10.4 & 9.8 & 2480 & 20.8 & 4.78 & 16.50 & protostellar  \\
G028.273--00.167 & ALMA2  &  10.9 & 5.1 & 6180 & 1.4 & 0.80 & 2.67 & prestellar  \\
G028.273--00.167 & ALMA3  &  10.9 & 5.1 & 6310 & 1.3 & 0.77 & 2.58 & prestellar \\
G327.116--00.294 & ALMA1  &  10.6 & 4.5 & 4950 & 2.6 & 1.22 & 7.06 & protostellar  \\
G331.372--00.116 & ALMA1  &  8.6 & 2.6 & 5780 & 1.4 & 0.73 & 2.66 & prestellar  \\
G332.969--00.029 & ALMA1  &  4.1 & 1.1 & 4600 & 1.3 & 0.55 & 2.50 & protostellar  \\
G337.541--00.082 & ALMA1  &  14.2 & 9.4 & 3210 & 13.0 & 3.88 & 19.20 & protostellar  \\
G340.179--00.242 & ALMA4  &  2.7 & 0.6 & 6160 & 0.3 & 0.20 & 0.77 & prestellar \\
G340.222--00.167 & ALMA2  &  8.5 & 4.2 & 7100 & 0.7 & 0.47 & 2.33 & prestellar  \\
G340.232--00.146 & ALMA1  &  30.4 & 9.3 & 9670 & 1.0 & 0.91 & 3.78 & protostellar  \\
G341.039--00.114 & ALMA6  &  5.0 & 2.0 & 5600 & 0.9 & 0.45 & 2.30 & protostellar \\
G343.489--00.416 & ALMA1  &  14.1 & 10.3 & 3170 & 13.4 & 3.95 & 32.90 & protostellar \\
\enddata
\tablecomments{This table includes all cores with masses larger than 10 \Msun. When a clump has no core above 10 \Msun, 
the most massive core is listed. $M_J$ is the clump Jeans mass (see Table~\ref{tbl-global}).} 
\end{deluxetable*}

\subsection{Fragmentation}

If clump fragmentation is governed by thermal Jeans instabilities, the initially homogeneous 
gas fragments into smaller objects defined by the Jeans length ($\lambda_J$) and the Jeans mass ($M_J$):
 \begin{equation}
\lambda_J = \sigma_{\rm th}\left(\frac{\pi}{G\rho}\right)^{1/2}~,
\label{jeanslength}
\end{equation}
 and
\begin{equation}
M_J = \frac{4\pi\rho}{3}\left(\frac{\lambda_J}{2}\right)^3=\frac{\pi^{5/2}}{6}\frac{\sigma_{\rm th}^3}{\sqrt{G^3\rho}}~,
\label{jeansmass}
\end{equation}
where $\rho$ is the mass density  and
 $\sigma_{\rm th}$ is the thermal velocity dispersion (or isothermal sound speed, $c_s$) given by
\begin{equation}
\sigma_{\rm th} = \left(\frac{k_{\rm B}T}{\mu m_{\rm H}}\right)^{1/2}~.
\label{sigma-th}
\end{equation}
The thermal velocity dispersion is mostly dominated by H$_2$ and He, and  it should be 
derived by using the mean molecular weight per free particle, $\mu=2.37$.  
The mean Jeans length for all clumps is 0.14 pc, ranging from 0.06 to 0.24 pc. The mean Jeans mass is 2.5 \Msun, ranging 
from 1.1 to 4.5 \Msun. 
If the fragmentation is driven by turbulence, the turbulent Jeans lengths and masses can be derived by replacing $\sigma_{\rm th}$ by the 
observed clump velocity dispersion listed in Table~\ref{tbl-clumps}. The  turbulent Jeans length ($\lambda_{\rm turb}$) for the whole clump sample 
has a mean of 0.87 pc, ranging from 0.3 to 1.6 pc.  The  turbulent Jeans mass ($M_{\rm turb}$) for all clumps has a mean of 950 \Msun, ranging from 
40 to 4710 \Msun. Therefore, turbulent Jeans lengths and masses are at least 2.5 times and 16 times larger than the corresponding thermal ones
(on average 7 and 440 times larger, respectively). Table~\ref{tbl-global} displays in Column (1) the clump name, in Column (2) the thermal velocity dispersion,
 in Column (3) the Jeans mass, in Column (4) the Jeans length, in Column (5) the turbulent Jeans mass, and in Column (6) the turbulent Jeans length. 

\subsubsection{Jeans Length and Core Separation}
\label{mst_sec}

To quantify core separations to compare with Jeans lengths, we have used the minimum spanning tree (MST) method, first developed for 
astrophysical applications by \cite{Barrow85}. MST determines a set of straight lines connecting a set of points (cores in this case) 
that minimizes the sum of the lengths. More details on this method can be found, for example, applied to simulations in \cite{Wu17} and to observations 
\cite{Dib18}.  

Figure~\ref{ALMA_cont1},~\ref{ALMA_cont2},~\ref{ALMA_cont3},~\ref{ALMA_cont4},~\ref{ALMA_cont5},~and~\ref{ALMA_cont6} display  the MST for each
 clump and Table~\ref{tbl-global} lists (Column 7) the average minimum separation ($L_{\rm av}$) between cores determined by MST for
 each clump. The mean $L_{\rm av}$ for all clumps is 0.11 pc, ranging from 0.07 to 0.17 pc. However, $L_{\rm av}$ is the measured separation projected on the sky and 
 the real (unprojected) value is equal or longer. On average, the observed separation will be $2/\pi$ times smaller than the unprojected one\footnote{The average value for $\cos (i)$, with $i$ the 
 angle between the core separation and the observed projected separation, is given by 
  \begin{equation} \frac{1}{\pi} \int^{\frac{\pi}{2}}_{-\frac{\pi}{2}} \cos (x) dx = \frac{2}{\pi}  \end{equation}}. We therefore divide $L_{\rm av}$ 
 by this factor to obtain $L_{\rm av\,corr}$ (Table~\ref{tbl-global}, Column 8). Given that $L_{\rm av\,corr}$ is comparable or slightly larger to the Jeans length by a factor 0.7 to 2, but consistent within the uncertainties, 
   the clump fragmentation is governed by thermal Jeans fragmentation. Turbulent Jeans lengths are a factor 2, up to 10, larger than $L_{\rm av\,corr}$. We therefore discard
    turbulence Jeans fragmentation as the controlling process of the early stages of high-mass star and cluster formation found in these IRDCs.
 
\subsubsection{Jeans Mass}

We find that $\sim$74\% of cores have masses lower than the Jeans mass, further indicating that turbulence does not play an important 
role in the global fragmentation of IRDCs. A large population of cores with masses $\lesssim M_J$ favors competitive accretion and global hierarchical collapse scenarios. 
The remaining 26\% of cores have masses on average 3 times the Jeans mass (up to 19 \x\ $M_J$).  If these relatively massive cores remain as single objects at higher angular 
resolution observations, they would need additional support, by for example turbulence and/or magnetic field, to avoid fragmentation. After accreting material 
from their surroundings, these super-Jeans cores are prime candidates to evolve into high-mass cores and form high-mass stars (see Table~\ref{tbl-most-massive-cores} for the most massive cores). 

\begin{deluxetable*}{lcccccccc}
\tabletypesize{\normalsize}
\tablecaption{Global Structure Parameters \label{tbl-global}}
\tablewidth{0pt}
\tablehead{
\colhead{IRDC}  & \colhead{$\sigma_{\rm th}$} & \colhead{$M_J$} & \colhead{$\lambda_J$}  & \colhead{$M_{\rm turb}$} & \colhead{$\lambda_{\rm turb}$} &  \colhead{$L_{\rm av}$} &  \colhead{$L_{\rm av\,corr}$}&  \colhead{$\mathcal{Q}$}        \\
\colhead{Clump} &  \colhead{(km s$^{-1}$)}      & \colhead{(\Msun)}      & \colhead{(pc)}           & \colhead{(\Msun)}            & \colhead{(pc)}                             &\colhead{(pc)}                &\colhead{(pc)}                       & \colhead{}          \\
\colhead{(1)} &  \colhead{(2)}      & \colhead{(3)}      & \colhead{(4)}           & \colhead{(5)}            & \colhead{(6)}                             &\colhead{(7)}                &\colhead{(8)}                       & \colhead{(9)}          \\
}
\startdata
G010.991-00.082 & 0.20 & 1.5 & 0.09 & 350 & 0.57 & 0.08 & 0.12 & 0.74 \\
G014.492-00.139 & 0.21 & 1.1 & 0.06 & 520 & 0.49 & 0.07 & 0.11 & 0.77 \\
G028.273-00.167 & 0.20 & 2.2 & 0.13 & 130 & 0.53 & 0.17 & 0.27 & 0.65 \\
G327.116-00.294 & 0.22 & 2.3 & 0.12 & 40 & 0.31 & 0.10 & 0.16 & 0.66 \\
G331.372-00.116 & 0.22 & 3.3 & 0.18 & 650 & 1.03 & 0.11 & 0.18 & 0.69 \\
G332.969-00.029 & 0.21 & 3.9 & 0.23 & 1180 & 1.55 & 0.10 & 0.16 & 0.63 \\
G337.541-00.082 & 0.20 & 1.5 & 0.09 & 1430 & 0.93 & 0.10 & 0.15 & 0.66 \\
G340.179-00.242 & 0.22 & 4.5 & 0.24 & 1370 & 1.63 & 0.16 & 0.25 & 0.76 \\
G340.222-00.167 & 0.23 & 2.0 & 0.10 & 4710 & 1.34 & 0.13 & 0.20 & 0.78 \\
G340.232-00.146 & 0.22 & 3.3 & 0.18 & 560 & 0.98 & 0.12 & 0.19 & 0.69 \\
G341.039-00.114 & 0.22 & 2.6 & 0.13 & 210 & 0.59 & 0.10 & 0.16 & 0.80 \\
G343.489-00.416 & 0.19 & 1.4 & 0.10 & 200 & 0.53 & 0.07 & 0.11 & 0.69 \\
\enddata
\tablecomments{Uncertainty ranges for the quantities above are: $\sigma_{\rm th}$, from 2 to 11\%; $M_J$, from 25 to 45\%; $\lambda_J$, from 24 to 27\%; $M_{\rm turb}$, from 25 to 36\%;
$\lambda_{\rm turb}$, from 24 to 26\%; $L_{\rm av}$ and $L_{\rm av\,corr}$ around 10\%. $\mathcal{Q}$ is distance independent and has negligible uncertainties. }
\end{deluxetable*}

\subsection{Spatial Core Distribution and Mass Segregation}
\label{spatial}

\subsubsection{Spatial Core Distribution}

Considering that the IRDC clumps in this study represent the earliest stages of high-mass and cluster formation, the spatial distribution of cores 
gives a characteristic imprint of the early fragmentation. Some clumps, for example G014.492--00.139, show a more centrally concentrated core 
distribution, while others, like G327.116--00.294, have a more widespread core distribution. 

To quantify the spatial distribution 
of cores, we follow the approach of \cite{Cartwright04} and define the $\mathcal{Q}$ parameter that can be used to distinguish between centrally 
peaked clusters of cores ($\mathcal{Q} > 0.8$) and hierarchical subclustering ($\mathcal{Q} < 0.8$). The $\mathcal{Q}$ parameter is defined as
 \begin{equation}
\mathcal{Q} = \frac{\bar{m}}{\bar{s}}~,
\label{QparamEq}
\end{equation} 
where $\bar{m}$ is the the normalized mean edge-length of the MST, given by
 \begin{equation}
\bar{m} = \sum^{N_c - 1}_{i=1} \frac{L_i}{\sqrt{(N_c A)(N_c - 1)}}~,
\label{mbar}
\end{equation} 
where $N_c$ is the number of cores, $L_i$ is the length of each edge, and $A$ is the cluster area, A = $\pi R_{\rm cluster}^2$, with $R_{\rm cluster}$ 
calculated as the distance from the mean position of all cores to the farthest core.  $\bar{s}$ is the normalized correlation length, i.e., the ratio of the mean core separation to the cluster radius, $R_{\rm cluster}$:
  \begin{equation}
\bar{s} = \frac{L_{\rm av}}{R_{\rm cluster}}~.
\label{QparamEq}
\end{equation} 
 Both $\bar{m}$ and $\bar{s}$ are 
  independent of the number of cores in the cluster-forming clump \citep[see further details in][]{Cartwright04}. 
  
  For  $\mathcal{Q} > 0.8$, $\mathcal{Q}$
   is correlated with centrally condensed spatial distributions with radial density profiles of the form 
 $n(r)\propto r^{-\alpha}$ (with $\alpha$ between 0 and 3), while for $\mathcal{Q} < 0.8$, $\mathcal{Q}$ is associated with the fractal dimension, $D$ 
 \citep[see Figure~5 in][]{Cartwright04}. A value of $\mathcal{Q}\simeq 0.8$ implies uniform density (i.e., $\alpha = 0$) and no subclustering ($D=3$). 
 The value of $D$ ranges from 3 (no subclustering) to 1.5 (strong subclustering, $\mathcal{Q}\simeq 0.45$).
 
As cluster-forming clumps evolve over time, it may be expected that the primordial distribution of cores dissolves due to dynamical relaxation to 
become radially concentrated. If this is the case, we may expect to see higher $\mathcal{Q}$ values toward more evolved clumps (those containing a 
larger fraction of protostellar cores). Table~\ref{tbl-global} (Column 9) summarizes the $\mathcal{Q}$ parameters measured toward each clump. The narrow range 
in $\mathcal{Q}$ values (0.63 to 0.80) may indicate that the evolutionary stage of the clumps is similar; indeed the embedded
  protostars have not significantly affected the clumps (all are 70 $\mu$m dark). Nevertheless,  we still find a weak correlation between the $\mathcal{Q}$ parameter and the fraction of protostellar cores in each clump (Figure~\ref{Qparam}), with a Spearman correlation coefficient $\rho_s=0.28$. The correlation 
  becomes stronger if we remove the ``outlier" clump with no protostellar cores (G340.222--00.167), with $\rho_s=0.59$. Those clumps with $\mathcal{Q} \sim$0.8 are consistent with spatial core distributions of uniform density ($\alpha=0$). 
  However, the whole sample shows $\mathcal{Q} \lesssim0.8$ (and thus $D \lesssim3$), indicating that the initial fragmentation in IRDCs favors (moderate) hierarchical subclustering over centrally peaked clustering.

\begin{figure}[h!]
\begin{center}
\includegraphics[angle=0,scale=0.38]{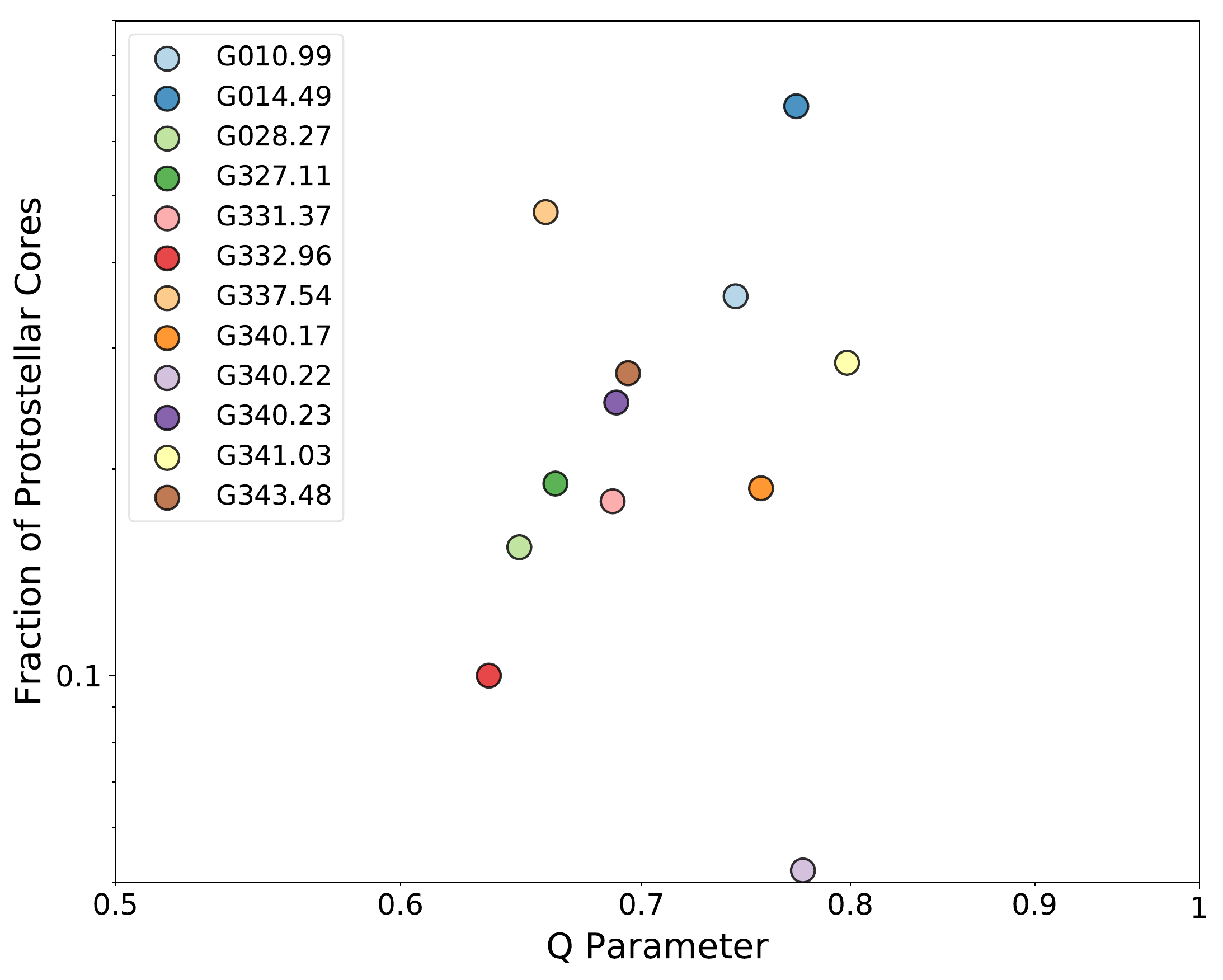}
\end{center}
\caption{Spatial distribution of cores (characterized by the $\mathcal{Q}$ parameter) versus the fraction of protostellar cores per clump, as a 
proxy of clump evolution. A weak correlation with a Spearman correlation coefficient $\rho_s=0.28$ is measured. After excluding the ``outlier" clump with no protostellar 
cores (G340.222--00.167), the correlation becomes stronger with $\rho_s=0.59$. For display reasons in the log-log plot, we have artificially assigned a protostellar 
fraction of 0.05 to G340.222--00.167. }
\label{Qparam}
\end{figure}

\subsubsection{Mass Segregation}

Mass segregation \cite[as defined by, e.g.,][]{Allison09,Parker15} refers to a different distribution (more concentrated) of 
massive objects with respect to lower mass objects than that expected by random chance. Mass segregation is observed in evolved
 clusters where it is believed to be produced by two-body relaxation (dynamical mass segregation), with some exceptions as described in \cite{Bonnell98}.  
 Primordial mass segregation is especially important because it has been predicted as a natural outcome of competitive accretion 
 models \citep{Bonnell98,Bonnell06}, in which the cores located at the center of the cluster accrete enough material to become massive and form high-mass stars.
 We note, however, that cluster formation simulations that are in agreement with the turbulent core accretion theory also find primordial mass segregation \citep{Myers14}. 
Considering (i) the early evolutionary stage of the IRDC clumps observed in this study, all are IR-dark with no signs of disruption from high-mass stars, and (ii) 
the large area mosaiced per clump that should cover most of the cluster members, this is an ideal sample in which to search for primordial mass segregation. 

To quantify mass segregation, we use the mass segregation ratio (MSR), $\Lambda_{MSR}$ as defined by \cite{Allison09} and 
$\Gamma_{MSR}$ as defined by \cite{Olczak11}, both based on the MST method. The first method ($\Lambda_{MSR}$) compares
the MSTs of random subsets of cluster members with the MST of the same number of most massive members. The value
  of $\Lambda_{MSR}$ is given by
  \begin{equation}
\Lambda_{MSR}(N_{\rm MST}) = \frac{\langle l_{\rm random}\rangle}{l_{\rm massive}} \pm \frac{\sigma_{\rm random}}{l_{\rm massive}}~,
\label{MSparamEq}
\end{equation} 
 where $ \langle l_{\rm random}\rangle$ is the average MST length of sets of N$_{\rm MST}$ random cores and $l_{\rm massive}$ is the MST length 
 of the N$_{\rm MST}$ most massive cores. A total of 1000 sets of random N$_{\rm MST}$ cores were used to derive the average MST length. 
 $\sigma_{\rm random}$ is the standard deviation associated to $ \langle l_{\rm random}\rangle$, i.e., the standard deviation of the 1000 sets of  $l_{\rm random}$. 
If the MST length of the most massive cores is shorter than the mean MST length of the random cores, the massive cores have a different, 
more concentrated distribution. Therefore,  $\Lambda_{MSR} \approx 1$ means massive cores are distributed in the same way than other cores 
(no mass segregation), $\Lambda_{MSR} > 1$ indicates massive cores are concentrated (mass segregation), and $\Lambda_{MSR} < 1$ implies 
more massive cores are spread out compared to other cores (inverse-mass segregation). The second method ($\Gamma_{MSR}$) uses an analogous 
approach with the difference that the geometric mean of the segments forming the MST length is used (instead of the arithmetic mean). The value of
 $\Gamma_{MSR}$ is given by
  \begin{equation}
\Gamma_{MSR} (N_{\rm MST})= \frac{\gamma_{\rm random}}{\gamma_{\rm massive}} (d\gamma_{\rm random})^{\pm1}~,
\label{MSparamEq}
\end{equation} 
where $\gamma_{\rm random}$ is the geometric mean of the MST segments for the N$_{\rm MST}$ random cores (1000 sets), $\gamma_{\rm massive}$ is the
 MST of the N$_{\rm MST}$ more massive cores, and $d\gamma_{\rm random}$ is the geometric standard deviation given by \citep{Olczak11}:
 \begin{equation}
d\gamma = \exp \left(\sqrt{\frac{\sum^n_{i=1} (\ln L_i - \ln \gamma_{\rm MST})^2}{n}}\right)~,
\label{dgamma2} 
 \end{equation} 
 where $L_i$ are the $n$ MST lengths.  
 The values obtained for $\Gamma_{MSR}$ are interpreted  in the same way as $\Lambda_{MSR}$, and 
 according to \cite{Olczak11}, $\Gamma_{MSR}$ would be more sensitive to finding weak mass segregation. 

There is no mass segregation for 8 clumps and only marginal departure from unity in four clumps. 
Figure~\ref{mass-seg} shows the derived $\Lambda_{MSR}$  and $\Gamma_{MSR}$ parameters in these four clumps at several N$_{\rm MST}$ values.
 For N$_{\rm MST}=2$ and N$_{\rm MST}=3$, there are three clumps with MSR values $\gtrsim$3 (weak mass segregation) and one with $\sim$0.4-0.5 (weak 
  inverse-mass segregation). Although the MSR values have a significance larger than 1$\sigma$ above or below unity, the results are not robust 
  considering the low number of cores (2 or 3). A different assumption for dust temperature on individual cores can modify the mass and completely change the output from 
  an MSR with small N$_{\rm MST}$. We have tested the effect of changing the temperature for the protostellar cores to 30 K and verified that the results are consistent with 
  using the lower clump temperature. The overall conclusion is that there is no significant evidence of primordial mass segregation, i.e.,  more massive
   cores are distributed in the same way than other cores in this IRDC sample.

\begin{figure}[h!]
\begin{center}
\includegraphics[angle=0,scale=0.4]{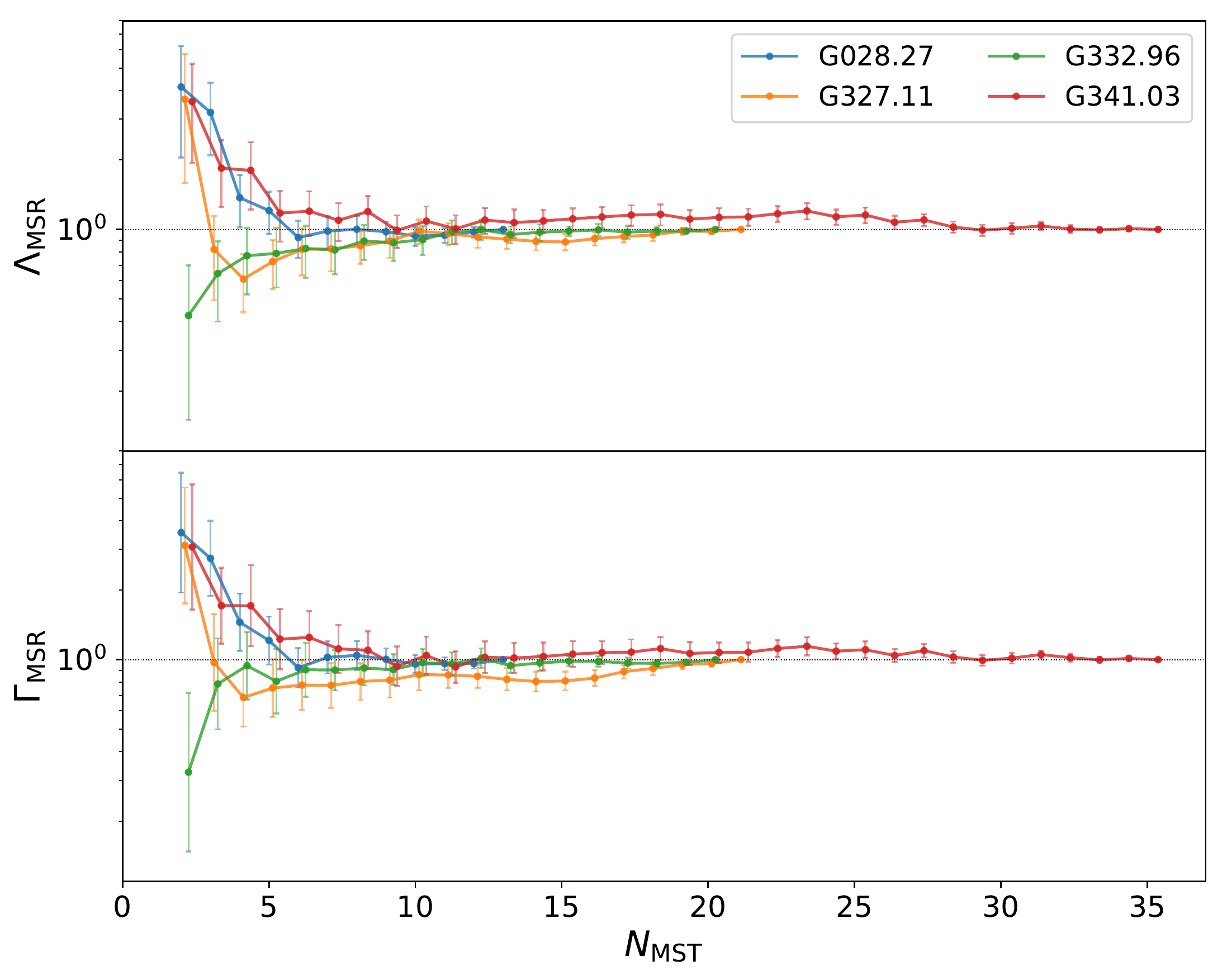}
\end{center}
\caption{Mass segregation ratios ($\Lambda_{MSR}$ and $\Gamma_{MSR}$) for different number of N$_{\rm MST}$ cores for the four clumps whose ratios have marginal departures
 from unity. For instance, if N$_{\rm MST} = 3$, $\Lambda_{MSR}$ 
and $\Gamma_{MSR}$ are calculated 1000 times from the ratio of the MST length derived from 3 random cores in the cluster and the MST length derived for the 3 most 
massive cores. A $\Lambda_{MSR} \approx 1$ (and $\Gamma_{MSR} \approx 1$) implies no mass segregation. }
\label{mass-seg}
\end{figure}

\subsection{Core Mass Function}
\label{CMF_section}

The initial mass function (IMF) is an empirical function that describes the initial distribution of masses of a stellar population and 
it is believed to be the result of star formation. 
The IMF has a shape similar to a log normal with a peak 
below 1 \Msun\ and a power law tail at the high-mass end of the form 
  \begin{equation}
\frac{dN}{d \log M} \propto M^{-\alpha} ~,
\label{MSparamEq}
\end{equation} 
with an index $\alpha$ = 1.35 \citep{Salpeter55} that is considered to be universal \cite[e.g.,][]{Bastian10}. In order to 
understand the origin of the universal IMF, the histogram of core masses (or core mass function, CMF) has been constructed 
mostly for nearby, low-mass star-forming regions. In this case, the CMF resembles the initial mass 
function (IMF) in shape, but apparently shifted to higher masses by an efficiency factor \citep[e.g.,][]{Alves07,Andre10,Konyves15}. This 
similarity has been interpreted as the IMF being for the most part determined  by the fragmentation process of molecular clouds. On the 
other hand, the few, more distant, high-mass star-forming regions observed with ALMA so far, point to a more dynamical picture. \cite{Motte18} find in the 
young massive cluster W43-MM1 a power law index $\alpha$ = 0.90 $\pm$ 0.06, much shallower than the Salpeter IMF. Their results can be interpreted as
 the CMF evolving over time from a shallower distribution to the universal IMF, likely producing low-mass objects continuously over the formation of the cluster while 
the massive objects were mostly formed early on. \cite{Cheng18} derived in the later-stage protocluster G286.21+0.17 located in Carina a power 
law index of 1.24 $\pm$ 0.17, slightly shallower but consistent with Salpeter within the uncertainties. More recently, in a combined CMF for clumps 
in seven IRDCs, \cite{Liu18b} find a power index of 0.86 $\pm$ 0.11. We note that these IRDCs, originally selected from \cite{Rathborne06} by
\cite{Butler09}, are in a more advanced evolutionary stage than those observed in this work, containing several embedded clumps with 
protostellar activity inferred from {\it Spitzer} images (see images in \citealt{Chambers09}; and an updated classification in \citealt{Sanhueza12}).

The IRDC clumps in ASHES are IR-dark from 3.6 to 70 $\mu$m in {\it Spitzer} and {\it Herschel} images (see Figure~\ref{IR-plots1}, ~\ref{IR-plots2},~\ref{IR-plots3}, 
and ~\ref{IR-plots4}) and the ALMA observations reveal that only 
29\% of the embedded cores have star formation activity. With a large population of prestellar core candidates, the CMF would likely represent a snapshot 
of the initial core mass distribution produced in massive clumps that will form high-mass stars. Figure~\ref{CMFplot} shows the CMF for the prestellar 
core population (in blue; 210 cores) and, as a reference, the whole core population (in black; 294 cores). The power law index for the prestellar core
 population is $\alpha$ = 1.17 $\pm$ 0.10 (blue dashed line), which is slightly, but significantly shallower than Salpeter ( $\alpha$ = 1.35; red solid line). 
 The power law fitting includes masses up to the peak of the CMF, $\gtrsim$0.6 \Msun. If the next bin is used instead, $\gtrsim$0.9 \Msun, the power law index is
  $\alpha$ = 1.24 $\pm$ 0.12. For the whole core population $\alpha$ = 1.07 $\pm$ 0.09 ($\gtrsim$0.6 \Msun; black dashed line). The effect of adding the
   protostellar cores, that are probably warmer than the assumed {\it Herschel} dust temperature, is to make the power index shallower. 
 
 In order to reconcile the power law indexes measured in the high-mass end of the CMF determined in massive clumps, we propose the following 
 scenario. The early fragmentation in 70 $\mu$m dark-IRDCs results in a power law index slightly shallower than Salpeter. The most massive cores can 
 accrete material, growing in mass quite quickly according to the recent finding of \cite{Contreras18}. They determine an accretion rate of 2 $\times$ 10$^{-3}$  
 \Msun\ yr$^{-1}$ in a relatively massive prestellar core. At this accretion rate, in a core free fall time of 3.3 $\times$ 10$^4$ yr, the core can accrete $\sim$4 times
  its mass (at a constant rate over the whole period.) This accretion rate is $\sim$2 orders of magnitude higher than those found in low-mass cores, which would cause massive cores 
 to accrete more compared to low-mass cores. Thus, the CMF would become shallower at more evolved stages of cluster formation, as observed 
 in the studies of  \cite{Liu18b} and  \cite{Motte18}. Later on, the high accretion rate cannot be maintained due to feedback and the continuum 
 clump fragmentation would catch up. Thus, at later stages of cluster formation, the high-mass end of the CMF would resemble the IMF, as found by
  \cite{Cheng18}. We however acknowledge the difficulty in comparing the results from different works, for example, using different methods for core determination, 
  inclusion/exclusion of the 7 m array, combining prestellar and protostellar cores. A uniform analysis of a large sample can definitively test the proposed
   scenario, as will be done with ALMA-IMF (ALMA large program; Motte et al., in prep.).

\begin{figure}[h!]
\begin{center}
\includegraphics[angle=0,scale=0.4]{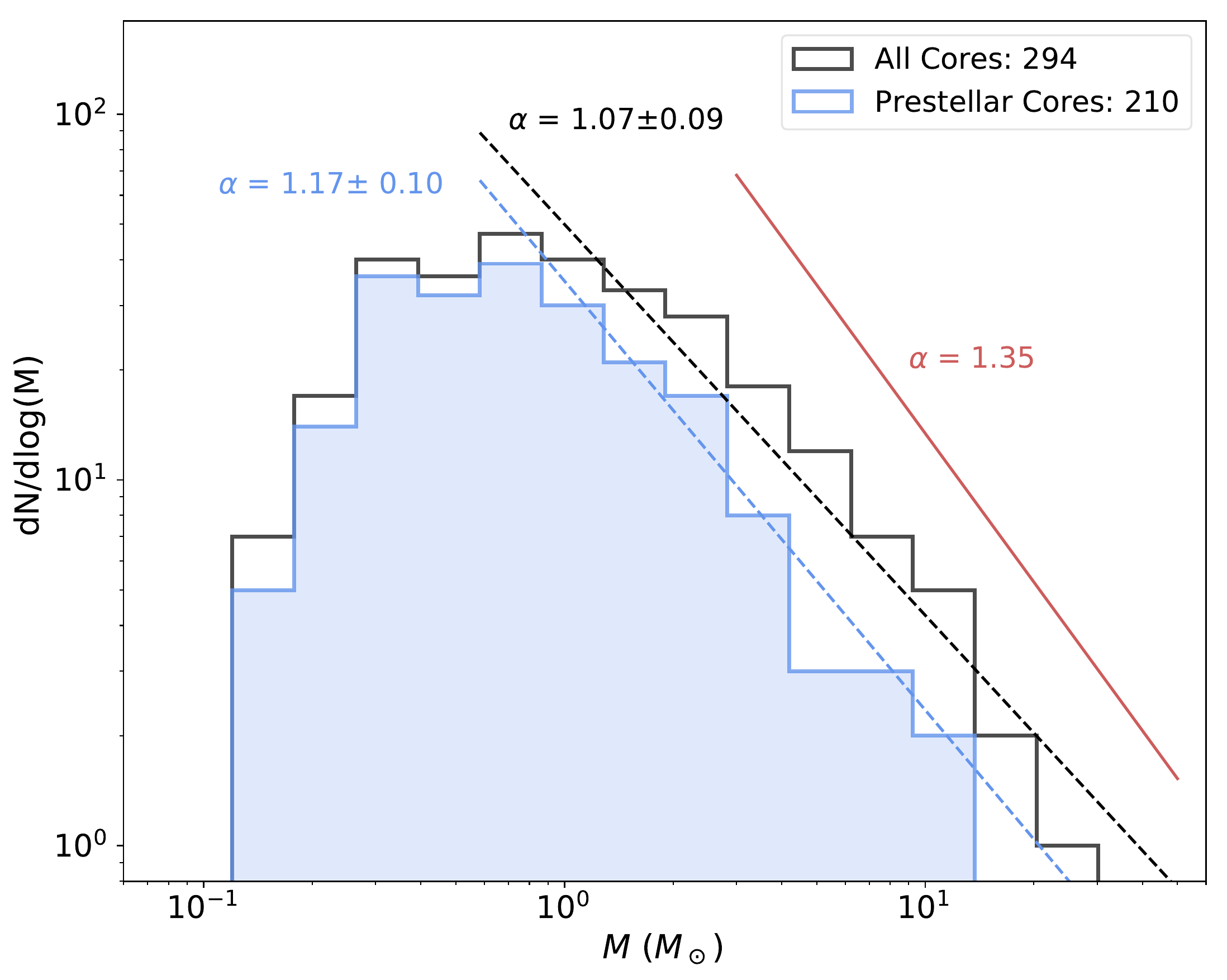}
\end{center}
\caption{Combined core mass function for the prestellar population, in blue, and for the whole core population, in black. The dashed blue line shows the power law 
fitting to the high-mass end for the prestellar population, $\alpha$ = 1.17 $\pm$ 0.10, including cores with masses $\gtrsim$0.6 \Msun. The dashed black line shows the power law 
fitting to the high-mass end for the complete core population, $\alpha$ = 1.07 $\pm$ 0.09, for the same range of core masses. Red solid line shows the Salpeter IMF with $\alpha$ = 1.35.}
\label{CMFplot}
\end{figure}

\subsection{High-Mass Star Formation Picture}

We have revealed the early fragmentation and discovered the first 
members of future stellar clusters that will host high-mass stars.
 Given the low degree of star formation activity (70 $\mu$m dark clumps and only 29\% of embedded 
 protostellar cores), protostellar feedback should only play a minor role in these IRDCs. To date, ASHES offers 
 the largest population of prestellar core candidates detected in high-mass clumps. Having characterized
 a large sample of cores, we are in position to constrain high-mass star formation models.
 
 High-mass prestellar cores, defined here as cores with masses $>$30 \Msun, are the cornerstone 
 of the turbulent core accretion model \citep{McKee03,Tan13,Tan14}. However, they have not been unambiguously found in IRDCs 
 \citep{Zhang09,Zhang15,Wang14,Ohashi16,Sanhueza17,Contreras18,Beuther18a,Kong18b}. The case is different in more 
 evolved high-mass star-forming regions. Whereas some studies find a few rare prestellar high-mass core candidates \citep{Cyganowski14,Liu17,Nony18}, 
  other studies find none \cite[e.g.,][]{Fontani18,Louvet19}. In ASHES, over 210 prestellar core candidates are detected, with no 
 high-mass prestellar cores detected. The most massive prestellar core has a mass of only 11 \Msun. The fact that the only high-mass prestellar 
 candidates found so far are near other high-mass protostellar objects suggest an environmental dependence, but this raises the question
  of how the earlier high-mass protostars formed. An alternative view can be that 
  the most massive prestellar cores found at the earliest stages of star formation, such as those IRDC cores in the mass range of
   10 to 20 \Msun\ found in the works mentioned above and in ASHES, take time to grow in mass. This is indeed possible in all of the clump-fed
    scenarios (competitive accretion scenario, \citealt{Bonnell04,Bonnell06,Smith09,Wang10}; global hierarchical collapse,
     \citealt{Vazquez09,Vazquez17,Vazquez19}), considering  the accretion rates determined, for now, in a single example \citep{Contreras18}.
      With accretion rates of $\sim$10$^{-3}$ \Msun\ yr$^{-1}$, cores can 
   significantly gain mass in a typical clump free fall time of few 10$^5$ yr \citep{Contreras16}. However, it is unclear if a prestellar 
   core of 10-20 \Msun\ can become massive and still remain starless, or if it will first form a low-mass protostar that will then be fed by the growing core.
   Nevertheless, the absence of high-mass prestellar cores in the early stages of fragmentation of high-mass star-forming regions constrains their formation 
   to only later stages of evolution. 
    
   \cite{Krumholz08} suggest that to allow the formation of high-mass stars and avoid ``excessive" fragmentation, cores 
   should have both surface densities $\Sigma \gtrsim$1 g cm$^{-2}$ and be heated by accreting surrounding protostars in order to increase the Jeans mass.
    Such heating has not been observed so far \citep{Zhang09,Wang12,Sanhueza17}. According to \cite{Tan13}, magnetically mediated high-mass
     star formation \citep[e.g.,][]{Commercon11,Myers13} would not require a minimum $\Sigma$ and lower values, e.g. $\sim$0.5 g cm$^{-2}$, would be sufficient. 
   We find that 89 (30\%) cores of all masses have $\Sigma >$ 0.5 g cm$^{-2}$, including the 16 most massive cores.

\begin{figure*}
\begin{center}
\includegraphics[angle=0,scale=0.6]{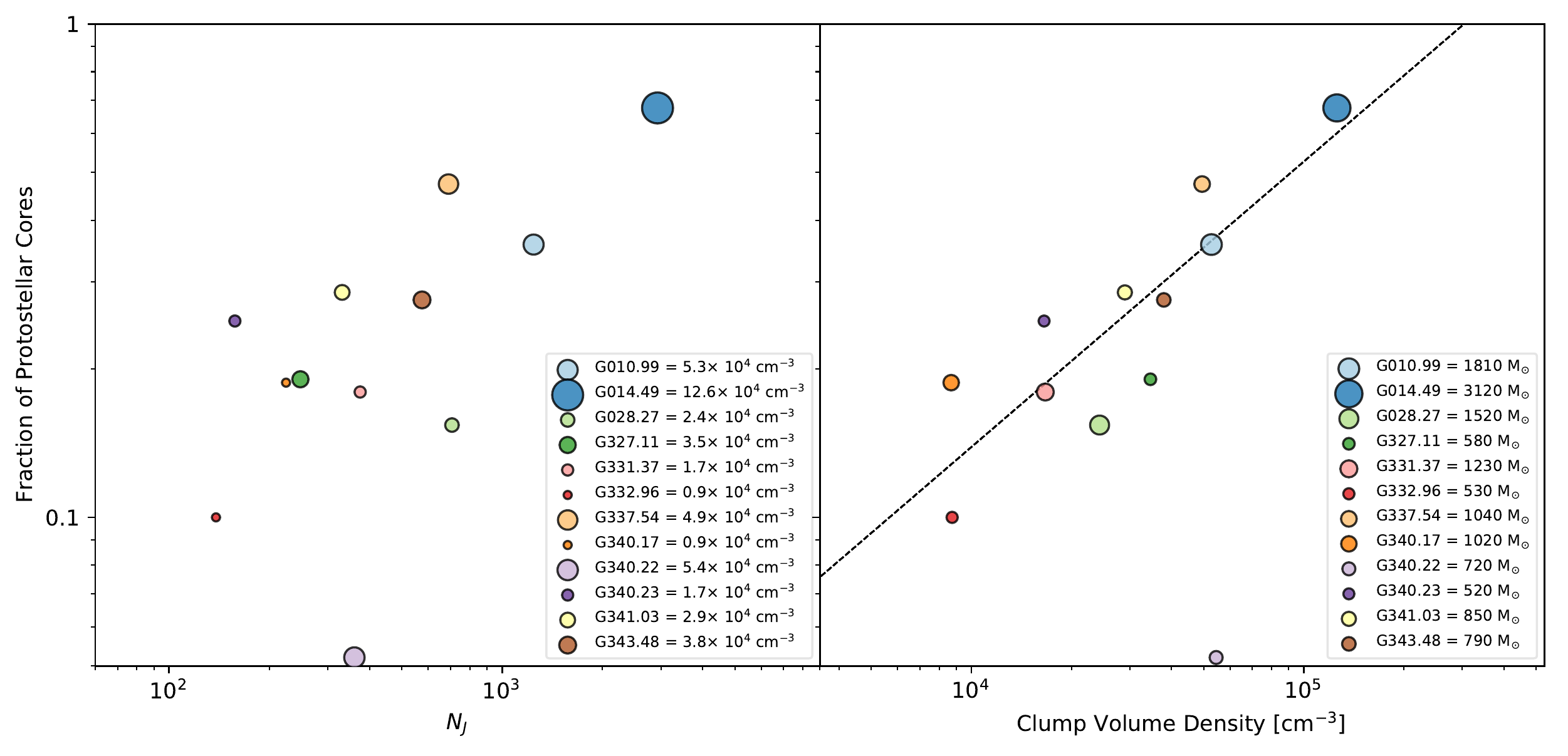}
\end{center}
\caption{Left: Jeans number, $N_J = M_{\rm cl}/M_{J}$, versus the fraction of protostellar cores per clump. A strong correlation with a Spearman correlation
 coefficient $\rho_s=0.52$ is measured. After excluding the ``outlier" clump with no protostellar 
cores (G340.222--00.167), the correlation becomes stronger with $\rho_s=0.6$. Right: Clump volume density against the fraction of protostellar cores per clump. 
The number of protostellar cores scale as $\propto n^{\beta}_{\rm cl}$ with $\beta = 0.57 \pm 0.11$. For display reasons in the log-log plot, we have artificially
 assigned a protostellar fraction of 0.05 to G340.222--00.167.}
\label{protofrac}
\end{figure*}

 We detect for the first time a large population of low-mass cores ($\lesssim$1 \Msun) evolving along with the seeds that will form high-mass stars.
 Studies of cluster formation made with a single-pointing or excluding the 7 m array may well miss a large portion of the low-mass cluster members 
 and, on occasion, even the most massive ones. The whole core population is quite consistent with thermal Jeans fragmentation with masses
  smaller or similar to the Jeans mass and separations comparable or larger (within a factor 2) than the Jeans length. Turbulent Jeans fragmentation cannot explain
  the initial fragmentation observed in these IRDCs.  We note that similar results have been also found in more evolved stages of high-mass star formation
   \citep{Palau15,Beuther18b,Svoboda19}, further confirming the irrelevance of turbulent Jeans fragmentation in the formation of high-mass stars and cluster formation.
   In general, a low-mass population of prestellar cores in clumps at early stages of evolution is more consistent with competitive
   accretion and global hierarchical collapse scenarios. Specifically, simulations run by \cite{Smith09} result in good agreement with our observations. They find in 
   the simulations an average prestellar core mass and radius of 0.7 \Msun\ and 
   2.4 $\times$ 10$^3$ AU at the moment the cores first become bound (total of 306 objects throughout the lifetime of the simulation). The maximum
    prestellar (bound) core mass reached that is able to form a 
   high-mass star, throughout the whole simulation, is 6.35 \Msun. In our survey, the mean prestellar core mass and radius are 1.2 \Msun\ and 
   2.9 $\times$ 10$^3$ AU, with a maximum prestellar core mass of 11 \Msun. The difference in the radii for the observations and simulations 
   is rather small ($\sim$20\%), while for the mass, the observed values (mean and maximum mass) are a factor 1.7 larger. The mean values 
   are measured in the simulations at the moment that cores become bound. In the competitive accretion scenario, it is expected that the cores will 
   grow in mass (and likely in size) and may be possible that the masses approach the observed values later on. On the other hand, the smaller 
    mass of the most massive prestellar core in the simulation may reflect the absence of magnetic fields in the simulation. Magnetic fields have been 
    suggested to halt fragmentation, making the mass of the fragments/cores larger \citep{Commercon11}. 
  
  \cite{Vazquez19} claim that a natural consequence of the global hierarchical collapse scenario is that the fragmentation level, measured as the
   total number of cores (protostellar plus prestellar cores), is directly proportional to the Jeans number or the root square of clump density ($N_J = M_{\rm cl}/M_{J} \propto n^{1/2}_{\rm cl}$). 
   Such a correlation has been observed in more evolved intermediate/high-mass star-forming regions by, for example, \cite{Palau14,Palau15}. However,
    \cite{Vazquez19} predict that this correlation would be present in an advanced stage after the global collapse has started, once most of the fragmentation
     epochs have occurred. Therefore, it is expected that the correlation would be weak or absent at the earliest stages of high-mass star formation. We have searched
      for this correlation and find a Spearman's coefficient $\rho_s$ = 0.33 for the number of cores versus $N_J$ and  $\rho_s$ = 0.44 for the number of cores
       versus $n^{1/2}_{\rm cl}$, indicating a moderate correlation, in agreement with the global hierarchical collapse scenario prediction. Interestingly, we find 
       a strong correlation between the fraction of protostellar cores and $N_J$. This correlation is shown in Figure~\ref{protofrac} and both 
       completely independent quantities correlate with  $\rho_s$ = 0.52 ($\rho_s$ = 0.6, excluding the prestellar clump). This correlation is in better agreement with
       the findings of  \cite{Palau14,Palau15}, 
       considering that their cores were mostly protostellar. According to  \cite{Vazquez19}, in the global hierarchical collapse scenario such a correlation 
       is expected. Clumps start gravitational collapse at a given density that defines a Jeans mass, but as time moves forward, the clump density increases, decreasing the 
       corresponding Jeans mass and increasing the number of Jeans masses over time. Therefore, clumps with a larger number of Jeans masses would be more
        evolved, presenting a larger fraction of protostellar cores that would be inversely proportional to the clump free-fall time. As a result, the fraction 
         of protostellar cores should scale with $\propto n^{1/2}_{\rm cl}$. We indeed find that the fraction of protostellar cores scale as $\propto n^{\beta}_{\rm cl}$ 
       with $\beta = 0.57 \pm 0.11$, consistent with the theoretical prediction, within the uncertainties.   
         
  The spatial core distribution, characterized by $\mathcal{Q}$ values from 0.63 to 0.80, is found to be consistent with hierarchical subclustering 
   rather than centrally peaked clustering. \cite{Maschberger10} analyze two cluster formation simulations, one of them of a 10$^3$ \Msun\ clump with 
   1 pc diameter made by \cite{Bonnell03}. Using the MST method in a similar fashion as done here, \cite{Maschberger10} find that the whole cluster spatial distribution is characterized 
    by a monotonic increase in $\mathcal{Q}$ values, starting at early times with $\sim$0.5 and evolving to $>$1 at the end of the simulation. The $\mathcal{Q}$ 
    values obtained in the simulation are consistent with our observations. Given that we are tracing only the very early stages of high-mass star formation, 
    the range of observed $\mathcal{Q}$ values is restricted to the values obtained at the beginning of the simulation. However, we do find a weak correlation (that becomes stronger after  
    removing the outlier) of increasing $\mathcal{Q}$ with the star formation activity, traced by the fraction of the protostellar cores. 
    
   Based on the premise that the cores near the center of the gravitational potential accrete more material than cores located at 
 other positions in the cluster, competitive accretion scenarios predict primordial mass segregation \citep{Bonnell98,Bonnell06}. \cite{Maschberger10} also 
 calculate the mass segregation ratio ($\Lambda_{MSR}$) finding values of 2-3 over the 
 10 most massive members by the end of the simulations ($\sim$0.5 Myr). They state that, because the simulation corresponds to a deeply embedded phase 
 of star formation, the mass segregation derived is primordial. They conclude that the most massive sinks are segregated for subcluters with over 30 members. 
 Mass segregation has also been found in simulations that are consistent with the analytical turbulent core accretion model \citep{Myers14}. 
  At least at the evolutionary phase traced in ASHES, we find no strong indication of primordial mass segregation produced by the fragmentation itself. However, 
 we cannot rule out the possibility that due to accretion, the members now near the center of the gravitational potential become the most massive cores 
 in the future because their privileged location in the forming cluster. 

 In the context of the turbulent core accretion model, the stellar mass is related by an approximately constant star formation efficiency 
 to the core mass. Consequently, the IMF is predicted to be the result of the prestellar CMF \citep{McKee03,Tan14,Cheng18} and the 
 efficiency factor is regulated mostly by outflow feedback \citep{Matzner00} and later on by radiative feedback from the 
 high-mass stars \citep{Tanaka17}. We therefore would expect to find a Salpeter power law index in the high-mass tail of the prestellar CMF.
 On the other hand, this mapping of the CMF into the IMF, i.e. a correspondence of core to star, neglects the influence of 
 environmental factors on the core during the accretion process \citep{Smith09}. Clump-fed scenarios would thus 
 have power law indexes different to Salpeter in the prestellar CMF that would evolve into Salpeter at the end of cluster formation
 \citep[e.g.,][]{Clark07}. We find a power law index of 1.17 $\pm$ 0.10 at the high-mass end ($>$0.6 \Msun), which is slightly, but significantly, shallower than 
 Salpeter. This may suggest some link between the early CMF and the final IMF. However, current evidence from 
 more evolved sites of high-mass star formation indicates that the power law index could evolve. Intermediate stage high-mass star-forming regions 
 \citep{Motte18,Liu18b} have power law indexes of $\sim$0.9, while in more evolved stages of protocluster formation \citep{Cheng18} the CMF appears
  Salpeter. The lack of a constant similarity between the CMF and the IMF over the lifetime of high-mass cluster formation may indicate that in high-mass star-forming
   regions the core masses are not the main gas reservoir to form stars, in opposition to predictions from the turbulent core accretion scenario. Instead, global 
   clump infall would increase core masses and provide most of the material that ultimately makes up stars. The growing evidence of global collapse observed over
    hundreds of massive clumps supports this hypothesis \citep{He15,He16,Jackson19}. For now, it is unclear 
 if competitive accretion scenarios can explain the few CMFs measured in high-mass star-forming regions observed so far. 
 \cite{Clark07} argue that due to different lifetimes for low- and high-mass cores, the CMF would need to be shallower to reproduce the IMF. However, the power law index
  would need to be much lower than it has been observed ($<$0.5) and it is unclear which core lifetime would be longer. \cite{Maschberger10} show that in the \cite{Bonnell03} 
  simulation, the IMF for sink particles has a power law index that smoothly decreases from 1.6 to 0.8 over 3 $\times$ 10$^5$ yr. This is partially consistent with the scenario 
  proposed here that early on the CMF at the high-mass end resembles the Salpeter IMF and then becomes shallower due to differential accretion depending on the core mass.
   However, the simulations do not show if at later times the 
  IMF for sink particles increases to reconcile with the Salpeter IMF. Although the CMF measurements in high-mass star-forming regions are still scarce, this is expected to 
  change with the surveys that are currently being observed with ALMA. A more complete understanding of the link or lack of connection between the CMF and the IMF 
  will also require simulations that can cover similar evolutionary stages than those observed. 
  
 Overall, based on the present study, a complete theory of high-mass star formation should reproduce the characteristics of the very early stages of evolution 
 discussed here: (i) absence of high-mass prestellar cores, (ii) large population of low-mass cores, (iii), hierarchical subclustering, (iv) absence of primordial mass 
 segregation, and (v) slightly shallower CMF than the Salpeter IMF slope in the high-mass tail. 
 
 \section{Conclusions}
 
 We have presented the first results of ASHES (the Alma Survey of 70 $\mu$m dark High-mass clumps in Early Stages), 
 a program aimed to characterize the elusive early stages of high-mass star formation to constrain high-mass star formation theories.
 In the pilot survey, we have mosaiced 12 massive IRDC clumps with ALMA in continuum and line emission, including both 12 and 7 m
  arrays and total power. In this study, we have presented the survey and analyzed the dust continuum emission to draw the following conclusions:

  1. We successfully detected cores in all 12 massive IRDC clumps. A total of 294 cores are discovered and classified 
  as protostellar (84; 29\%), if they are associated with outflow activity or warm line emission, and as prestellar (210; 71\%), if they lack of 
  any star formation signatures. We conclude that 11 of 12 70 $\mu$m dark clumps have nascent, but deeply embedded, star formation activity. 
  However, the revealed star formation activity is from low-mass protostars likely forming along with the seeds that will eventually become high-mass 
  protostars. These seeds could be in the form of prestellar cores or growing low-mass protostars. The number of detected cores is
   independent of the 3.5$\sigma$ threshold used to define a core. On average, the most massive clumps tend to form more cores. 
  
  2. A large population of low-mass cores ($<$1 \Msun) is detected evolving along with the seeds that will form high-mass stars, which is 
  consistent with the competitive accretion and the global hierarchical collapse scenarios. No high-mass prestellar cores ($>$30 \Msun) are detected, 
  constraining the formation of high-mass prestellar cores predicted in the turbulent core accretion scenario to only later times in the cluster formation. The most massive
   prestellar cores have 11 \Msun, which corresponds to 5 times the Jeans mass. The most massive prestellar cores in each clump are likely to continue
    accreting material and growing in mass to finally form a high-mass star \citep[e.g.,][]{Contreras18}, as suggested theoretically by the competitive accretion
     and global hierarchical collapse scenarios, and the growing observational evidence of large numbers of massive clumps under global collapse. 
    Therefore, it is likely that the seeds that will form high-mass stars form early on, but the high-mass star itself forms later as the whole clump evolves. 
    However, it is unclear if the cores will reach a ``high-mass status" as prestellar or with an embedded low-mass protostar located at their centers. We also find that the most
     massive cores have surface densities ($>$0.5 g cm$^{-2}$) consistent with the predictions of turbulent core accretion. 
  
  3. To characterize the core separation, we have used the minimum spanning tree (MST) technique. The average minimum separation between cores, 
  as defined by the MST, is comparable or larger (within a factor 2) than the derived Jeans length for each clump. While the observations of these clumps at early evolutionary
   stages reveal a large range of core masses and core separations, the mean masses and mean separations
    are consistent with the thermal Jeans fragmentation. Turbulent Jeans lengths are typically larger than the observed core
    separations and the turbulent Jeans masses are orders of magnitude higher. Turbulent Jeans fragmentation is therefore ruled out by these observations.
  
  4. Making use of the MST and the $\mathcal{Q}$ parameter, we found that the spatial core distribution follows hierarchical subsclustering rather than 
  centrally peaked clustering. With $\mathcal{Q}$ values ranging from 0.63 to 0.80, we find a weak correlation between the $\mathcal{Q}$ value and star 
  formation activity in the clumps (traced by the fraction of protostellar cores). The range of $\mathcal{Q}$ values and the trend are both 
  consistent with competitive accretion simulations.
  
  5. Using mass segregation ratios ($\Lambda_{MSR}$  and $\Gamma_{MSR}$), we have searched for primordial mass segregation. Eight clumps are 
  fully consistent with an equal spatial distribution of low/massive cores ($\Lambda_{MSR} \approx \Gamma_{MSR} \approx 1$). The other four clumps 
  have segregation ratios departing from unity, but only with a low number (2 or 3) of massive members clustered together. The low number of clustered 
  massive cores makes the results strongly sensitive to the temperature assumption used for mass determination. We conclude there is no solid evidence of
   primordial mass segregation, in direct contrast to the predictions of competitive accretion theory. However, we cannot rule out that 
   later in the evolution of the clumps, accretion into the cores rather than dynamical effects may produce core mass segregation.
  
  6. We have constructed the CMF combining all prestellar cores detected in each clump. The high-mass end has a power law index of 1.17 $\pm$ 0.10, which 
  is slightly shallower than the Salpeter index for the initial mass function. Placing in context this work with (scarce) previous works in other more evolved
   high-mass star-forming regions, we propose that 
  the CMF at early times is nearly Salpeter (but shallower), then it evolves into a significantly shallower CMF due to larger accretion rates of the 
  most massive members, to then become Salpeter again once accretion for the massive members has ceased and due to a continuous clump fragmentation 
  producing new (mostly) low-mass cores. This scenario and the current observational evidence on the variations of the power law index over the clump 
  evolution suggest a dynamical high-mass star formation picture. The core masses are not the main gas reservoir to form stars and accretion plays an important 
  role shaping the final IMF, which is in opposition to the prediction from the turbulent core accretion theory.  Competitive accretion and global hierarchical collapse 
  theories predict variations on the power law index, but it is unclear if they agree with the proposed scenario. Larger samples over different evolutionary stages and 
  more simulations tracing the evolution of the power law index are necessary to fully understand the origin of the IMF.
  
  In this study, we have put firm constraints on the earliest stages of high-mass star formation and we expect to refine them once the whole survey is analyzed. 
  We finally conclude that a complete high-mass star formation theory should reproduce the general features presented in this work, as well as the core dynamics 
  (virial equilibrium, non-thermal component, Mach number, core-to-core velocity dispersion) presented in \cite{Contreras19}. We acknowledge that the whole
   observational picture will not be complete until we constrain the magnetic field at the early stages of high-mass star formation.
  
  \acknowledgements
  
P.S. and B.W. were financially supported by Grant-in-Aid for Scientific Research (KAKENHI Number 18H01259) 
of Japan Society for the Promotion of Science (JSPS). P.S. is grateful for the comments from the anonymous referee and the valuable discussion with 
Javier Ballesteros-Paredes.  
P.S., Y.C., and A.E.G. gratefully acknowledge the support from the NAOJ Visiting Fellow Program that enabled a
 visit to the National Astronomical Observatory of Japan on 2016 November--December. G.G. acknowledges support from CONICYT Project AFB-170002. 
 H.B. acknowledges support from the European Research Council under the Horizon 2020 Framework Program via the ERC Consolidator Grant CSF-648505. 
 H.B. also acknowledges support from the Deutsche Forschungsgemeinschaft via SFB 881, The Milky Way System (sub-project B1).
 This paper makes use
  of the following ALMA data: ADS/JAO.ALMA\#2015.1.01539.S and ADS/JAO.ALMA\#2016.1.01246.S. ALMA is a
   partnership of ESO (representing its member
   states), NSF (USA) and NINS (Japan), together with NRC (Canada), MOST and ASIAA (Taiwan), and KASI (Republic of Korea),
    in cooperation with the Republic of Chile. The Joint ALMA Observatory is operated by ESO, AUI/NRAO and NAOJ.
The ATLASGAL project is a collaboration between the Max-Planck-Gesellschaft, the European Southern Observatory 
(ESO) and the Universidad de Chile. It includes projects E-181.C-0885, E-078.F-9040(A), M-079.C-9501(A), 
M-081.C-9501(A) plus Chilean data. Data analysis was in part carried out on the open use data analysis computer
 system at the Astronomy Data Center (ADC) of the National Astronomical Observatory of Japan. 
 
 \facilities{ALMA}


\software{CASA}

 \appendix
 \section{Derivation of the maximum stellar mass using the IMF}
\label{app}

Here we refine the derivation presented in \cite{Sanhueza17}. Compared with the previous derivation, we have added the 
lowest mass regime of the IMF from \cite{Kroupa01} as follows:  $\xi(m) \propto$ $m^{-0.3}$
 for 0.01 \Msun\ $\leq$ $m$ $<$ 0.08 \Msun, $\xi(m) \propto$ $m^{-1.3}$
 for 0.08 \Msun\ $\leq$ $m$ $<$ 0.5 \Msun, and $\xi(m) \propto$ $m^{-2.3}$ for 
 $m$ $\geq$ 0.5 \Msun, where $m$ corresponds to the star's mass and 
 $\xi(m) dm$ is the number of stars in the mass interval $m$ to $m + dm$. Therefore 
 Equations (A4) and (A5) from \cite{Sanhueza17} are updated to:
 \begin{equation}
m_{\rm max} = \left(\frac{0.3}{\epsilon_{\rm sfe}}\frac{21.0}{(M_{\rm clump}/\Msun)} + 1.5\times 10^{-3}\right)^{-0.77}~\Msun~,
\label{eqn-IMF-m-max-ap}
\end{equation}
and 
\begin{equation}
M_{\rm clump} = \frac{0.3}{\epsilon_{\rm sfe}}\,\, \frac{21.0}{((m_{\rm max}/\Msun)^{-1.3} - 1.5\times10^{-3})}~\Msun~.
\label{eqn-IMF-m-clump}
\end{equation}
where $m_{\rm max}$ is the maximum stellar mass  (assuming $m_{\rm max}\ge0.5$ \Msun), $\epsilon_{\rm sfe}$ is the star formation efficiency with a 
fiducial value of 30\%, and $M_{\rm clump}$ is the clump mass. For $m_{\rm max}$ = 8 \Msun, the necessary 
clump mass to form a high-mass star is 320 \Msun.

 \section{Additional figures}
\label{appimages}

Figure~\ref{four_rad_mass} gives more details on Figure~\ref{four_plots}. In Figure~\ref{four_rad_mass}, the data is color-coded by clump to show that 
the correlation is found per individual clump. 

\begin{figure*}
\begin{center}
\includegraphics[angle=0,scale=0.65]{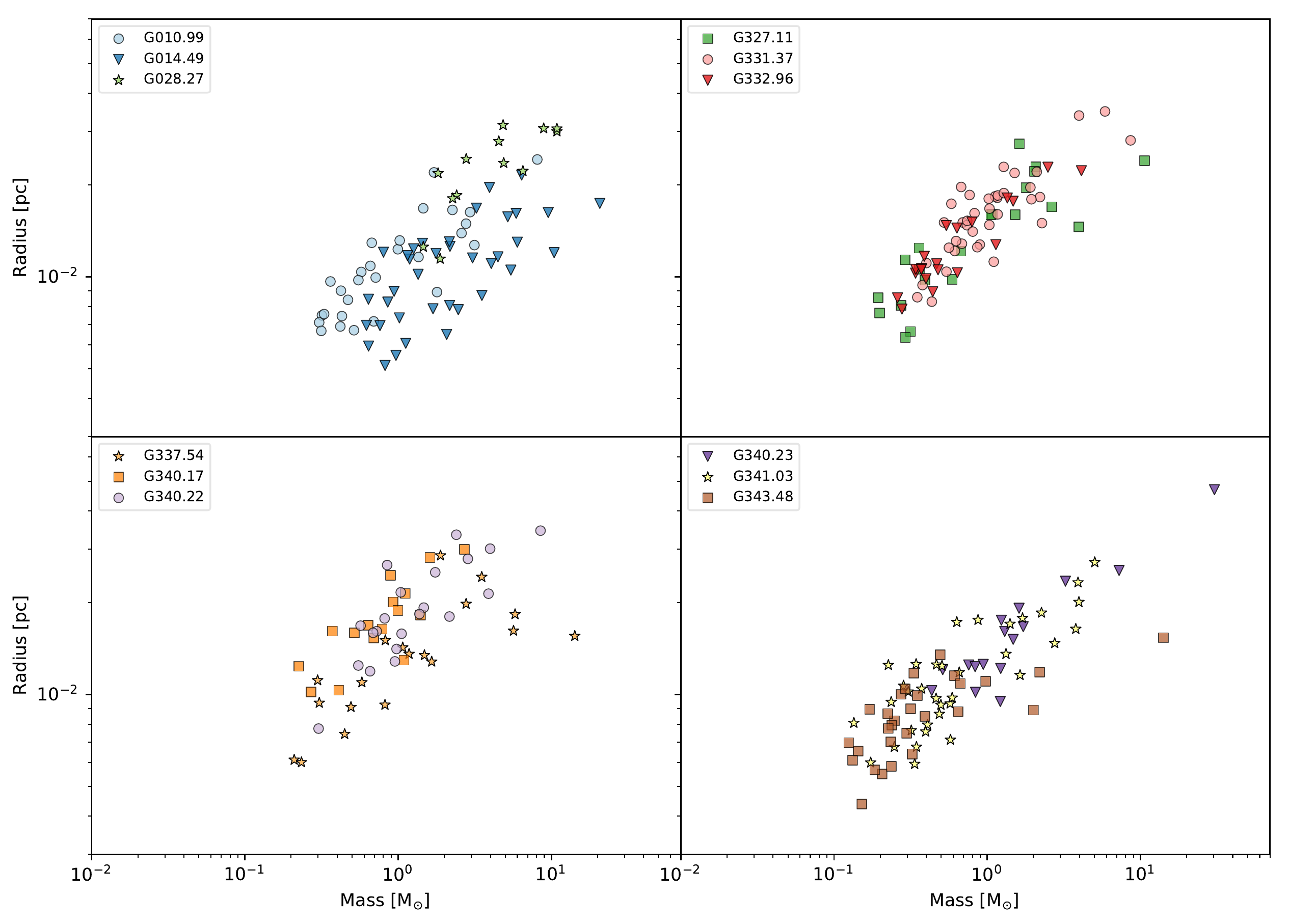}
\end{center}
\caption{Core radius against the core mass color-coded by clump. The purpose of these scatter plots is to show that for all cores embedded in a clump, i.e., at the 
same distance, the core radius correlates with the core mass.}
\label{four_rad_mass}
\end{figure*}

 On this work, core temperatures have been assumed to be the same as their host clump. Figure~\ref{four_plots} shows the distribution of core properties and differences
  can be seen among the evolutionary stages. However, protostellar cores are likely warmer and differences in Figure~\ref{four_plots} could be produced 
 by the assumed temperature. In Figure~\ref{four_30K}, we test the effect of temperature by assuming 30 K for protostellar cores. Differences in cores at different evolutionary stages
  almost disappear.  
 
\begin{figure*}
\begin{center}
\includegraphics[angle=0,scale=0.75]{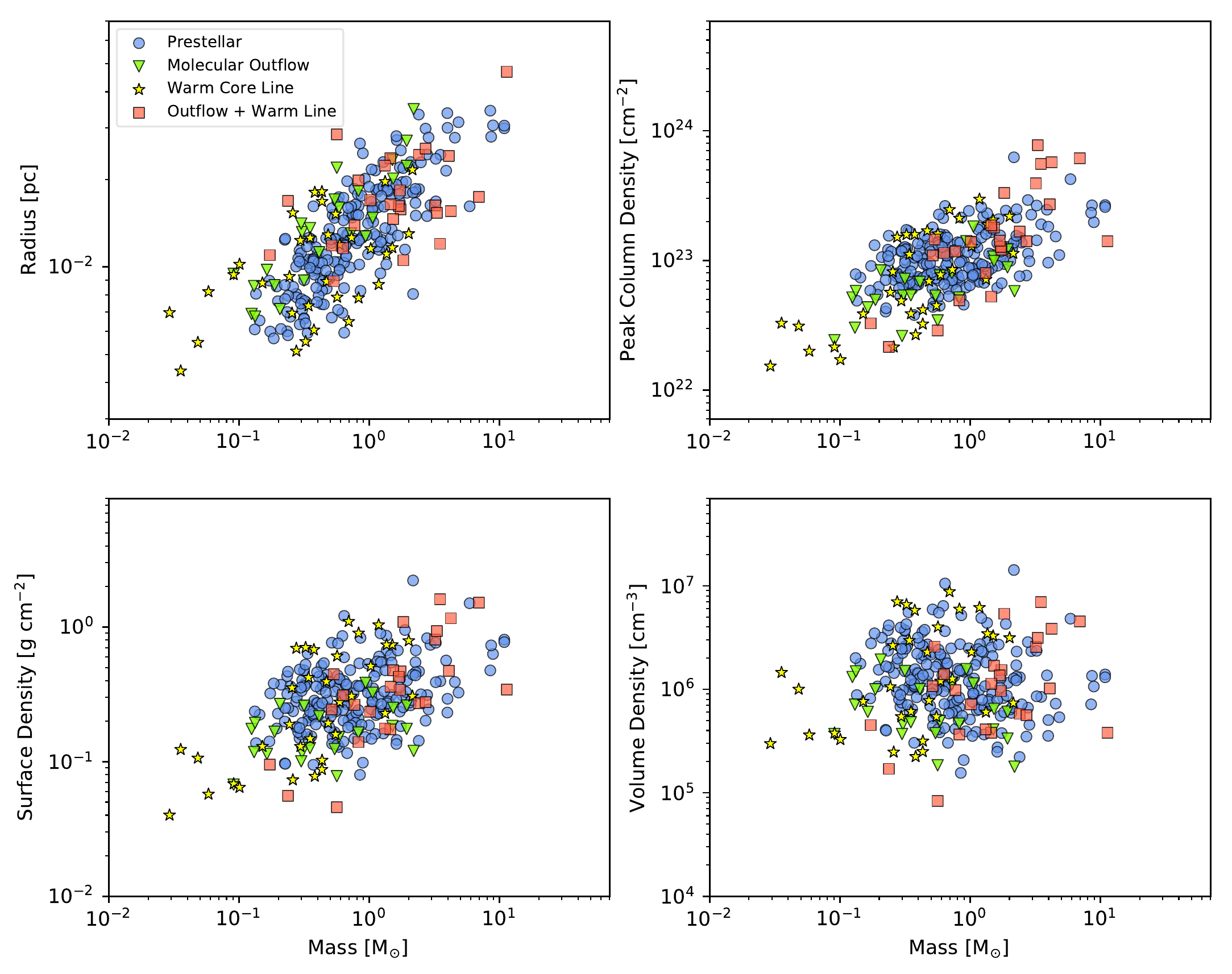}
\end{center}
\caption{Radius, peak column, surface density, and volume density of cores against the core mass color-coded by protostellar activity (prestellar, molecular outflow only, warm core line only, and 
both protostellar indicators; see Section~\ref{evo_stage}). The purpose of these scatter plots is to show the distribution of core properties by increasing the protostellar core temperatures to 30 K, which 
is about a factor 2 larger than the originally assumed clump temperature.}
\label{four_30K}
\end{figure*}

\end{document}